\documentclass[11pt,a4paper]{article}
\usepackage{jheppub}

\usepackage{bm}
\usepackage{wrapfig}

\usepackage{subcaption}
\usepackage{bm}
\usepackage{physics}

\usepackage{csquotes}
\numberwithin{equation}{section}
\usepackage{appendix}
\usepackage{bbold}
\usepackage{MnSymbol}
\usepackage{framed}
\usepackage{mdframed}
\usepackage{float}

\def\be#1\ee{\begin{align}#1\end{align}}
\def\pint{-\hskip-0.41cm \int}

\renewcommand{\Re}{\text{Re}\,}
\renewcommand{\Im}{\text{Im}\,}

\title{Injecting the UV into the Bootstrap: Ising Field Theory}
\author[a,b]{Miguel Correia,}
\author[b]{João Penedones}
\author[b]{and Antoine Vuignier}
\affiliation[a]{CERN, Theoretical Physics Department,\\
CH-1211 Geneva 23, Switzerland}
\affiliation[b]{Fields and Strings Laboratory, Institute of Physics, Ecole Polytechnique Federale de Lausanne (EPFL),\\
CH-1015 Lausanne, Switzerland}

\abstract{We merge together recent developments in the S-matrix bootstrap program to develop a dual setup in 2 space-time dimensions incorporating scattering amplitudes of massive particles and matrix elements of local operators. In particular, the stress energy tensor allows us to input UV constraints on IR observables in terms of the central charge $c_{UV}$ of the UV Conformal Field Theory. We consider two applications: (1) We establish a rigorous lower bound on $c_{UV}$ of a class of $\mathbb{Z}_2$ symmetric scalar theories in the IR (including $\phi^4$); (2) We target Ising Field Theory by, first, minimizing $c_{UV}$ for different values of the magnetic field and, secondly, by determining the allowed range of cubic coupling and one-particle form-factor for fixed $c_{UV} = 1/2$ and magnetic field. 

\begin{figure}[h]
    \centering
    \includegraphics[scale=0.57]{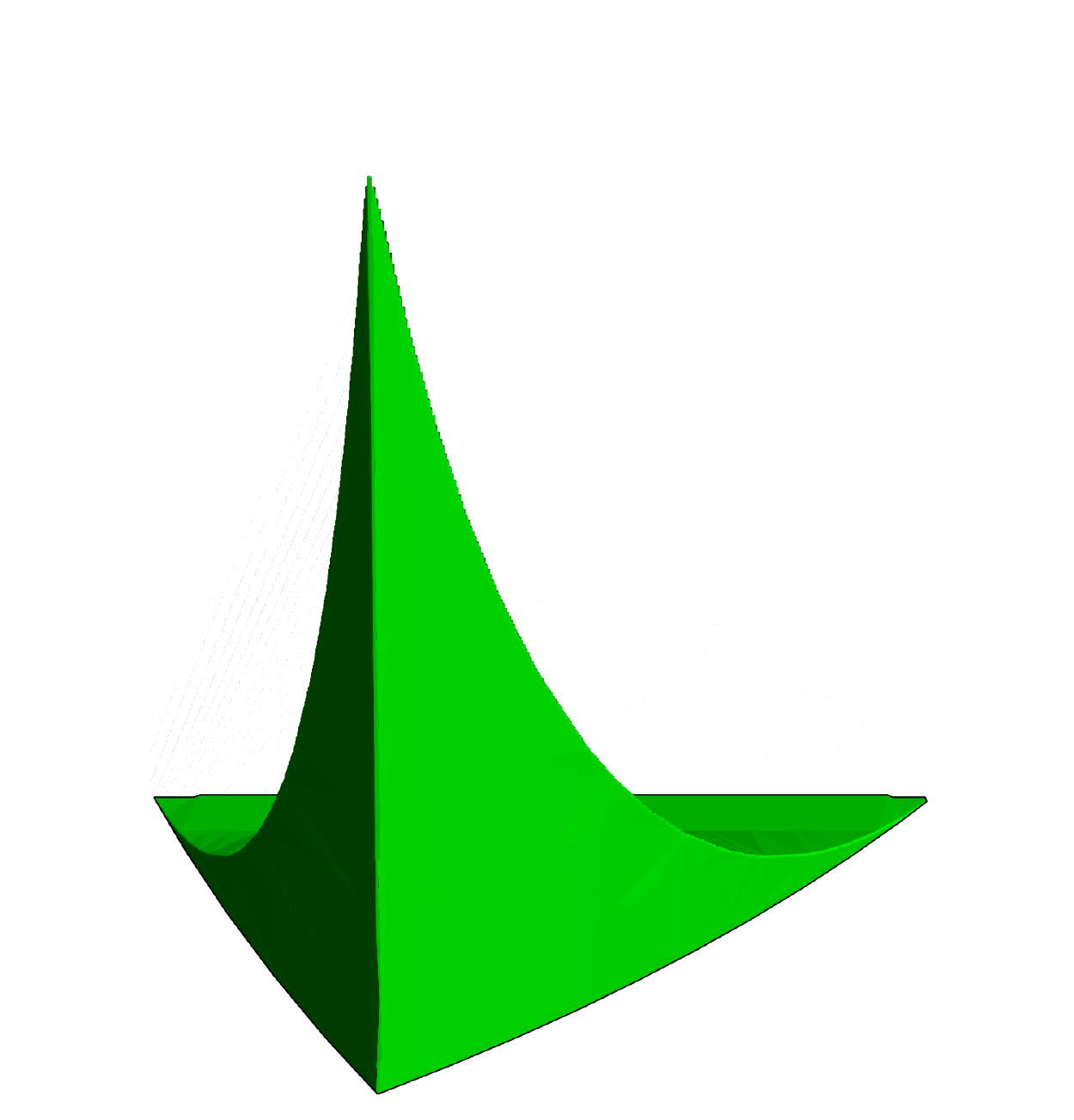}
    \captionsetup{labelformat=empty}
    \caption{\emph{In this pyramid lies Ising Field Theory}}
    \label{nicepyramid}
\end{figure}}

\begin{document}
\maketitle

\noindent

\numberwithin{equation}{section}

\setcounter{page}{1}
\renewcommand{\thefootnote}{\arabic{footnote}}
\setcounter{footnote}{0}

\setcounter{tocdepth}{2}

 \def\nref#1{{(\ref{#1})}}

\newpage

\parskip 5pt plus 1pt   \jot = 1.5ex

\newpage

\section{Introduction}
\par
The S-matrix bootstrap aims at constraining  scattering amplitudes in quantum field theory (QFT) based on general principles, such as unitarity, causality  and Lorentz invariance. The S-matrix program arose in the 60s as a way to tackle the problem of the strong interactions. However, imposing all the constraints in a predictive way proved difficult and, eventually, Quantum Chromodynamics (QCD) rose to prominence with the observation of deep inelastic scattering and the discovery of asymptotic freedom, which allowed the use of Feynman perturbation theory at high energies.
\par
Today, finding efficient ways to compute observables in low energy QCD and, more generally, in non-perturbative QFTs, remains an unsolved problem at large. Besides the S-matrix bootstrap, which has seen a revival in recent years, other modern techniques include Monte-Carlo simulations of lattice discretizations, Hamiltonian truncation and tensor networks.  The latter techniques typically involve an UV cutoff whose extrapolation to infinity is a non-trivial computational problem, whereas the S-matrix bootstrap is directly set up in the continuum, as required by Poincaré invariance.
\par
The primal approach to the S-matrix bootstrap constructs an explicitly analytic and crossing-symmetric ansatz for the amplitude and then constrains its parameters by imposing unitarity on the physical scattering region \cite{Paulos:2016but,Doroud:2018szp,Paulos:2017fhb,He:2018uxa,Cordova:2018uop,Guerrieri:2018uew,Homrich:2019cbt,EliasMiro:2019kyf,Paulos:2018fym,Bercini:2019vme,Guerrieri:2020bto,Hebbar:2020ukp,Guerrieri:2021ivu,Chen:2021pgx,a_anomaly,Chen:2022nym,Miro:2022cbk}. In practice, the infinite-dimensional space of amplitudes is truncated to finite dimension $N$ (the number of parameters in the ansatz).  By maximizing a given observable (e.g. the residue of a pole) one directly explores the space of amplitudes from the ``inside" with the true boundary of the allowed space for the observable (presumably) reached asymptotically, i.e. as $N$ is taken to infinity. This means that the primal approach is incapable of producing rigorous bounds at finite $N$.
\par
Conversely, the dual approach, as the name suggests, approaches the boundary of the allowed space from the ``outside" by excluding disallowed regions of parameter-space using a finite number $N$ of parameters. Therefore, for each $N$, there is a rigorous bound on the observable. As $N$ is increased the excluded region becomes larger, and the bound becomes tighter.  Dual formulations of the S-matrix bootstrap were first developed in 70s \cite{LOPEZ1975358,Lopez:1975wf,LOPEZ1975437,BONNIER197563,Lopez:1976zs} and recently revived in $d=2$ \cite{Cordova:2019lot,Guerrieri_2020,Kruczenski:2020ujw,EliasMiro:2021nul} and in $d = 4$ \cite{He:2021eqn,Guerrieri:2021tak}. Reference \cite{Guerrieri:2021tak}, in particular, showed that the dual approach can be formulated as a linear optimization problem amenable to implementation in SDPB  \cite{simmonsduffin2015semidefinite,landry2019scaling}. 
\par
The S-matrix, which dictates how asymptotic one-particle states scatter, is an IR observable. Nonetheless, if the  S-matrix originates from a UV complete QFT,\footnote{A UV complete QFT can be defined non-perturbatively as a CFT in the UV deformed by relevant deformations that trigger a renormalization group (RG) flow to the IR. The CFT in the IR is assumed to be empty, such that the QFT has a mass gap and the S-matrix is a well-defined object.} knowledge of the UV conformal field theory (CFT) from which the QFT flows from should further constrain the S-matrix. 
This was the idea behind \cite{Karateev_2020} which, besides scattering states, considered states given by the action of local operators, such as the stress-energy tensor, on the vacuum.  Concretely, in $d=2$, UV information can be included via the $c$-sum rule \cite{PhysRevLett.60.2709} which relates the spectral density of the trace of the stress-energy tensor to $c_{UV}$, the central charge of the UV CFT,

\begin{equation}
    c_{UV} = 12\pi \int_{m^2 \,>\, 0}^\infty \frac{\rho(s)}{s^2} \; ds \,.
\end{equation}

\par

\par

Here, we make the next logical step in this story. We merge these ideas together to develop a dual bootstrap approach in $d = 2$ that encompasses S-matrix elements, form factors and spectral densities. Our method, which is described in section \ref{sec:dual}, produces a linear optimization problem that can be tackled with SDPB and which, moreover, converges appreciably faster than the primal approach of \cite{Karateev_2020}. 
\par
With our method we can address the following question:
\emph{Given a gapped unitary QFT with no bound states how small can the central charge $c_{UV}$ of the UV CFT be?} Our method outputs $c_{UV}^{(min)} = 1/2$, the central charge of the Ising CFT, which is the smallest among the unitary conformal minimal models \cite{PhysRevLett.52.1575}. Indeed, our optimal S-matrix is given by $S(s) = - 1$ corresponding to a free massive Majorana fermion in the IR, which can originate from the pure thermal deformation of the Ising CFT.
\par
We can refine the previous question by fixing a parameter $\Lambda$ in the IR. For example: the amplitude at the crossing-symmetric point, $T(s=2m^2) = -\Lambda$, which plays the role of a `quartic coupling'. In this case we find a minimal $c_{UV}$ for a given $\Lambda$. The result is plotted below in fig. \ref{cUV_fixed_Lambda}.

\begin{figure}[h]
    \centering
\includegraphics[scale=0.8]{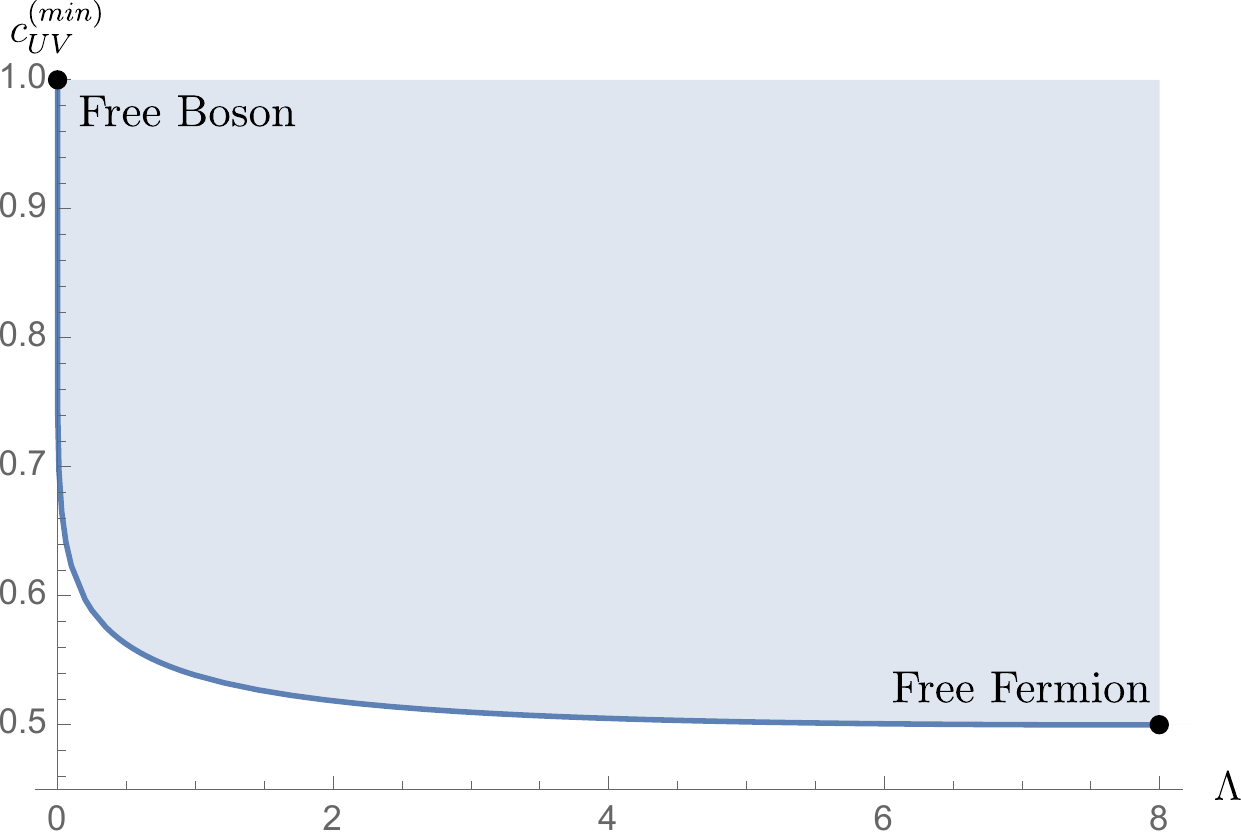}
    \caption{Rigorous lower bound on the UV central charge in $\mathbb{Z}_2$ symmetric QFTs with a single stable particle, which is $\mathbb{Z}_2$ odd.  The parameter $\Lambda$ is defined as (minus) the value of the amplitude at the crossing-symmetric point $s = 2m^2$. The lower bound on $c_{UV}$ goes from $c_{UV}^{(min)} = 1$ at  $\Lambda = 0$, where the S-matrix, form factor and spectral density become that of a free massive boson, to $c_{UV}^{(min)} = 1/2$ at the other end $\Lambda = 8$, where the matrix elements are that of a free massive fermion. }
    \label{cUV_fixed_Lambda}
\end{figure}
\par
In section \ref{sec:leaf}, we further refine the previous studies. Following \cite{Chen_2022}, we parameterize the theory space by $(\Lambda, \Lambda^{(2)})$, where $\Lambda^{(2)}$ is the second derivative of the amplitude at the crossing symmetric point. As in the case above, we assume no cubic coupling, i.e. no poles on the physical sheet for the amplitude nor for the form factor. In this way, we target a class of $\mathbb{Z}_2$ symmetric theories, in which $\phi^4$ theory is included. Using the numerical dual approach, we rigorously bound $c_{UV}$ across the allowed range of couplings $(\Lambda, \Lambda^{(2)})$ generating the 3-dimensional plots in fig. \ref{dual_leaf_cUV}. Remarkably, we also found analytical expressions (described in appendix \ref{sec:analytic_bootstrap}) that match the extremal solutions  (see e.g. figs. \ref{leaf_up} to \ref{leaf_int} for a comparison).

\par
In section \ref{sec:IFT} we consider the Ising Field Theory (IFT) which can be thought of as the QFT describing the $d=2$ Ising model near the critical point. Above the critical temperature $T>T_c$ and for zero magnetic field $h=0$ IFT reduces to the theory of a free massive Majorana fermion. For $h > 0$ the theory becomes fully interacting making it an appropriate playground for non-perturbative methods.\footnote{The parameter space of IFT is very rich (see e.g. \cite{Delfino:2003yr} for a short review). For $T > T_c$ and as $h$ increases the number of stable particles goes from 1 to 3 \cite{Zamolodchikov:1989fp,Zamo_2013,gabai2019smatrix}, and then jumps to 8 (5 resonances become stable) at the $h \to \infty$ integrable point where IFT becomes equivalent to the $E_8$ affine Toda theory \cite{Zamolodchikov:1989fp,1989PhLB..226...73H,Braden:1989bu}. For $T < T_c$ the spectrum of IFT consists of a tower of ``mesons" 
\cite{Fonseca:2006au}. Further interesting phenomena occur when $h$ is taken to be complex, namely the existence of Lee-Yang edge singularities \cite{Fonseca:2001dc}.}

We have at our disposal several pieces of information that we can use to target IFT:
\begin{enumerate}

     \item The UV central charge of IFT has the value $c_{UV} = 1/2$ given that IFT is, by definition, a deformation of the Ising CFT. Specifically, both relevant deformations are turned on, namely the operators conjugate to the temperature $(T - T_c)$ and the magnetic field $h$.
    \item At sufficiently low $h$ (but still outside the perturbative regime) there is only one stable particle in the spectrum \cite{Zamo_2013}. This particle self-interacts via a cubic interaction, meaning that both S-matrix and the two-particle form factor contain a pole at the location of the particle's own mass. 
    \item As shown in \cite{Zamo_2013,gabai2019smatrix}, the S-matrix has a real zero which slides towards the two-particle threshold as $h$ is increased.\footnote{If $h$ is increased past a certain value the zero makes its way across the two-particle branch cut and the associated pole pops up on the physical sheet, as the second lightest particle becomes stable. See e.g. \cite{gabai2019smatrix}. } The position of the zero provides a non-perturbative IR handle on the value of the magnetic field $h$. 
   
\end{enumerate}

We first implement points 2. and 3. and minimize the central charge $c_{UV}$ over a range of magnetic fields $h$ (parameterized by the position of the zero). We find that the lower bound on $c_{UV}$ drops below $c_{UV} = 1/2$ for non-zero $h$ (see fig. \ref{cuv2}). We then fix $c_{UV} = 1/2$ and find the allowed range of cubic coupling and one-particle form factors for a range of values of the magnetic field. We carve a 3-dimensional `pyramid' in these parameters inside which IFT must lie (see fig. \ref{fig:pyramid}). We conclude with section \ref{sec:conclusion} where we discuss our results in further detail and point out some potential future directions. 
\par
Let us now briefly outline the remaining appendices. In appendix \ref{sec:SG} we present a further application of our dual method targeting the Sine-Gordon model where we noticeably improve on the primal result of \cite{Karateev_2020}. Technical and numerical details regarding the dual optimization problems are collected in appendices \ref{sec:dual_problems} and \ref{sec:numerics}. Appendix \ref{sec:perturbation_theory} contains the perturbative computations of the one particle form factor and cubic coupling used to place IFT within the dual bounds (fig. \ref{fig:pyramid}). Finally, appendices \ref{sec:csum} and \ref{sec:normF} respectively review the c-sum rule and the normalization of the 2 particle form factor of the stress energy tensor.

\par

\section{Dual S-matrix and form factor Bootstrap}
\label{sec:dual}

In this section we develop our setup by recalling some definitions and results from standard massive QFT. We will be complete but concise, more details can be found in \cite{Karateev_2020, karateev2020twopoint}.

The first ingredient in our setup is the 2 to 2 scattering amplitude defined by
\begin{equation}
   _{out} \braket{p_1 p_2}{k_1 k_2}_{in} \equiv (2\pi)^2 \delta^{(2)}(k_1+k_2-p_1-p_2) \mathcal{N}_2S(s), \qquad 
    \mathcal{N}_2 \equiv 2 \sqrt{s}\sqrt{s-4m^2},
    \label{S}
\end{equation}
with $s=-(p_1+p_2)^2$.
We also define $\mathcal{T}(s)$, the interacting part of the scattering amplitude, by 
\begin{equation}
\label{eq:ST}
    S(s) \equiv 1+i \mathcal{N}_2^{-1}\mathcal{T}(s).
\end{equation}
We are also interested in some scalar local operator $\mathcal{O}(x)$, which leads us to consider its $n$ particles form factors
\begin{equation}
     \mathcal{F}_n^\mathcal{O}(p_1,p_2,..,p_n) \equiv _{out}\bra{ p_1 p_2...p_n} \mathcal{O}(0) \ket{0},
      \label{F}
\end{equation}
and its spectral density
\begin{equation}
    2\pi \rho_\mathcal{O}(s) \equiv \int d^2x e^{-ipx} \bra{0} \mathcal{O}^\dagger(x) \mathcal{O}(0) \ket{0},
    \label{rho}
\end{equation}
where we used Lorentz invariance to write $\rho_\mathcal{O} = \rho_\mathcal{O}(s=-p^2) $.
\par

We focus on massive QFTs whose Hilbert space $\mathcal{H}$ is spanned by asymptotic multi particle states $\ket{p_1,...,p_n}$ with completeness relation
\begin{equation}
\label{completenessH}
\begin{split}
    \mathbb{1}_{\mathcal{H}} &= \sum_{n=0}^\infty \frac{1}{n!}\int \frac{d \bm{p}_1}{ (2\pi) 2 E_{\bm{p}_1}}...\frac{d\bm{p}_n}{ (2\pi) 2 E_{\bm{p}_n}} \ket{p_1...p_n}_{in}\ _{in}\bra{p_1...p_n} \equiv \sumint \ket{p_1...p_n}_{in}\ _{in}\bra{p_1...p_n},
\end{split}
\end{equation}
 where $E_{\bm{p}} \equiv \sqrt{\bm{p}^2+m^2}$ and $\bm{p}$ is the spatial part of the 2-momentum $p$.
The complete set of states can be inserted in the two point function \eqref{rho} to get the relation between the spectral density and the form factors
\begin{equation}
    2\pi \rho_\mathcal{O}(s) = \sumint (2\pi)^2 \delta^{(2)}(p-p_n) |\mathcal{F}_n^\mathcal{O}|^2,
    \label{SD_FF}
\end{equation}
where $p_n \equiv p_1+...+p_n$ is the total momentum of the form factor. Note that the first contribution is a delta function at $s=m^2$ and the second contribution starts at $s=4m^2$. Explicitly 
\begin{equation}
    \rho_{\mathcal{O}}(s) = |\mathcal{F}_1^\mathcal{O}|^2 \delta(s-m^2) +  \frac{|\mathcal{F}_2^\mathcal{O}(s)|^2}{2\pi \mathcal{N}_2} \theta(s-4m^2) + ...
    \label{rho_parts}
\end{equation}

In this work we are interested in the trace of the stress energy tensor, so we introduce the simplified notation
\begin{equation}
    \mathcal{F}(s) \equiv \mathcal{F}_2^\Theta(s), \qquad \rho(s) \equiv \rho_\Theta(s),
\end{equation}
where the $s$ dependence of the 2 particle form factor comes from Lorentz invariance.
The normalization of the stress energy tensor acting on one particle states implies the following normalization of the 2 particle form factor (see appendix \ref{sec:normF})
\begin{equation}
    \mathcal{F}(s=0) = -2m^2.
\end{equation}

The value of the UV central charge can be computed from the spectral density of $\Theta$ as \cite{Karateev_2020} (in alternative, see appendix  \ref{sec:csum})
\begin{equation}
    c_{UV} = 12\pi \left( m^{-4} |\mathcal{F}_1^\Theta|^2 + \int_{4m^2}^\infty ds \frac{\rho(s)}{s^2} \right).
\end{equation}

\begin{figure}[h]
\centering
\includegraphics[scale=1.4]{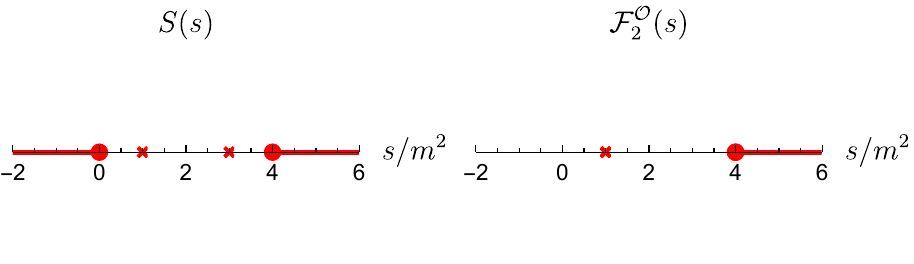}
\caption{Complex planes of the two particles form factor $\mathcal{F}_2^\mathcal{O}(s)$ (right) and the scattering amplitude $S(s)$ (left) in the presence of an asymptotic particle of mass $m = 1$. The red crosses symbolize the presence of a pole and the red lines represent the branch cuts.}
\label{CPFF}
\end{figure}

Let us now turn to the analytic structure of these functions depicted on Figure \ref{CPFF}.  
Assuming the existence of only one massive particle of mass $m$, the scattering amplitude is analytic except for a pole at $s=m^2$ and a cut starting from the two particles production threshold $s=4m^2$, and also crossing symmetric. This can be written through the dispersion relation\footnote{In writing this dispersion relation, we assumed that $ \lim_{|s|\to \infty} \frac{\mathcal{T}(s)}{s} =0 $. Our dual optimization problem will however not depend on this assumption, i.e. the behavior of the amplitude at infinity is unconstrained. See equation \eqref{eq:Lag} and footnote \ref{foot6}.}
\begin{equation}
\begin{split}
    \mathcal{T}(s)-\mathcal{T}(2m^2) &= -g^2\left( \frac{1}{s-m^2} + \frac{1}{3m^2-s}-\frac{2}{m^2} \right) \\&+ \int_{4m^2}^\infty \frac{dz}{\pi} \text{Im} \mathcal{T}(z)\left( \frac{1}{z-s} + \frac{1}{z+s-4m^2}- \frac{2}{z-2m^2} \right).
    \end{split}
\end{equation}
The 2-particle form factor $\mathcal{F}(s)$ has a similar analytic structure except that it does not satisfy crossing. The dispersion relation reads\footnote{We have used the fact that both functions are real analytic functions, i.e. $\mathcal{T}^*(s)=\mathcal{T}(s^*)$ and $\mathcal{F}^*(s)=\mathcal{F}(s^*)$ , which can be traced back to LSZ and CPT invariance \cite{olive}.}
\begin{equation}
\mathcal{F}(s)-\mathcal{F}(0) = -g_F\left( \frac{1}{s-m^2} + \frac{1}{m^2} \right)+ \int_{4m^2}^\infty \frac{dz}{\pi} \text{Im} \mathcal{F}(z)\left( \frac{1}{z-s} - \frac{1}{z} \right).
\label{FF_dispersion}
\end{equation}

\par
As shown in \cite{Karateev_2020}, the form factor residue $g_F$ is given by 
\begin{equation}
    g_F = g \mathcal{F}_1^\Theta,
\end{equation}
where $g$ is the (square root of the) residue of the scattering amplitude.

For future convenience, we define the analyticity and crossing constraints for the S-Matrix and form factor, denoted respectively by $\mathcal{A}_\mathcal{T}(s)=0$ and $\mathcal{A}_\mathcal{F}(s)=0$, with
\begin{equation}
\begin{split}
    \mathcal{A}_\mathcal{T}(s) &\equiv   \mathcal{T}(s)-\mathcal{T}(2m^2)  +g^2\left( \frac{1}{s-m^2} + \frac{1}{3m^2-s}-\frac{2}{m^2} \right) \\ &- \int_{4m^2}^\infty \frac{dz}{\pi} \text{Im} \mathcal{T}(z)\left( \frac{1}{z-s} + \frac{1}{z+s-4m^2}- \frac{2}{z-2m^2} \right), \\
    \mathcal{A}_\mathcal{F}(s) &\equiv \mathcal{F}(s)-\mathcal{F}(0)  +g_F\left( \frac{1}{s-m^2} + \frac{1}{m^2} \right)- \int_{4m^2}^\infty \frac{dz}{\pi} \text{Im} \mathcal{F}(z)\left( \frac{1}{z-s} - \frac{1}{z} \right).
\end{split}
\end{equation}

Following \cite{Karateev_2020}, the unitarity constraints in presence of local operators such as $\Theta(x)$ can be implemented by defining 3 states 
\begin{equation}
    \begin{split}
        \ket{\Psi_1} &\equiv \ket{p_1,p_2}_{in},\\
        \ket{\Psi_2} &\equiv \ket{p_1,p_2}_{out}, \\
        \ket{\Psi_3} &\equiv m^{-1} \int d^2x e^{i (p_1+p_2)x} \Theta(x) \ket{0}.
    \end{split}
\end{equation}
We now define a $3\times 3$ matrix by taking inner products between those states and extracting an overall delta function, which reads 
\begin{equation}
    B_{ij} \times (2\pi)^2\delta^{(2)}(p_1+p_2-p_1'-p_2') \equiv \braket{\Psi_i'}{\Psi_j}.
\end{equation}
Using equations \eqref{S}, \eqref{F}, \eqref{rho} and the definition of $B$ as inner products between states, the unitarity of the theory imposes the semi positive definite constraint\footnote{Here we used the relation $_{in}\!\bra{p_1,p_2} \Theta(0) \ket{0} = \bra{0} \Theta^\dagger(0) \ket{p_1,p_2}_{out} = \mathcal{F}^*(s) $ that comes from CPT invariance \cite{Karateev_2020}.}
\begin{equation}
    B(s) \equiv \begin{pmatrix} 1 & S^* & m^{-1} \omega \mathcal{F}^* \\S & 1 & m^{-1} \omega \mathcal{F} \\ m^{-1} \omega \mathcal{F} & m^{-1} \omega \mathcal{F}^* &  2\pi m^{-2} \rho  \\ \end{pmatrix} \succcurlyeq 0, \qquad \omega \equiv \mathcal{N}_2^{-1/2}.
\end{equation}
The problem we want to solve is the minimization of some parameter, say the central charge $c_{UV}$, under the bootstrap constraints which can be implemented via Lagrange multipliers.
More precisely, we write
\begin{equation}
    c_{UV} \ge 
     \underset{\mathcal{T},\mathcal{F},\rho}{\rm inf}\ 
     \underset{w,\mathbb{\Lambda}\succcurlyeq 0}{\rm sup}
    \mathcal{L}
    \ge 
    \underset{w,\mathbb{\Lambda}\succcurlyeq 0}{\rm sup}
    \  \underset{\mathcal{T},\mathcal{F},\rho}{\rm inf}
    \mathcal{L}
\end{equation}
\begin{equation}
    \mathcal{L} = c_{UV}(\mathcal{F}_1^\Theta,\rho) + \int_{4m^2}^\infty ds \left[ w_\mathcal{T}(s) \mathcal{A}_\mathcal{T}(s) + w_\mathcal{F}(s) \mathcal{A}_\mathcal{F}(s) - \operatorname{Tr} \mathbb{\Lambda}(s) B(s) \right] + ...
    \label{L_general}
\end{equation}
where $\mathcal{A}_\mathcal{T}(s)$ and $\mathcal{A}_\mathcal{F}(s)$ are the analyticity and crossing constraints, $w_\mathcal{T}$ and $w_\mathcal{F}$ are Lagrange multipliers, $\mathbb{\Lambda}$ is a hermitian and positive semidefinite $3\times 3$ matrix of Lagrange multipliers and "..." stands for any other constraint we would like to implement (e.g. fixing some parameter). 

The positive semidefiniteness of $\mathbb{\Lambda}$ is the direct generalisation of positiveness in the pure S- Matrix (non-linear) dual problem from \cite{Guerrieri_2020}. Here we are searching for the lower bound instead of the upper bound, which accounts for the sign difference in the trace term.

We parametrize $\mathbb{\Lambda}$ as 
\begin{equation}
    \mathbb{\Lambda} \equiv \begin{pmatrix} \lambda_1 & \lambda_4 & \lambda_6 \\ \lambda_4^* & \lambda_2 & \lambda_5 \\ \lambda_6^* & \lambda_5^* & \lambda_3 \end{pmatrix}.
\end{equation}

With the dual approach, we will first extremize \eqref{L_general} over the primal variables $\mathcal{T}, \mathcal{F}$ and $\rho$ analytically to get the dual Lagrangian, and then over the dual variables $w_\mathcal{T}, w_\mathcal{F}$ and $\lambda_i$, $i=1,...,8$, numerically. The problem being linear in the dual variables can be tackled down with SDPB \cite{simmonsduffin2015semidefinite,landry2019scaling}.

Having the solution for the optimal dual variables we can get optimal primal variables if the duality gap closes, which is guaranteed because the primal problem is convex.\footnote{Since the primal problem can be implemented in SDPB \cite{simmonsduffin2015semidefinite,landry2019scaling} both the objective and the constraints must be convex. To ensure strong duality, i.e. that the duality gap closes, we must further require that there exists an interior point, i.e. a non-optimal primal solution. This is known as Slater's condition \cite{RePEc:cwl:cwldpp:80} (see appendix A of \cite{Guerrieri_2020} for a proof in the S-matrix language). Slater's condition can be shown to be satisfied by explicitly constructing a non-optimal primal solution, as it usually happens in the primal bootstrap at finite $N$ \cite{Karateev_2020}. Physically, moreover, the existence of such an interior point is guaranteed since we can always have inelastic S-matrices and form factors which do not saturate unitarity. }
Therefore, at the optimum we have $\operatorname{Tr} \mathbb{\Lambda}(s)  B(s)=0$ and, since $\mathbb{\Lambda}$ and $B$ are both semipositive definite, the product must vanish by itself\footnote{Indeed we have
\begin{equation}
\begin{split}
    \operatorname{Tr} \mathbb{\Lambda} \cdot B &=  \operatorname{Tr} \sqrt{\mathbb{\Lambda}}^\dagger \sqrt{\mathbb{\Lambda}} \cdot \sqrt{B} \sqrt{B}^\dagger = \operatorname{Tr} (\sqrt{\mathbb{\Lambda}}\sqrt{B})^\dagger (\sqrt{\mathbb{\Lambda}}\sqrt{B}) \equiv \lVert\sqrt{\mathbb{\Lambda}}\sqrt{B}  \rVert_F,
\end{split}
\end{equation}
where $  \lVert \cdot  \rVert_F$ is the Frobenius norm. It follows that $ \operatorname{Tr} \mathbb{\Lambda} \cdot B = 0 \Rightarrow \mathbb{\Lambda} B = 0$.}
\begin{equation}
    \mathbb{\Lambda}(s) B(s)=0.
\end{equation}

Solving this equation we get constraints between dual variables 
\begin{equation}
    \lambda_5 = \lambda_6^*, \qquad \lambda_1 = \lambda_2, \qquad 2\lambda_1 |\lambda_6|^2 - \lambda_3 \lambda_1^2 - 2 \operatorname{Re}(\lambda_4^* \lambda_6^2)+ \lambda_3|\lambda_4|^2 = 0.
\end{equation}
We would like to stay with entries that are linear in the dual variables in the matrix $\mathbb{\Lambda}$, and therefore we do not use the last constraint to eliminate one of the variables. 
\par
Since the duality gap closes we can find the optimal primal solution:
\begin{equation}
\label{primalfromdual}
    S = \frac{\lambda_6^* \lambda_1 - \lambda_6\lambda_4^*}{\lambda_6 \lambda_1 - \lambda_6^*\lambda_4}, \qquad \omega \mathcal{F} = \frac{|\lambda_4|^2-\lambda_1^2}{\lambda_6 \lambda_1 - \lambda_6^*\lambda_4}, \qquad 2\pi \rho = \frac{(|\lambda_4|^2-\lambda_1^2)^2}{|\lambda_6 \lambda_1 - \lambda_6^*\lambda_4|^2}.
\end{equation}

Note that we automatically saturate the unitarity bounds
\begin{equation}
\label{eq:unisat}
    |S|^2-1=0, \qquad 2\pi \rho - |\omega \mathcal{F}|^2 = 0, \qquad 2\pi \rho (1-|S|^2) - 2 |\omega \mathcal{F}|^2 + 2 \operatorname{Re}(\omega^2 \mathcal{F}^2 S^*) = 0.
\end{equation}
Watson's equation $\mathcal{F}/\mathcal{F}^* = S$ follows from using the first equation on the last equation.

\subsection{Example: Minimization of $c_{UV}$ for fixed quartic coupling $\Lambda$}
\label{sec:quartic}

We will now present an explicit and detailed example of a dual linear bootstrap formulation. We want to find the lower bound on the central charge $c_{UV}$ when there is only one particle that is $\mathbb{Z}_2$ odd and the scattering amplitude obeys $\mathcal{T}(2) = - \Lambda$. For simplicity we work in units where $m=1$. The Lagrangian reads
\begin{equation}
\label{eq:Lag1}
    \begin{split}
        \mathcal{L} = 12\pi  \int_4^\infty ds \frac{\rho(s)}{s^2} &+ \int_4^\infty ds w_{\mathcal{T}}(s)\left[ \mathcal{T}(s)-\mathcal{T}(2) - \int_4^\infty \frac{dz}{\pi} \text{Im} \mathcal{T}(z)\left( \frac{1}{z-s} + \frac{1}{z+s-4}- \frac{2}{z-2} \right) \right] \\
        &+ \int_4^\infty ds w_{\mathcal{F}}(s) \left[\mathcal{F}(s)-\mathcal{F}(0) - \int_4^\infty \frac{dz}{\pi} \text{Im} \mathcal{F}\left( \frac{1}{z-s} - \frac{1}{z} \right) \right]\\
        &- \int_4^\infty ds \operatorname{Tr}\mathbb{\Lambda}(s) \cdot B(s).
    \end{split}
\end{equation}
The subtractions are chosen so that we can easily implement the constraint $\mathcal{T}(2)=-\Lambda$ and the normalization $\mathcal{F}(0)=-2$.
\par
Following \cite{Guerrieri_2020} we define the dual scattering function
\begin{equation}
    \begin{split}
        &W_{\mathcal{T}}(s) \equiv \int_4^\infty \frac{dz}{\pi} w_{\mathcal{T}}(z) \left( \frac{1}{z-s} - \frac{1}{z+s-4} +\frac{2}{s-2}\right), \\
        & \operatorname{Im}W_\mathcal{T}(s) = w_\mathcal{T}(s), \qquad \operatorname{Re}W_\mathcal{T}(s) = -P\int_4^\infty \frac{dz}{\pi} w_{\mathcal{T}}(z) \left( \frac{1}{s-z} + \frac{1}{s-(4-z)} -\frac{2}{s-2} \right),
    \end{split}
    \label{WTZ2}
\end{equation}
and the dual form factor function 
\begin{equation}
        W_{\mathcal{F}}(s) \equiv \int_4^\infty \frac{dz}{\pi} w_{\mathcal{F}}(z)  \left(\frac{1}{z-s} + \frac{1}{s} \right).
        \label{dualWF}
\end{equation}
The analyticity constraints can then be considerably simplified. We get\footnote{Eq. \eqref{eq:Lag} can be derived directly without introducing $w_\mathcal{T}$ or $w_\mathcal{F}$. The analyticity constraint for $\mathcal{T}(s)$, say, can be written as $\oint W_\mathcal{T}(s) \mathcal{T}(s) ds = 0$  where $W_\mathcal{T}(s)$ is an analytic function, for an arbitrary closed cycle enclosing no singularity. Now, we blow up this contour and, seeing $W_\mathcal{T}(s)$ as infinitely many Lagrange multipliers (one for each $s$), we are free to choose $W_\mathcal{T}(s)$ so that the contour only picks up the contributions of $\mathcal{T}(s)$ we wish to constrain, which is the unitarity cut $s \geq 4$ and the value at $\mathcal{T}(2) = -\Lambda$. Therefore, we let $W_\mathcal{T}$ have a cut for $s \geq 4$, a pole at $s = 2$, and sufficiently fast decay at infinity (faster than $\mathcal{T}(s)$). Blowing up the contour will then lead to 
\begin{equation}
\label{eq:arcTW}
    0 = \oint \mathcal{T}W_\mathcal{T} \, ds = 4i\int_4^\infty \mathrm{Im} (\mathcal{T}W_\mathcal{T}) ds - 2\pi i\left[\mathrm{Res}_{s=2} W_\mathcal{T} \right] \mathcal{T}(2) 
    .
\end{equation}

Note that the residue on the pole is fixed by blowing up $0 = \oint W_\mathcal{T} ds = -2\pi i \mathrm{Res}_{s=2} W_\mathcal{T} + 4i \int_4^\infty \Im W_\mathcal{T} ds$. Including crossing symmetry, repeating the argument for the form factor  will lead us to eq. \eqref{eq:Lag}. 
\label{foot6}
}
\begin{equation}
\label{eq:Lag}
    \begin{split}
        \mathcal{L} =12\pi  \int_4^\infty ds \frac{\rho(s)}{s^2} &+ \int_4^\infty ds \left[ \operatorname{Im}(\mathcal{T} W_\mathcal{T}) + \operatorname{Im}(\mathcal{F} W_\mathcal{F})+ 2 \operatorname{Im}W_\mathcal{F} +\Lambda \operatorname{Im}W_\mathcal{T}  - \operatorname{Tr} \mathbb{\Lambda}\cdot B\right]
    \end{split}
\end{equation}

We are now ready to eliminate the primal variables. Varying with respect to $\mathcal{F}, \mathcal{T}$ and $\rho$ we get
\begin{equation}
    \lambda_4 = -\frac{\mathcal{N}_2}{2}W_\mathcal{T}, \qquad \lambda_6 = -i\frac{\sqrt{\mathcal{N}_2}}{4}W_\mathcal{F}, \qquad \lambda_3 = \frac{6}{s^2}.
\end{equation}
The dual Lagrangian then reads
\begin{equation}
    \mathcal{L} =  \int_4^\infty ds \left( -2\lambda_1 + 2 \operatorname{Im}W_\mathcal{F} -\mathcal{N}_2 \operatorname{Re}W_\mathcal{T} + \Lambda \operatorname{Im}W_\mathcal{T}\right).
\end{equation}

All the primal variables are eliminated and we are ready to extremize over dual variables. 
We can finally formulate the dual problem that can be implemented in SDPB: 

\begin{framed}
\underline{Dual Problem ($c_{UV}$ minimization)}
\begin{equation}
\underset{\{\lambda_1, W_\mathcal{T}, W_\mathcal{F}\}}{\text{Maximize}}\left[ \int_4^\infty ds \left( -2\lambda_1 + 2 \operatorname{Im}W_\mathcal{F} - \mathcal{N}_2 \operatorname{Re}W_\mathcal{T} + \Lambda \operatorname{Im}W_\mathcal{T}\right)\right]
\end{equation}
Subject to
\begin{equation}
    \begin{pmatrix}
    \lambda_1& \frac{\mathcal{N}_2}{2} W_\mathcal{T} & i \frac{\sqrt{\mathcal{N}_2}}{4} W_\mathcal{F} \\ \frac{\mathcal{N}_2}{2} W_\mathcal{T}^* & \lambda_1 & -i \frac{\sqrt{\mathcal{N}_2}}{4} W_\mathcal{F}^* \\ -i \frac{\sqrt{\mathcal{N}_2}}{4} W_\mathcal{F}^* &  i \frac{\sqrt{\mathcal{N}_2}}{4} W_\mathcal{F} & \frac{6}{s^2}
    \end{pmatrix} \succcurlyeq 0, \quad \forall s \in [4,\infty).
\end{equation}
\end{framed}

The result is given in figure \ref{cUV_fixed_Lambda}.

\section{Applications}
\label{sec:applications}

\subsection{
$\mathbb{Z}_2$ symmetric theories}
\label{sec:leaf}
A first application of our dual formalism is to explore the allowed region in the subspace spanned by $(\Lambda, \Lambda^{(2)},c_{UV})$ where $c_{UV}$ is the UV central charge and 
\begin{equation}
\label{eq:LL2}
    \Lambda \equiv -\mathcal{T}(2), \qquad \Lambda^{(2)} \equiv \lim_{s\to 2} \frac{\partial^2 }{\partial s^2} \mathcal{T}(s).
\end{equation}
Before considering the central charge we look at the space spanned by $\Lambda$ and $\Lambda^{(2)}$, which is commonly called the "leaf", that was already explored using a primal approach in \cite{Chen_2022} and also analytically in \cite{https://doi.org/10.48550/arxiv.2207.12448}. Exploring the same region with the dual formalism we get Figure $\ref{dual_leaf}$ on which we also plotted the analytical bounds derived in the appendix \ref{sec:analytic_bootstrap} and given by \eqref{eq:leaf_an} that we repeat here for convenience
\begin{equation}
    \Lambda^{(2)}_{-}(\Lambda) = \frac{1}{32} \Lambda^2, \qquad  \Lambda^{(2)}_{+}(\Lambda) = \frac{1}{32} (16 \Lambda -\Lambda^2), \qquad \Lambda \in [0,8],
    \label{analytic_leaf}
\end{equation}
where $\Lambda^{(2)}_+$ and $\Lambda^{(2)}_-$ denote respectively the analytical upper and lower bounds for $\Lambda^{(2)}$ for a given $\Lambda$.
We reproduce these bounds numerically with our dual approach purely applied to the S-matrix sector (see figure \ref{dual_leaf} below).
\begin{figure}[h]
    \centering
    \includegraphics[scale=0.9]{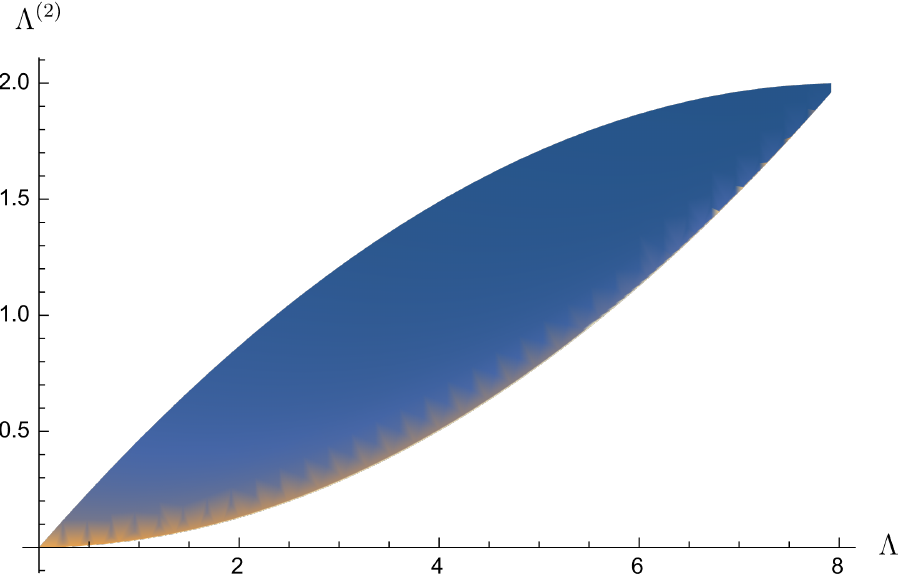}
    \caption{Allowed region in the $(\Lambda, \Lambda^{(2)}$) plane. The boundary is obtained from the dual bootstrap problem and matches perfectly the analytical bounds, with the color gradient corresponing to the value of the minimal central charge discussed below. Using the non linear dual approach given by eq.\eqref{nonlindual1} to \eqref{nonlindual2}, the numerics for the points on the boundary were simple enough to be done on Mathematica with only $N=2$ parameters in the Ansatz. }
    \label{dual_leaf}
\end{figure}

We now minimize the UV central charge $c_{UV}$ for a given value of  $(\Lambda, \Lambda^{(2)})$ , which we do over the allowed region. The minimal $c_{UV}$ is represented in color in fig. \ref{dual_leaf} over the allowed parameter space for $(\Lambda, \Lambda^{(2)})$ and also in 3D in fig. \ref{dual_leaf_cUV} below.

\begin{figure}[h]
\centering
\begin{subfigure}{0.49\textwidth}
\centering
\includegraphics[scale=0.75]{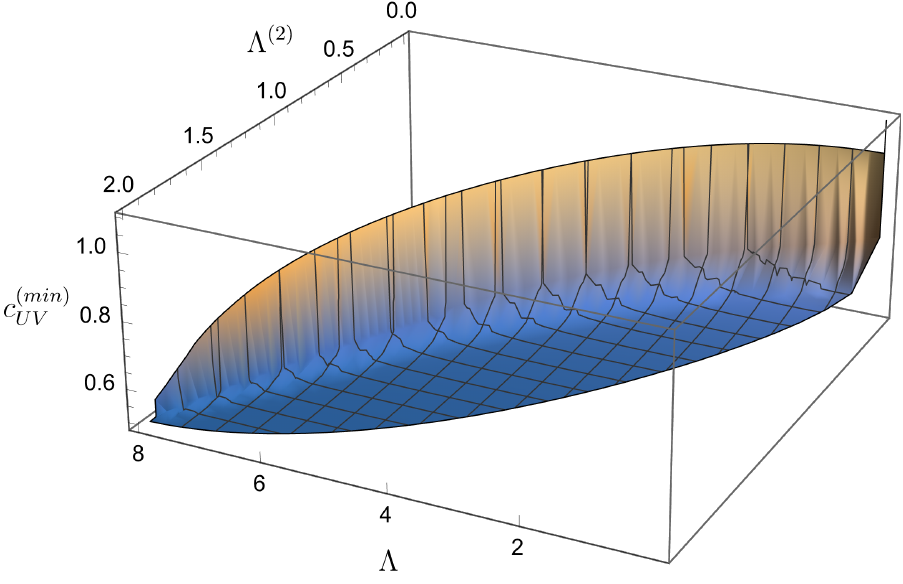}
\end{subfigure}
\hfill
\begin{subfigure}{0.49\textwidth}
\centering
\includegraphics[scale=0.75]{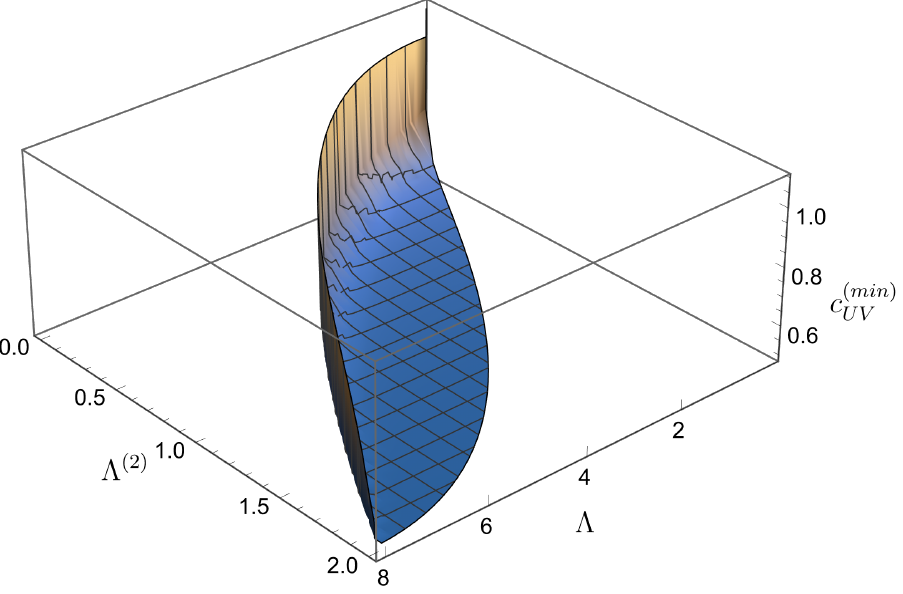}
\end{subfigure}
\caption{Allowed region in the $(\Lambda, \Lambda^{(2)}, c_{UV})$ space from different angles. We used $N=50$. We find perfect agreement with the analytical bootstrap result which assumes that the optimal S-matrix minimizes $\Lambda^{(4)} = \lim_{s\to 2} \frac{\partial^4 }{\partial s^4} \mathcal{T}(s)$ (see appendix \ref{sec:analytic_bootstrap}).}
\label{dual_leaf_cUV}
\end{figure}

On the edge corresponding to the lower bound $\Lambda^{(2)}_-$, the numerical bounds on the central charge $c^{(min)}_{UV}(\Lambda, \Lambda^{(2)})$ at the tips are 
\begin{equation}
    c_{UV}^{(min)}(0,0) = 0.99999..., \qquad c_{UV}^{(min)}(8,2) = 0.49985..,
\end{equation}
corresponding to the central charges of the free boson and free fermion that are respectively 1 and 1/2. We now check that we also recover the scattering amplitudes, form factors and spectral densities of these theories. The analytical results for the free Majorana fermion can be found in \cite{Mussardo:1281256} and those for the free boson can be found in \cite{Karateev_2020}. We repeat them here for convenience, they  are 
\begin{equation}
    \begin{split}
        & S(s)=1, \qquad \mathcal{F}(s) = -2, \qquad \rho(s) = \frac{1}{\pi \sqrt{s} \sqrt{s-4}}, \qquad \text{(free boson)},\\
        & S(s)=-1, \qquad \mathcal{F}(s) = - \sqrt{4-s}, \qquad \rho(s) = \frac{\sqrt{s-4}}{4\pi \sqrt{s} }, \qquad \text{(free fermion)}.
    \end{split}
\end{equation}

\begin{figure}[h]
    \centering
    \includegraphics[scale=0.6]{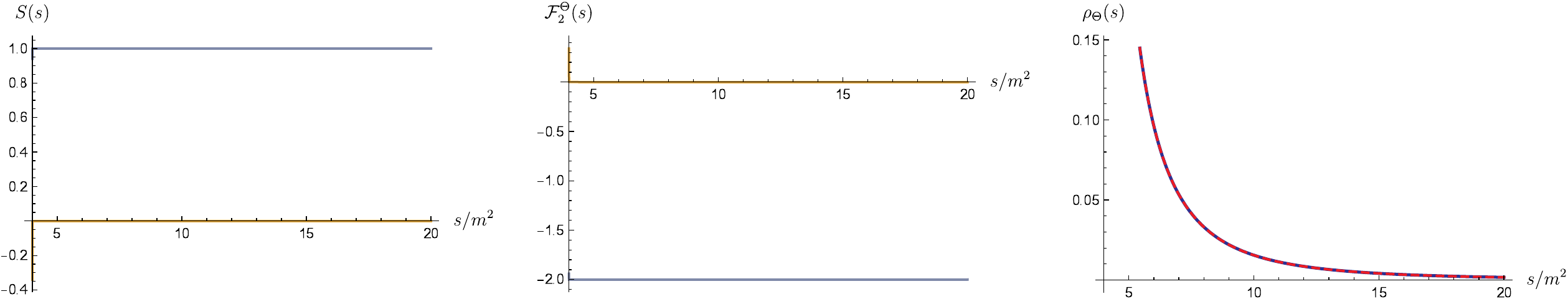}
    \caption{Scattering amplitude $S(s)$ (on the left), form factor $\mathcal{F}(s)$ (in the middle) and spectral density $\rho(s)$ (on the right) at the free boson point ($\Lambda =0, \Lambda^{(2)}=0)$). The blue lines are the real parts, the orange lines are the imaginary parts, and the dashed red line is the analytical solution for the free boson. We used $N=50$.} 
    \label{boson_dual_leaf_cUV}
\end{figure}

\begin{figure}[h]
    \centering
    \includegraphics[scale=0.6]{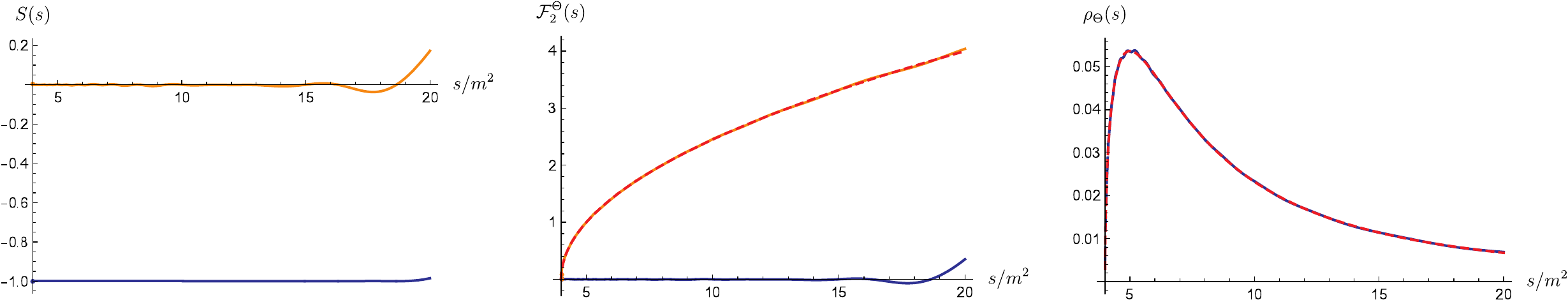}
    \caption{Scattering amplitude $S(s)$ (on the left), form factor $\mathcal{F}(s)$ (in the middle) and spectral density $\rho(s)$ (on the right) at the free fermion point ($\Lambda =8, \Lambda^{(2)}=2)$). The blue lines are the real parts, the orange lines are the imaginary parts, and the dashed red line is the analytical solution for the free fermion. We used $N=50$. }
    \label{fermion_dual_leaf_cUV}
\end{figure}

The numerical results for these quantities at the tips of the leaf are plotted on Figure $\ref{boson_dual_leaf_cUV}$ (at $\Lambda =0$) and Figure $\ref{fermion_dual_leaf_cUV}$ (at $\Lambda = 8)$. We observe a nice convergence of our numerics toward the analytical results.

Even with $N=50$ we observe some wiggles in the functions, even if the optimal bound for $c_{UV}$ is reached quickly with a lower $N$ (see next section for a convergence study and comparison with the primal). We believe that it is due to the non linear map between primal variables and dual variables (\ref{primalfromdual}) and the fact that our ansätze for the dual functions might not be optimal.

\begin{figure}[h]
    \centering
    \includegraphics[scale=0.9]{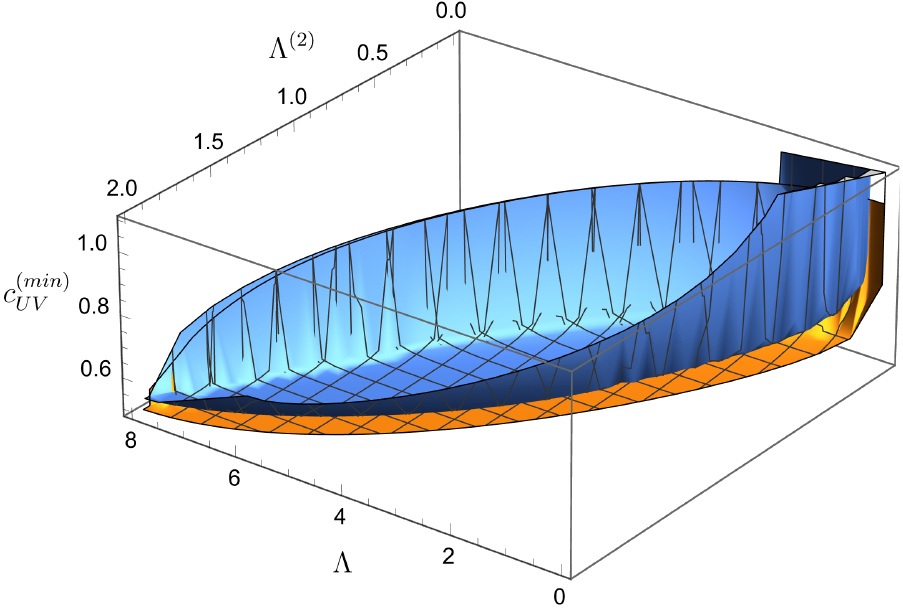}
    \caption{Minimal central charge in the $(\Lambda, \Lambda^{(2)}, c_{UV}$) space computed with the dual approach (in orange) and the primal (in blue). We used $N=50$. }
    \label{dualprimal_leaf_cUV}
\end{figure}

\begin{figure}[h]
\centering
\begin{subfigure}{0.49\textwidth}
\centering
\includegraphics[scale=0.8]{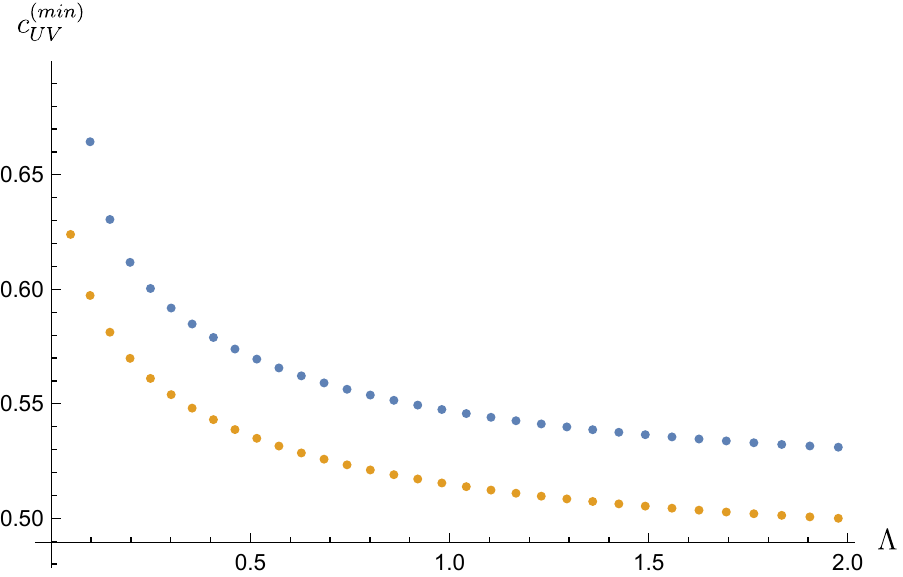}
\end{subfigure}
\hfill
\begin{subfigure}{0.49\textwidth}
\centering
\includegraphics[scale=0.8]{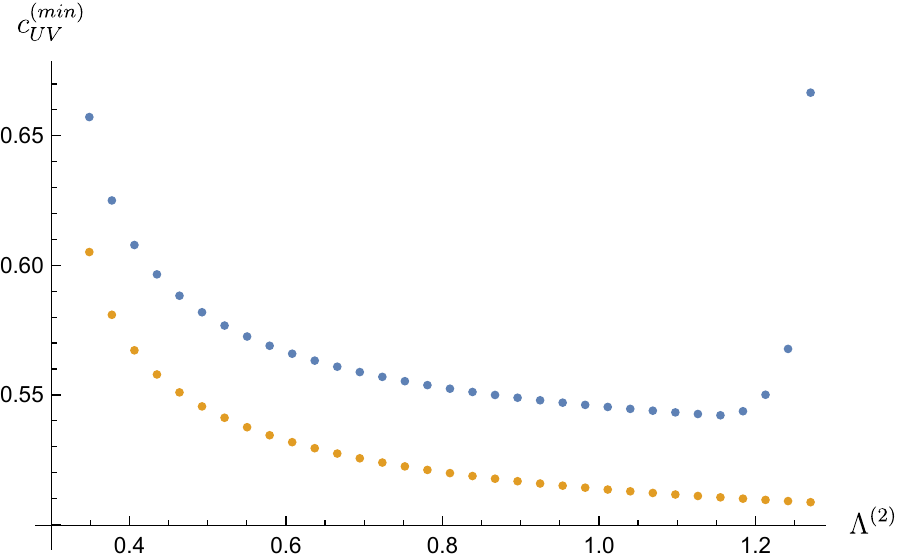}
\end{subfigure}
\caption{Minimal central charge in the $(\Lambda, c_{UV}$) space (left) and in the $(\Lambda^{(2)}, c_{UV}$) space (right) computed with the dual approach (in orange) and the primal (in blue). The value of $\Lambda^{(2)}$ is defined by $\Lambda^{(2)} = \frac{1}{5} \frac{\Lambda^2}{32} + \frac{1}{5}\Lambda$ (left) and on the one of $\Lambda$ by $\Lambda=3.2$ (right).  We used $N=50$.}
\label{dualprimal_leaf_cUV2}
\end{figure}

It is interesting to compute the allowed region with a primal approach and to compare with the dual bounds. We do it\footnote{We use the numerical primal setup from \cite{Karateev_2020} and the ansatz for the scattering amplitude given by eq. (4.4) and (4.5) in \cite{Chen_2022}.}  in Fig. \ref{dualprimal_leaf_cUV}, and various sections of this 3D plot are shown on Figs. \ref{dualprimal_leaf_cUV2}. It is reassuring to see that the primal lower bound is always greater than the dual, and that the two shapes seem to converge toward each other. We believe that the reason for the gap is the slow convergence of the primal formalism, which we also observe on Figure \ref{cuv2}, where the dual reaches the analytical bound. 

\begin{figure}[h]
    \centering
    \includegraphics[scale=1]{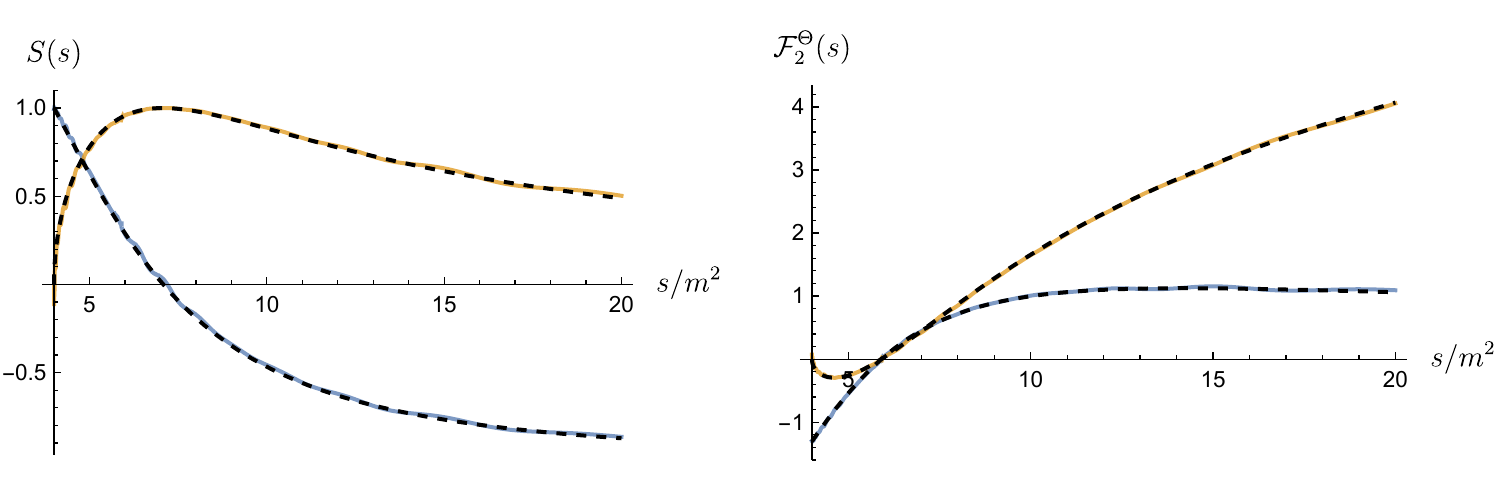}
    \caption{Comparison of the real part (in blue) and imaginary part (in orange) of the S-Matrix (left) and the form factor (right) between the numerical (plain line) and analytical (black dashed line) results, on the upper edge of the leaf with $\Lambda = 2.4$. We used $N=50$. The numerical and analytical central charges are $c_{UV}^{(num,min)}=0.514419...$, $c_{UV}^{(an,min)}=0.514451...$ }
    \label{leaf_up}
\end{figure}

\begin{figure}[h]
    \centering
    \includegraphics[scale=1]{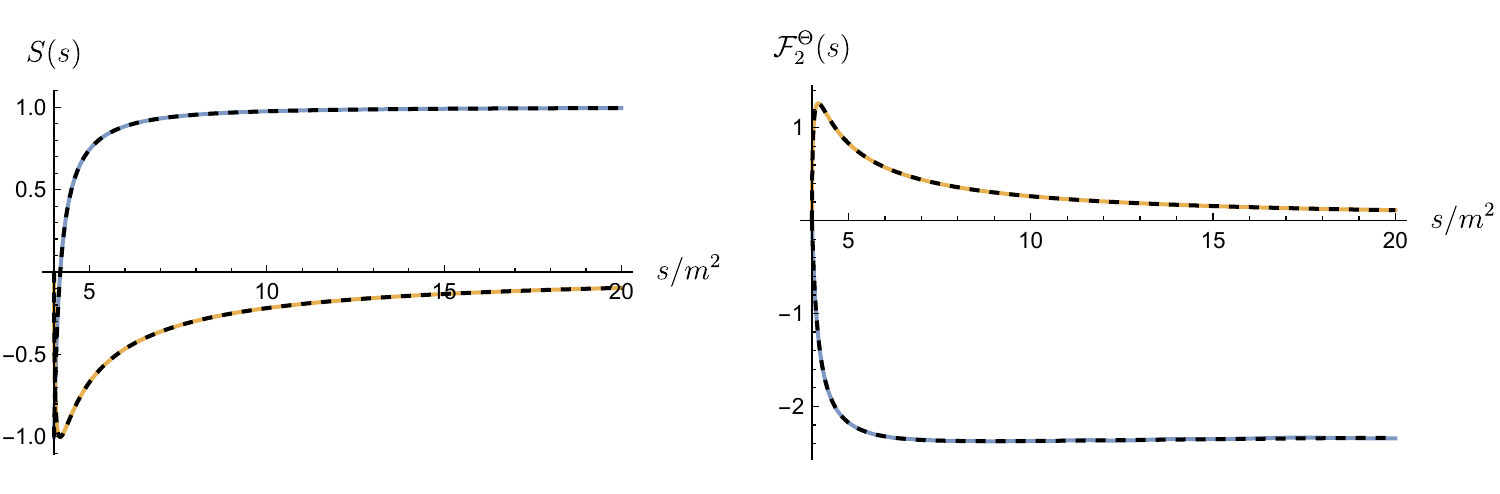}
    \caption{Comparison of the real part (in blue) and imaginary part (in orange) of the S-Matrix (left) and the form factor (right) between the numerical (plain line) and analytical (black dashed line) results, on the lower edge of the leaf with $\Lambda = 2.4$. We used $N=50$. The numerical and analytical central charges are $c_{UV}^{(num,min)}=0.993662...$, $c_{UV}^{(an,min)}=0.993812...$ }
    \label{leaf_down}
\end{figure}

\begin{figure}[H]
    \centering
    \includegraphics[scale=1]{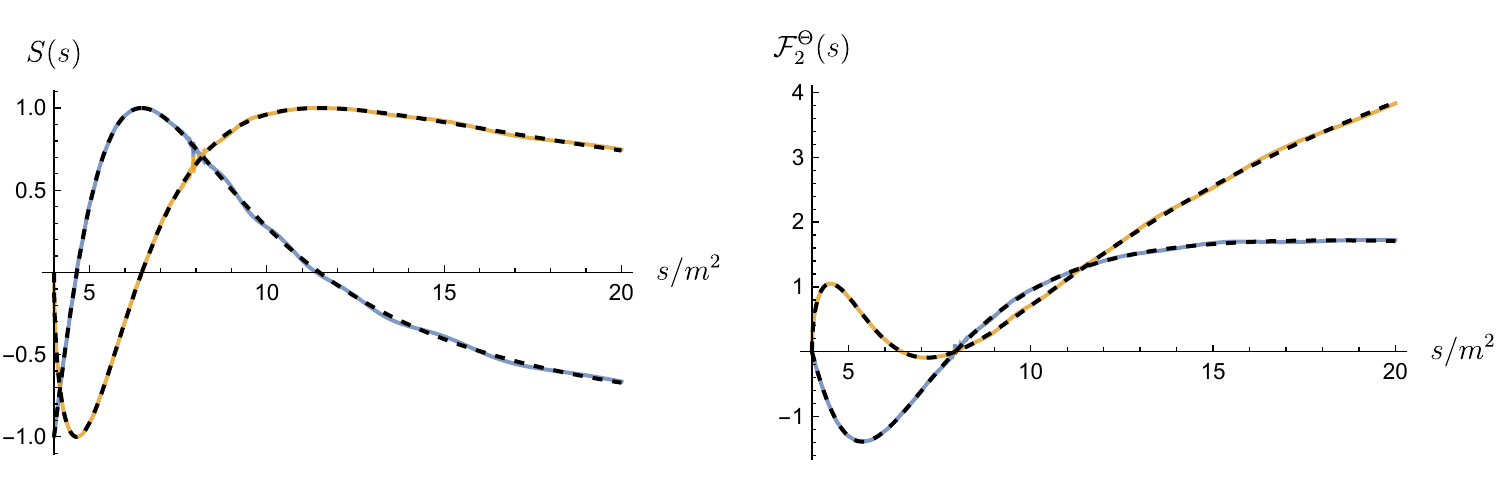}
    \caption{Comparison of the real part (in blue) and imaginary part (in orange) of the S-Matrix (left) and the form factor (right) between the numerical (plain line) and analytical (dashed line) results, in the interior of the leaf with $\Lambda = 4.56$ and $\Lambda^{(2)}=0.85$. We used $N=50$. The numerical and analytical central charges are $c_{UV}^{(num,min)}=0.526354...$, $c_{UV}^{(an,min)}=0.526591...$.}
    \label{leaf_int}
\end{figure}

We now compare our numerical results with the analytical results from the appendix \ref{sec:analytic_bootstrap} where we assume that $c_{UV}$ minimization for fixed $(\Lambda, \Lambda^{(2)})$  minimizes $\Lambda^{(4)} = \lim_{s\to 2} \frac{\partial^4 }{\partial s^4} \mathcal{T}(s)$. For the scattering amplitude on the edges the analytical result is given by \eqref{eq:TS} and \eqref{eq:A}. The form factor on the upper edge is given by \eqref{eq:leaf_upperedge_an} and \eqref{eq:bmin_upperedge}, and on the lower edge by \eqref{eq:F_loweredge}. Figure \ref{leaf_up} shows a point on the upper edge of the leaf,  Figure \ref{leaf_down} shows one on the lower edge and Figure \ref{leaf_int} shows a point in the middle. The agreement is perfect.

\subsection{Ising Field Theory}
\label{sec:IFT}
Our second application is to target the Ising Field Theory. We use the fact that the S-Matrix has a zero whose position varies with the magnetic field 
\begin{equation}
\label{eq:zero}
    S(m^2(1-x))=0, \qquad x=x(\eta), \qquad \eta = \frac{m}{h^{8/15}}.
\end{equation}

The limit $x\to 0$ corresponds to $h \to 0$ and then should give back the free massive Majorana fermion. When $x \to 1$ we are in the region just before the second lightest particle emerges from the two particle cut.

\subsubsection{Minimization of $c_{UV}$ for fixed magnetic field}

\begin{figure}[h]
    \centering
    \includegraphics[scale=1]{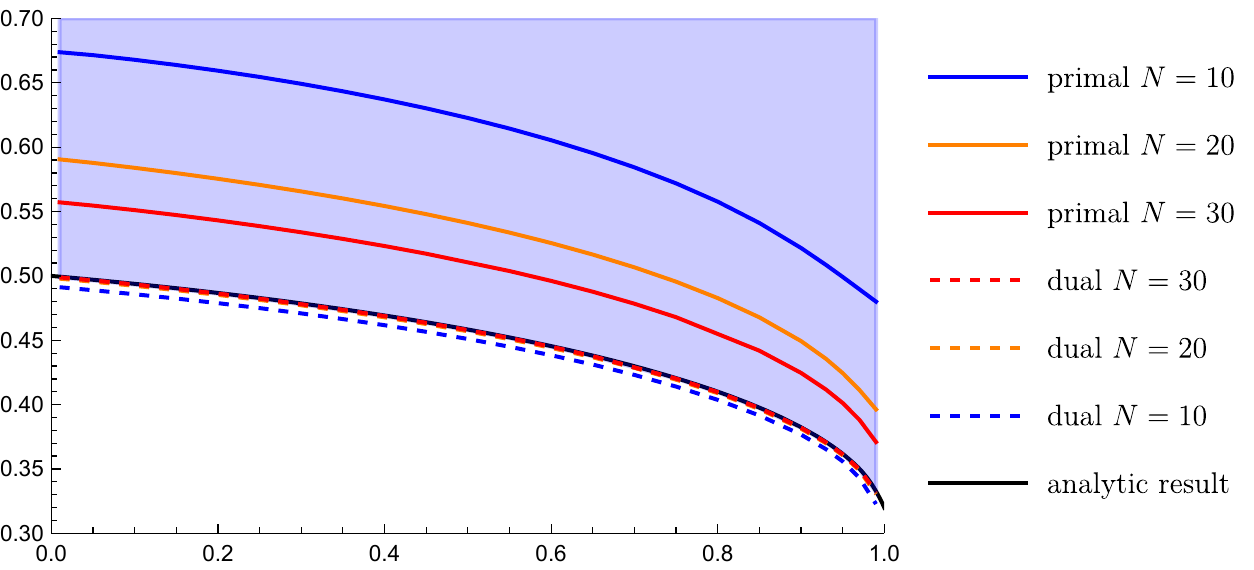}
    \caption{Lower bound on the UV central charge as function of $x$, the position of the zero of the scattering amplitude $S(m^2(1-x)) = 0$. The blue region is allowed. For $x=0.01$ and $N=30$ with the dual we obtained $c_{UV}^{(min)} = 0.498814...$. The black line corresponds to the analytical result \eqref{eq:cUVmin_analytic} assuming that the S-matrix has maximal cubic coupling.  }
    \label{cuv2}
\end{figure}

The first exercise is to minimize the UV central charge for different values of $x$ and the result is shown on Figure $\ref{cuv2}$. We performed this analysis with both primal\footnote{We use the primal numerical implementation described in \cite{Karateev_2020}} and dual formulations and for different $N$  to compare them.  We also plotted the analytical central charge given by \eqref{eq:cUVmin_analytic} which assumes that optimal S-matrix maximizes the cubic coupling $g$ (and has the zero \eqref{eq:zero}). 
\par
As a simple check, when $x\to 0$ the zero cancels the pole, so we are back to asking for the minimal $c_{UV}$ for an S-matrix with no poles. With the dual for $N=30$ we get
\begin{equation}
    x=0.01, \qquad \text{min}c_{UV} = 0.498814...,
\end{equation}
which numerically approaches the result for the free Majorana fermion where $c_{UV}=1/2$.

We now compare primal and dual methods in more detail. First, it is clear that the duality gap closes as $N \to \infty$. We also note the difference in the rate of convergence between the two formalisms. 
For the primal we parametrize the three functions $\mathcal{T}, \mathcal{F}$ and $\rho$ while for the dual we parametrize $\lambda_1, W_\mathcal{T}$ and $W_\mathcal{F}$. Therefore the number of numerical parameters scales like $3N$ for both problems and it makes sense to compare them for a given $N$. Hence we claim that the rate of convergence for the dual formalism is much better than the one for the primal.

\begin{figure}[h]
    \centering
    \includegraphics[scale=0.4]{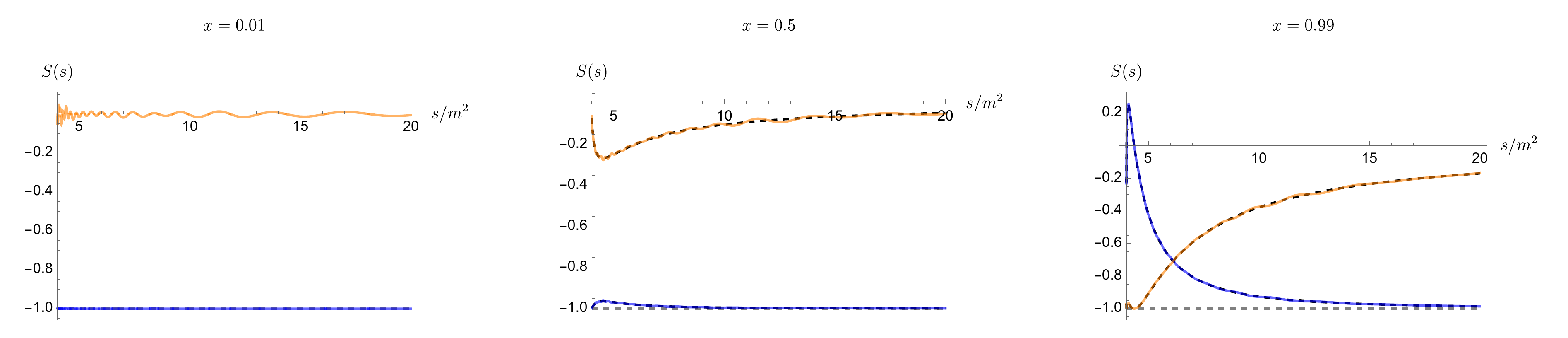}
    \caption{Scattering amplitude $S(s)$ for different values of $x$ when there is a zero at $S(m^2(1-x))$. The imaginary part (in orange) and the real part (in blue) tend toward the free fermion (in dashed gray) when $x \to 0$.  The dashed black line is the analytical result given by \eqref{eq:maximalS}.  We used $N$=50.}
    \label{S0v2}
\end{figure}

\begin{figure}[h]
    \centering
    \includegraphics[scale=0.4 ]{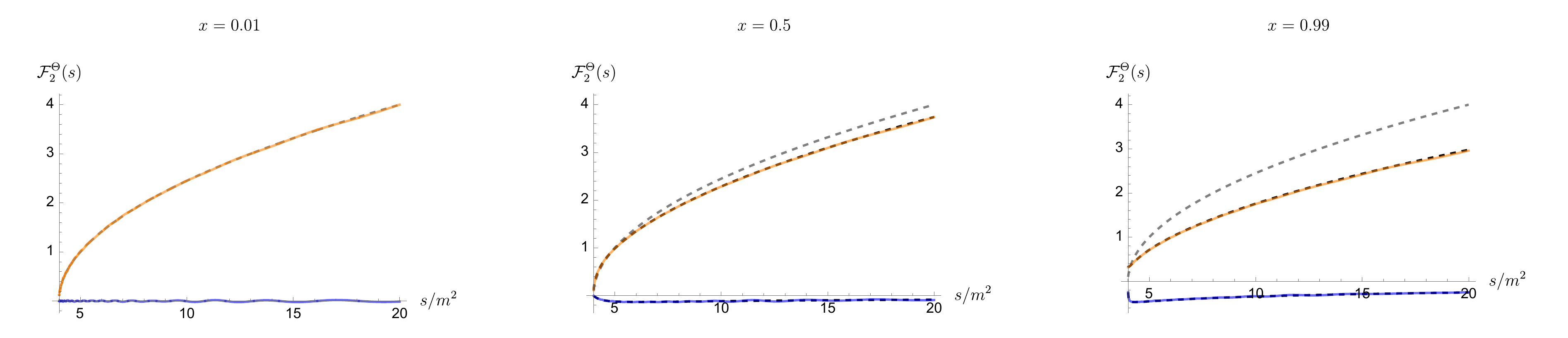}
    \caption{Two particles form factor $\mathcal{F}^\Theta_2(s)$ for different values of $x$ when there is a zero at $S(m^2(1-x))$. The imaginary part (in orange) and the real part (in blue) tend toward the free fermion (in dashed gray) when $x \to 0$. The dashed black line is the analytical result given by \eqref{eq:Fan2} and \eqref{eq:cUVmin_analytic}. We used $N$=50.}
    \label{FFv2}
\end{figure}

\begin{figure}[h]
    \centering
    \includegraphics[scale=0.38]{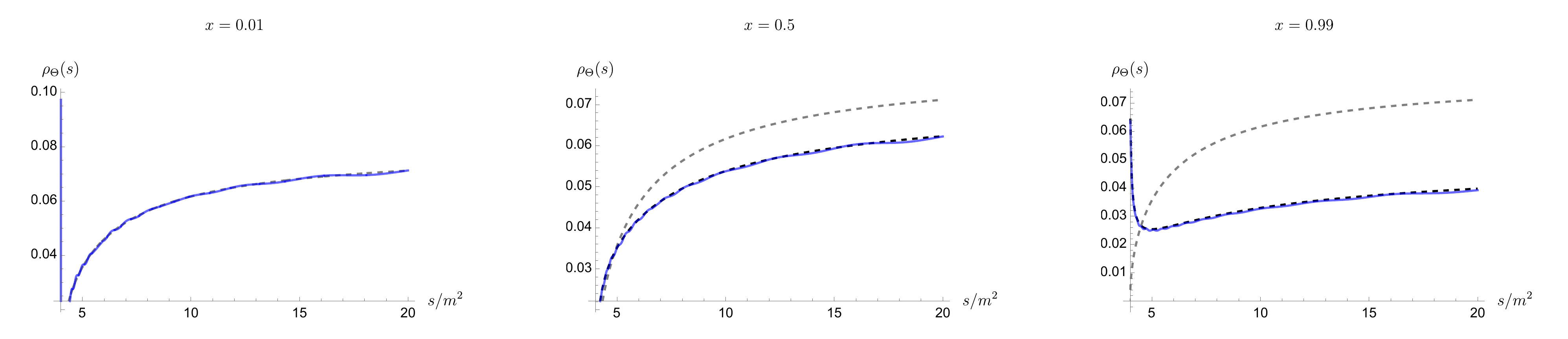}
    \caption{Spectral density $\rho_\Theta(s)$ for different values of $x$ when there is a zero at $S(m^2(1-x))$. The bootstrap result (in blue) tends toward the free fermion (in dashed gray) when $x\to 0$. We also observe the delta function appearing at $s=4m^2$ when $x\to 1$ which corresponds to the second lightest particle that will become stable. The dashed black line is the analytical result given by \eqref{eq:Fan2} and \eqref{eq:cUVmin_analytic} and the elasticity relation \eqref{eq:unitot}. We used $N$=50.}
    \label{Sdv2}
\end{figure}

We can now look at the observables. The scattering amplitude, the two particles form factor and the spectral density are plotted respectively on Figures $\ref{S0v2}, \ref{FFv2}$, and $\ref{Sdv2}$.

 We also observe perfect agreement between the dual numerical results and the analytical results derived in appendix \ref{sec:analytic_bootstrap}. They are given by \eqref{eq:maximalS}, \eqref{eq:Fan2} and \eqref{eq:cUVmin_analytic}.

\begin{figure}[h]
    \centering
    \includegraphics[scale=1]{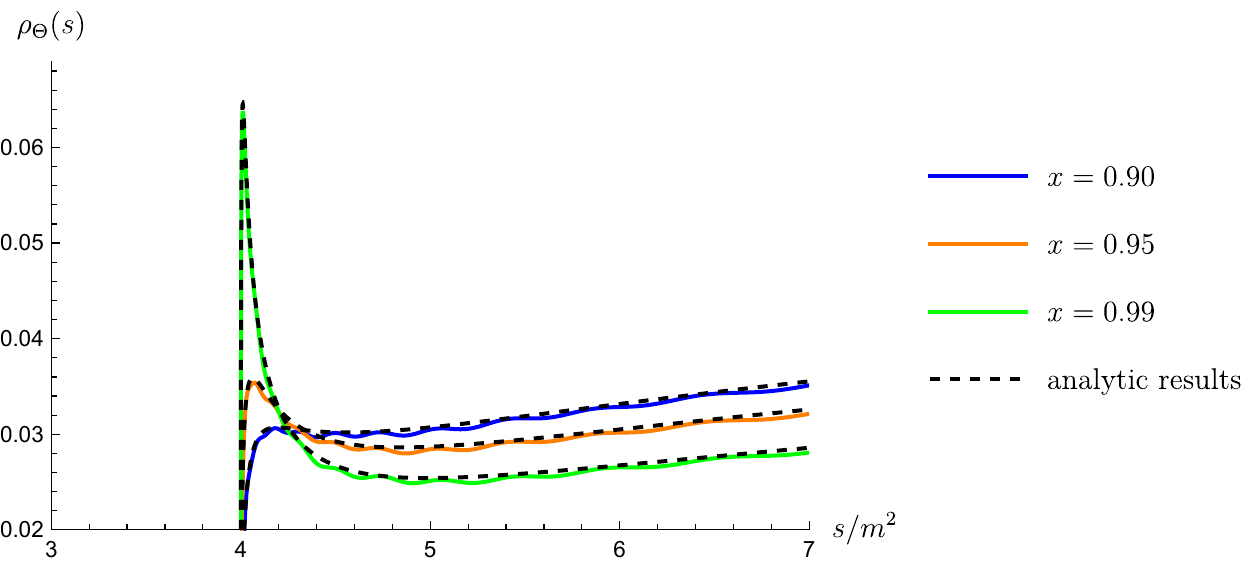}
    \caption{Spectral density $\rho_\Theta(s)$ for different values of $x$ when there is a zero at $S(m^2(1-x))$. We used $N$=50.}
    \label{Sd}
\end{figure}

There is an interesting observation for the spectral density when $x\to 1$. We know that $x=1$ is exactly the point where the pole associated to the second lightest particle enters  the first sheet and exchanges  place with the zero. We also know that the spectral density has delta functions at the positions of the masses of the stable asymptotic particle states. On the figure we see a peak forming $s=4m^2$, which is exactly where the zero and the pole exchange their places. This is illustrated more precisely in Figure \ref{Sd}, where we see the position of the peak moving toward $4m^2$.

Note however that the central charge of this theory is $c_{UV} \approx 0.34$ and therefore it can not be the IFT. Nevertheless it is still interesting to see that those two theories share the presence of the zero in the S-Matrix and also seem to share the resonance associated to an unstable particle when $x\to 1$.

\subsubsection{Carving out IFT parameter space: $c_{UV} = 1/2$}

We now target the IFT by fixing the central charge to $c_{UV}=1/2$ and explore the allowed region in the $(g \mathcal{F}^\Theta_1, g^2,x)$ 3-dimensional space, where we recall
\begin{equation}
    \mathcal{T}(s) = -\frac{g^2}{s-m^2} + ..., \qquad \mathcal{F}(s) = - \frac{\mathcal{F}^\Theta_1 g}{s-m^2} + ...
\end{equation}

\begin{figure}[H]
\centering
\begin{subfigure}{0.49\textwidth}
\centering
\includegraphics[scale=0.75]{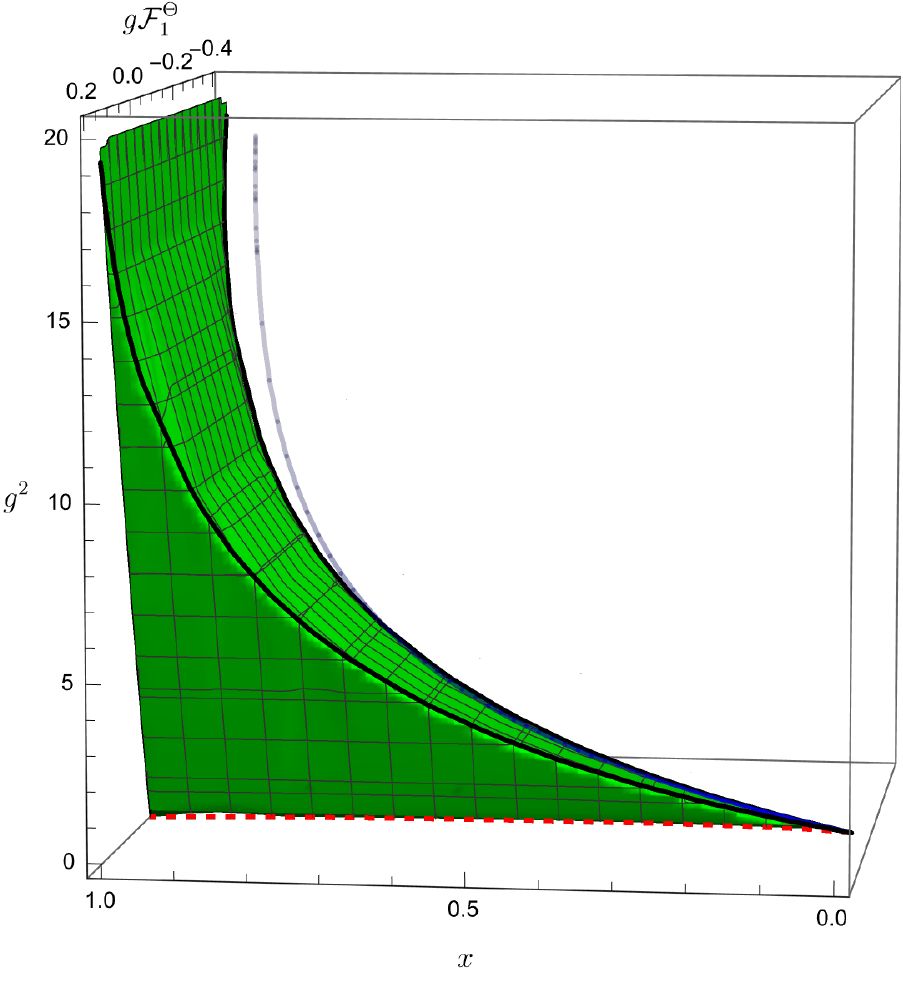}
\end{subfigure}
\hfill
\begin{subfigure}{0.49\textwidth}
\centering
\includegraphics[scale=0.8]{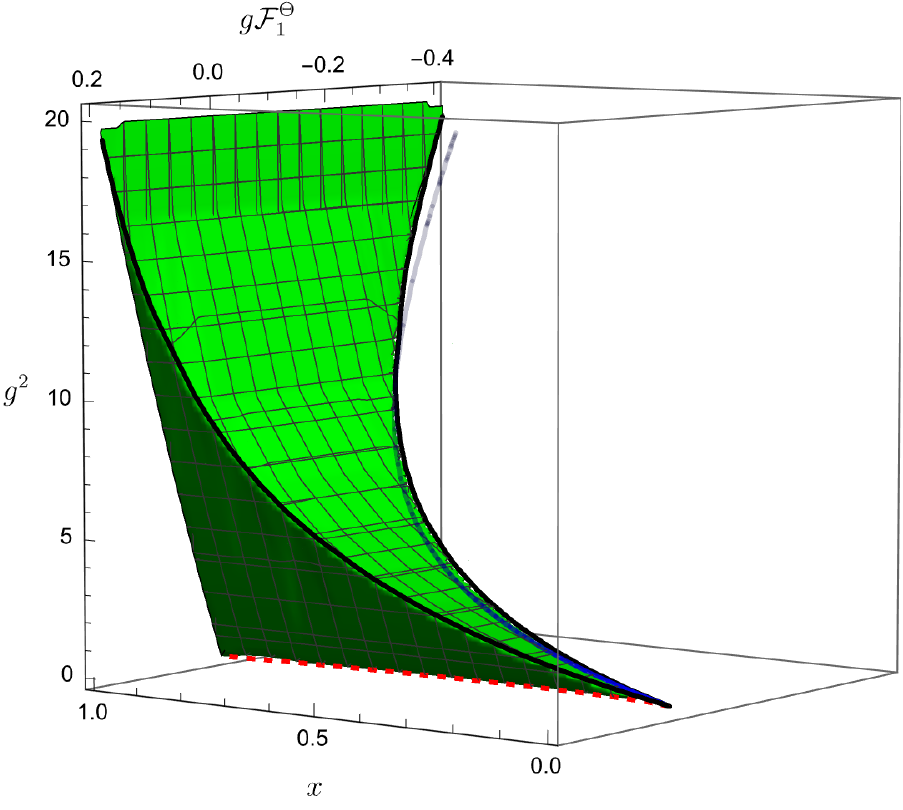}
\end{subfigure}
\begin{subfigure}{0.49\textwidth}
\centering
\includegraphics[scale=0.8]{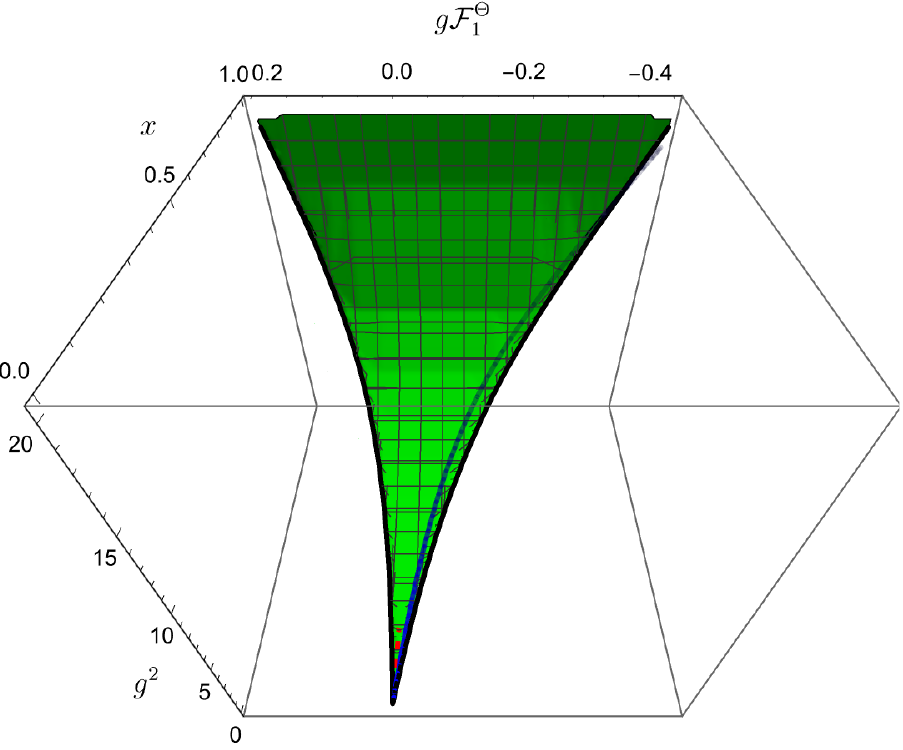}
\end{subfigure}
\hfill
\begin{subfigure}{0.49\textwidth}
\centering
\includegraphics[scale=0.75]{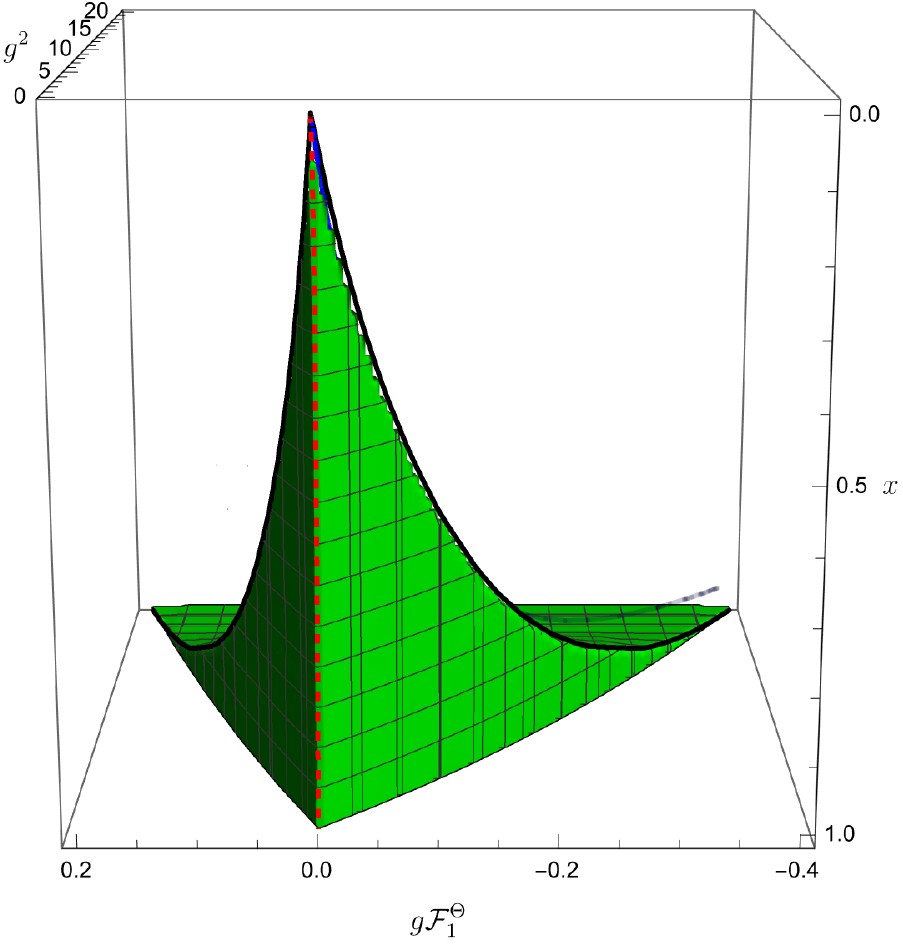}
\end{subfigure}
\caption{Allowed region (in green) in the $(g\mathcal{F}_1^\Theta,x,g^2)$ space.  The blue line is the perturbative trajectory for $g^2(x)$ and $g(x) \mathcal{F}_1^\Theta(x)$ in IFT (computed in appendix \ref{sec:perturbation_theory}), the black lines are the analytical result \eqref{eq:gFroots} and the dashed red line is the minimal cubic coupling discussed in Figure \ref{mingsq}. } 
\label{fig:pyramid}
\label{pyramid}
\end{figure}

We compute the bounds for $g^2$ and $g \mathcal{F}_1^\Theta$ for $x \in (0,1)$, which corresponds to the range in magnetic field with only one stable particle. The result is the pyramid shown in Figure \ref{pyramid} inside which IFT must lie.

\begin{figure}[h]
    \centering
    \includegraphics[scale=1]{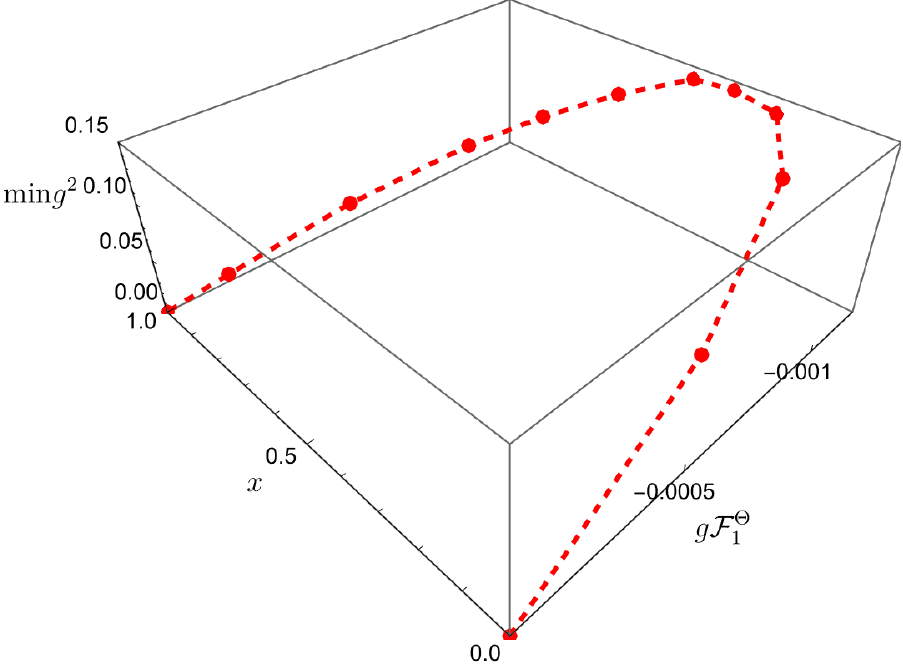}
    \caption{Lower edge of the pyramid corresponding to the lowest bounds on $g^2$. This is computed with $N=70$.}
    \label{mingsq}
\end{figure}

We noticed that the lower edge of the pyramid, corresponding to the lowest bounds on $g^2$ for different $x$, is slightly in-curved toward the negative $g \mathcal{F}_1^\Theta$. The effect is tiny and shown on Figure \ref{mingsq}.

\begin{figure}[h]
    \centering
    \includegraphics[scale=1.1]{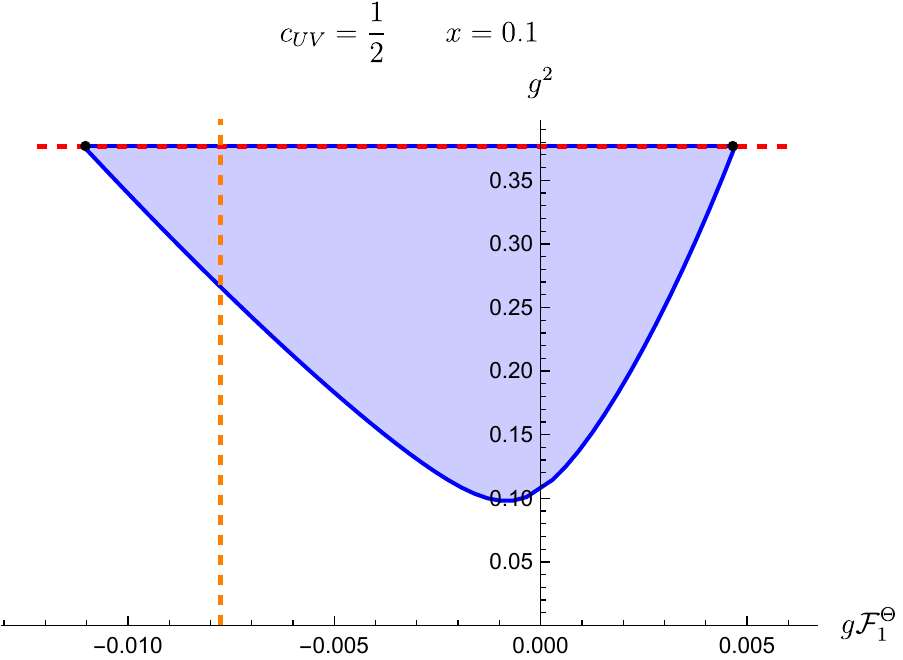}
    \caption{Allowed region in the $(g \mathcal{F}_1^\Theta, g^2)$ plane when the central charge is $c_{UV}=1/2$ and there is a zero in the S-Matrix at $x=1/10$. The red and orange dashed line are the values of $\mathcal{F}_1^\Theta$ and $g^2$ computed with perturbation theory.  This is computed with $N=70$. The black dots are the analytical solutions \eqref{eq:gFroots}.}
    \label{region_IFT}
\end{figure}

 The slice at $x=1/10$ of the pyramid is shown in Figure \ref{region_IFT}, where we also place IFT using form factor perturbation theory: the cubic coupling $g$ was computed in \cite{Zamolodchikov_2011} and $\mathcal{F}_1^\Theta$ is an original computation (both calculations are in appendix  \ref{sec:perturbation_theory}).

The figure is generated by first computing the bounds on $g\mathcal{F}_1^\Theta$ for which we got
\begin{equation}
   g \mathcal{F}_1^{\Theta,(min)} = -0.0110843..., \qquad  g \mathcal{F}_1^{\Theta,(max)} = 0.0047159...,
\end{equation}
and then computing the bounds on $g^2$ with fixed values of $g \mathcal{F}^\Theta_1$.

\begin{figure}[h]
    \centering
    \includegraphics[scale=1]{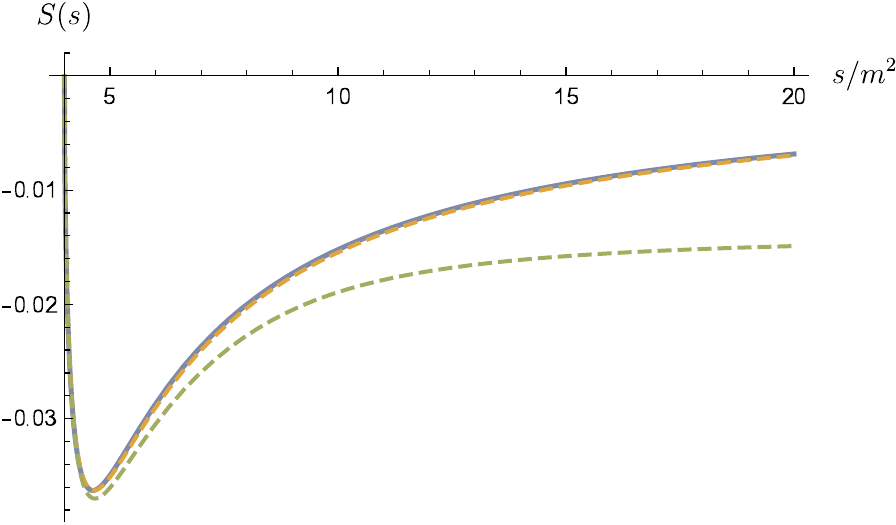}
    \caption{Scattering amplitude obtained from the bootstrap (in blue) at the point matching the perturbation theory results compared to one from perturbation theory for $x=0.1$. The green dashed line is the full perturbative S-Matrix while the orange dashed line is only the elastic contribution. We used $N=20$.}
    \label{S_PT_IFT}
\end{figure}

Note that the perturbative result for $g^2$ saturates the higher bound for the cubic coupling allowed by the bootstrap. The perturbative result for $\mathcal{F}_1^\Theta$ however does not saturate the bounds but still is inside the allowed region. At the points where the red and orange lines meet we can compare the optimal scattering amplitude to the one known in perturbation theory from \cite{Zamolodchikov_2011}. This is shown in Figure $\ref{S_PT_IFT}$. We observe that we match perfectly  with the elastic contribution of the perturbative S-Matrix. This is expected since our bootstrap setup saturates unitary at all energies and does not account for theories with particle production. Therefore we are not surprised that inelastic contributions are not obtained.

We note the absence of wiggles in this numerical optimal function, showing that the slow convergence of the optimal dual functions is not a systematic feature of our formalism. This supports the idea that our ansätze are not optimal and can only easily reproduce a certain subclass behaviours.

\begin{figure}[h]
    \centering
    \includegraphics[scale=1]{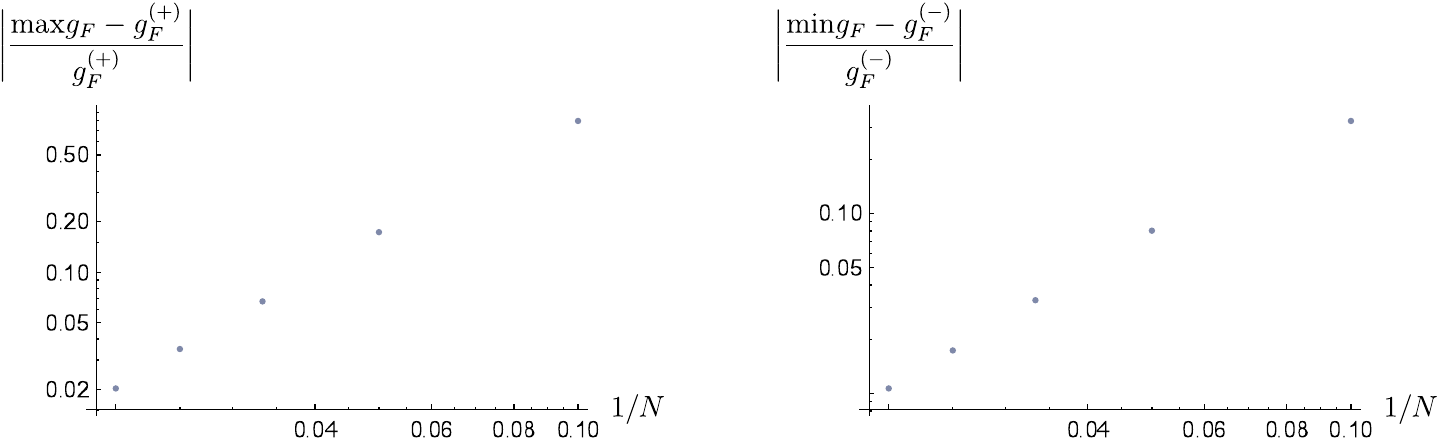}
    \caption{Convergence study (on a logarithmic scale) for the upper corners of the triangle approaching the analytical roots $g_F^{(\pm)}$.}
    \label{conv_edges}
\end{figure}

In Figure $\ref{conv_edges}$ we perform a convergence study for the upper corners of the triangle with respect to the analytical roots given by \eqref{eq:gFroots}. The corners of the allowed region coming from the dual bootstrap approach the analytical roots

\begin{figure}[h]
    \centering
    \includegraphics[scale=1]{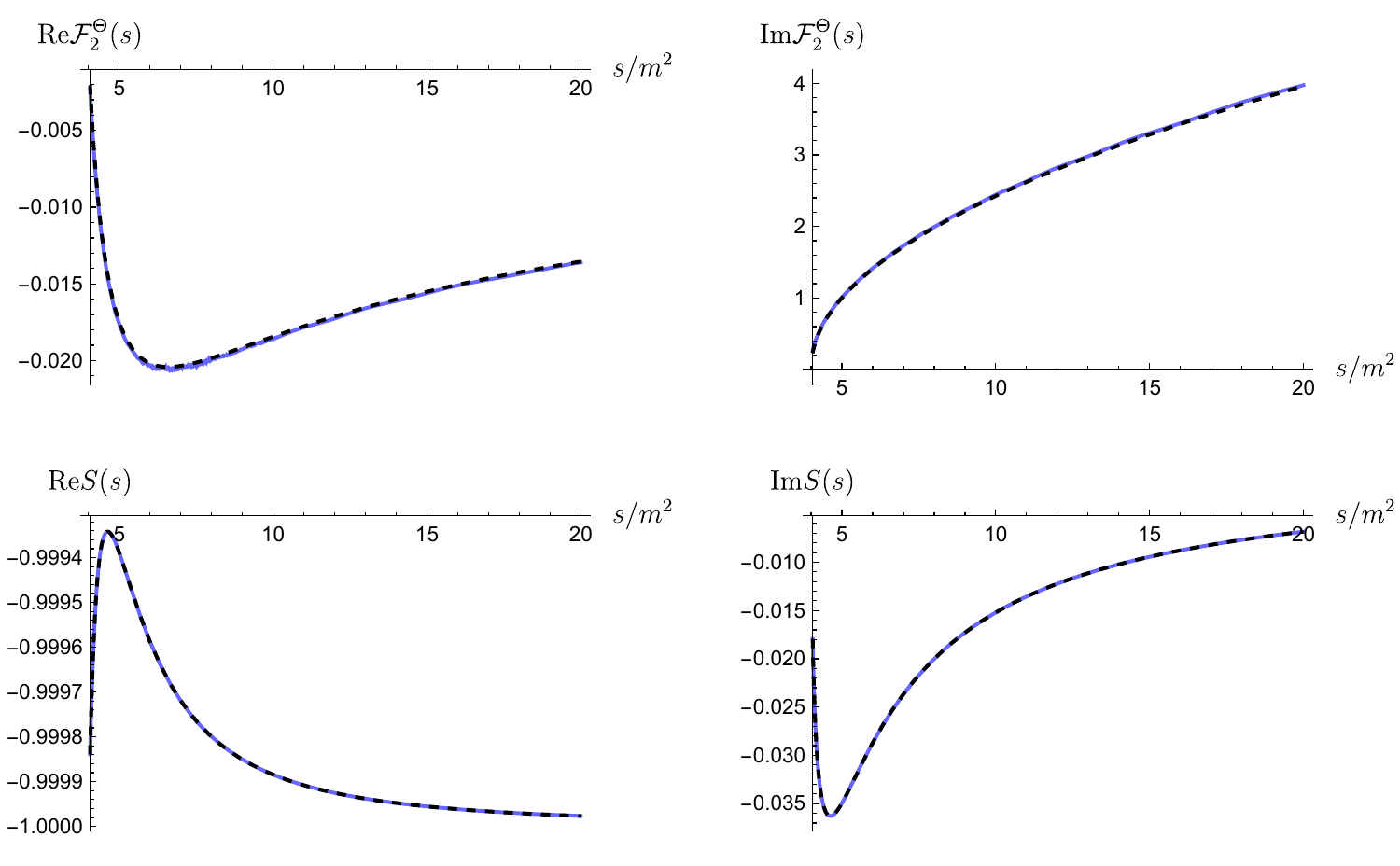}
    \caption{Numerical scattering amplitude and form factor on the upper edge of the triangle (in blue) compared to the analytical result (in dashed). We took the point $g \mathcal{F}_1^\Theta=0$  for reference but the data looks identical on the whole upper edge. We used $N=50$.}
    \label{upperedge}
\end{figure}

\begin{figure}[h]
\centering
\begin{subfigure}{0.49\textwidth}
\centering
\includegraphics[scale=0.8]{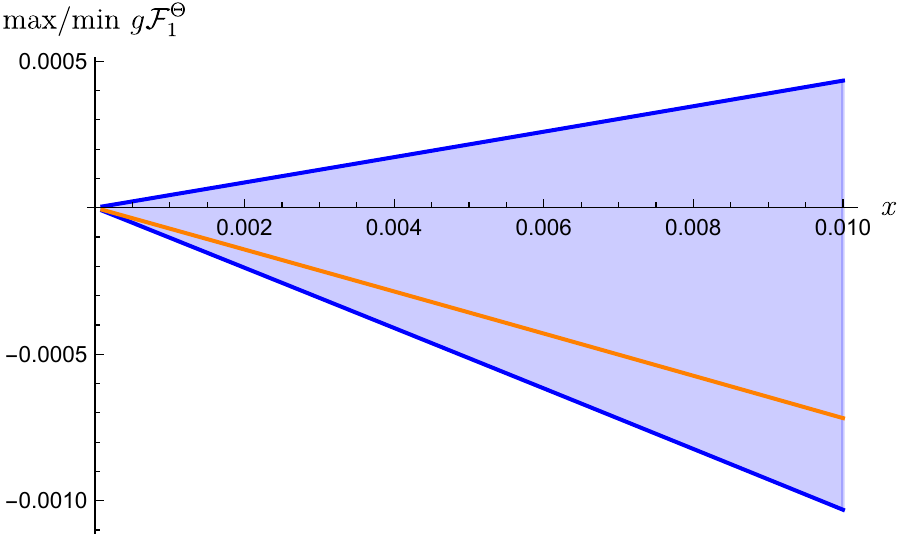}
\end{subfigure}
\hfill
\begin{subfigure}{0.49\textwidth}
\centering
\includegraphics[scale=0.8]{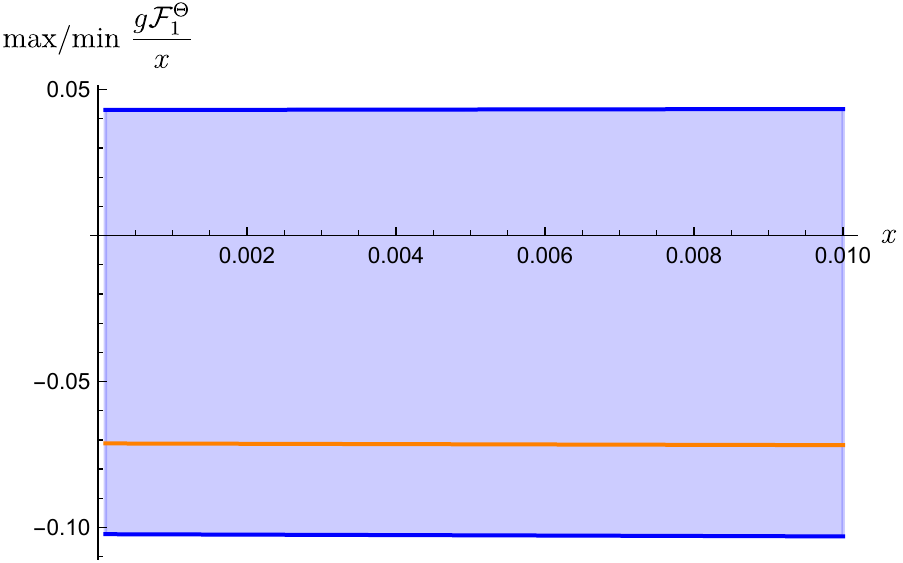}
\end{subfigure}

\caption{Evolution of the upper boundary of the allowed region (in blue) via the numerical bootstrap compared to the evolution of the perturbative result (in orange) in the perturbative regime for small values of $x$. We observe that the bootstrap bounds and perturbation theory both scale linearly with $x$, or quadratically with $h$. We used $N=70$.} 
\label{BvsPT}
\end{figure}

Since the perturbative result for $\mathcal{F}_1^\Theta$ does not saturate the bootstrap bounds we can investigate what happens when the position of the zero (or equivalently the magnetic field) is changed. We plot the perturbative result and the bootstrap bounds for different magnetic fields in Figure \ref{BvsPT}. We observe that both perturbation theory and bootstrap bounds scale with $x$ or, equivalently, with $h^2$, but at a different rate,
\be
{g}_F^{(\pm)} \; = \;  \left(\frac{1}{2 \sqrt{3}} -\frac{1}{\pi } \pm {1 \over \sqrt{3} \pi} \mp {1 \over 9}\right) 
 x \;\approx \; ( - \, 0.1023, + \,0.0430 \,) \, x \; \approx \;( - \, 11.7601, + \,4.9466 \,) \, h^2.
\ee
whereas
\be
g_F^{(P.T)} &= -\frac{1}{3\sqrt{2}}\left(\frac{15}{8}+ \frac{1}{2\pi}-\sqrt{3} \right)x \approx -0.0712 x  \approx -8.1857h^2.
\ee
These numbers were computed in appendices \ref{sec:analytic_bootstrap} and \ref{sec:perturbation_theory}, respectively.
\par

\begin{figure}[h]
    \centering
    \includegraphics[scale=1.1]{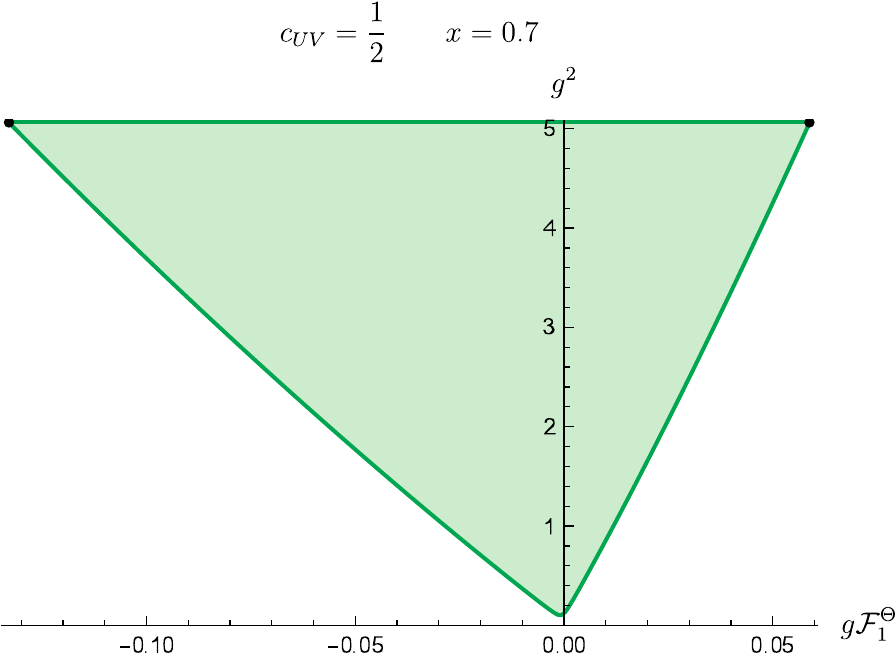}
    \caption{Allowed region in the $(g \mathcal{F}_1, g^2)$ plane when the central charge is $c_{UV}=1/2$ and there is a zero in the S-Matrix at $x=7/10$. This is computed with $N=70$. The black dots are the analytical roots \eqref{eq:gFroots}.}
    \label{region_IFT_2}
\end{figure}

In Figure \ref{region_IFT_2} we show the bounds on $g^2$ and $g \mathcal{F}_1^\Theta$ in the strongly interacting regime $x=0.7$. The upper boundary is again saturated by the analytical results \eqref{eq:gFroots} and \eqref{eq:g2}. In particular the roots of eq.\eqref{eq:gFroots} correspond to the corners.

\newpage

\section{Conclusion}
\label{sec:conclusion}
In this work we have merged the primal S-Matrix and form factor bootstrap proposed in \cite{Karateev_2020} with the dual S-Matrix bootstrap developed in \cite{Guerrieri_2020}. With our method we can relate UV data, namely the central charge $c_{UV}$ of the UV CFT, with couplings of the gapped QFT in the IR, and place rigorous bounds on any of these parameters. We observe a vastly improved numerical convergence compared to the primal bootstrap of \cite{Karateev_2020}. 
\par
We established a lower bound on $c_{UV}$ in a class of $\mathbb{Z}_2$ symmetric QFTs (fig. \ref{dual_leaf_cUV}) and also for QFTs for which the particle self-interacts with a cubic coupling and where the S-matrix has a real zero (a dial for the magnetic field in IFT) - see fig. \ref{cuv2}. In all these cases we find that the optimal S-matrix extremizes some coupling. This allows us to find the optimal S-matrix via pure analytical S-matrix bootstrap (viz. CDD bootstrap \cite{Castillejo:1955ed}). Then, in appendix \ref{sec:analytic_bootstrap}, we showed how the form factor and the minimal $c_{UV}$ can be obtained analytically if the S-matrix is known. We find perfect agreement between the dual optimization and the analytical bootstrap.
\par
To specifically target Ising Field Theory we fix $c_{UV} = 1/2$ and find the allowed space in cubic couplings and magnetic field, see fig \ref{pyramid}. We insert IFT inside this `pyramid' for small magnetic field - close to the tip of the pyramid - using form factor perturbation theory (see appendix \ref{sec:perturbation_theory}). For larger magnetic fields we enter the non-perturbative regime of IFT and this is where our results can be useful: IFT \emph{must} lie somewhere inside the pyramid.

We know however that IFT cannot lie on the boundary of the pyramid because the boundary is elastic: the optimal S-matrix does not have inelasticity,  and we know that IFT has particle production \cite{Zamolodchikov_2011}.
But even for small enough magnetic field $h$ (where particle production is negligible) IFT is still somewhat away from the boundary, as the bounds on the one particle form factor go to zero as $h^2$ with a different numerical coefficient than the perturbative form factor  (see fig. \ref{BvsPT}).

We can think of several ways to improve our setup and `shrink' the pyramid wherein IFT must lie. One way is to include more UV information. So far we are only fixing the value of the central charge, but we are not `telling' the bootstrap that the RG flow is being triggered by the thermal and magnetic deformations, $\epsilon$ and $\sigma$, respectively. The central charge is straightforward  to include  in $d=2$ due to the $c$-sum rule which gives an \emph{integrated} relation over the spectral density of the trace of the stress energy tensor $\Theta$, and not just via some asymptotic constraint at $s \to \infty$, which is difficult to implement numerically.\footnote{ Moreover, the normalization of the two particle form factor of $\Theta$ is fixed (see appendix \ref{sec:normF}), which is one less parameter to vary over in the bootstrap.} So, can we find similar integral relations that involve somehow $\epsilon$ or $\sigma$ (besides the combination $\Theta$)? One possibility is the sum rule derived in Eq. (29) of \cite{Delfino_1996} to fix the scaling dimension of an operator. However, this sum rule requires knowing the vacuum expectation value of that operator. For the case of $\sigma$, its expectation value should be related to the magnetic field and therefore to the zero of the S-matrix, but the precise relation is unknown to us.
It should be possible to measure it using other methods like the lattice or  Hamiltonian truncation.

Another possibility is to use the twist property of the order field $\sigma$. Twist fields are reviewed in \cite{Cardy_2007, Castro_Alvaredo_2018} and references therein. The fact that $\sigma$ acts as a twist field is derived in \cite{SCHROER197880}. This could be coupled to the input from lattice measurements, for example using tensor networks.
 In order to measure the two point function $\langle \sigma \Theta \rangle$ on the lattice, one needs   to map the continuum stress tensor to the lattice stress tensor. This  might be done by generalizing some ideas developed in \cite{Bastiaansen_1998} and \cite{Giusti_2015}. We plan to explore  this direction in the near future.

In addition, we can include more IR information. IFT is known to interpolate between the free massive Majorana fermion and the $E_8$ Toda theory, which has 8 particles. While we are studying the region close to the free fermion, where there is only one particle, it is natural to expect, due to analyticity, that the other particles exist in other `sheets', i.e. as resonances (corresponding to zeros on the physical sheet). We already impose one zero, which eventually becomes the second lightest particle as the magnetic field is increased. It is possible that the other particles imply the presence of further zeros on the physical sheet  \cite{gabai2019smatrix}. 
\par
If we are feeling brave, we can dial the magnetic field a bit higher and go to the regime where more particles are stable. If we want to go beyond lightest particle scattering we would have to import the multiple amplitude bootstrap of \cite{Homrich:2019cbt} and extend it in several directions. Namely, to dualize it, to include form factors and, on a more fundamental level, to understand the singularity structure of the scattering amplitude of heavier particles, where anomalous thresholds are expected \cite{Coleman:1978kk,Correia:2022dcu}.\footnote{Anomalous thresholds are indeed present at the $E_8$ integrable point, where in $d=2$ they are known as Coleman-Thun poles \cite{Delfino:2003yr}.} 
\par
In a similar vein, we can also consider multi-particle scattering, i.e. to not only consider 2-particle states but also states of 3- or more particles. In this way we could try target the inelasticity of IFT. However, a systematic understanding of the analyticity structure of multi-particle amplitudes remains a big open problem dating back to the 60s.
\par
The latter directions we have mentioned are of course relevant to many other QFTs besides IFT. We have put a lower bound on the $c_{UV}$ of a class of $\mathbb{Z}_2$ symmetric theories but we could have chosen to be more specific. For example, it would be interesting to target $\phi^4$ which has been the focus of many recent nonperturbative studies \cite{Coser:2014lla,Rychkov:2014eea,Bajnok:2015bgw,Bosetti:2015lsa,Anand:2017yij,Serone:2018gjo,Chen:2021pgx,Chen:2021bmm}. This theory is a deformation of the free massless boson, so we could fix $c_{UV} = 1$ in our $\mathbb{Z}_2$ symmetric setup to target it. Moreover, we could place $\phi^4$ inside our bounds, not only using perturbation theory but also with Hamiltonian truncation data \cite{Chen:2021pgx}.

\par
Finally, it would be interesting to generalize our setup to higher dimensions. A primal approach incorporating UV input via the $a$-anomaly was developed in \cite{a_anomaly}. Can we reproduce these bounds more rigorously and more efficiently with a dual method?

\textbf{Acknowledgements} 
We are grateful to Denis Karateev and Kamran Salehi Vaziri for collaboration at the early  stages of this project. We thank Luc\'{i}a C\'{o}rdova, Barak Gabai, Alessandro Georgoudis, Kelian Häring, Yifei He, Alexandre Homrich and Pedro Vieira for useful discussions. M.C. is grateful to the Institute for Advanced Study in Princeton and to the Perimeter Institute in Waterloo for the kind hospitality while this work was being completed. This project has received funding from the European Research Council
(ERC) under the European Union’s Horizon 2020 research and innovation programme (grant agreement number 949077).
The authors are also supported by the Simons Foundation grant 488649 (Simons Collaboration on the Nonperturbative Bootstrap)  and the Swiss National Science Foundation through the project
200020\_197160 and through the National Centre of Competence in Research SwissMAP.
\newpage
\begin{appendices}

\section{Analytical Bootstrap}
\label{sec:analytic_bootstrap}

There are three objects we want to constrain. The $2 \to 2$ scattering matrix $S(s)$ or amplitude $\mathcal{T}(s)$, the 2-particle form factor $\mathcal{F}(s)$ and the spectral density $\rho(s)$. The first two are real analytic functions,\footnote{A real analytic function $f(s)$ satisfies $f^*(s) = f(s^*)$.} and the S-matrix further satisfies crossing symmetry, $S(s) = S(4m^2 - s)$.
There is still the constraint of unitarity which relates all of them. We observe from the optimization problem that unitarity is ``saturated", eq. \eqref{eq:unisat}, meaning
\be
\label{eq:unitot}
|S(s)|^2 = 1, \qquad { \frac{\mathcal{F}(s)}{\mathcal{F}^*(s)}} = S(s), \qquad \rho(s) = { \frac{|\mathcal{F}(s)|^2}{ 4 \pi \sqrt{s(s-4m^2)}}},
\ee
for $s \geq 4m^2 $ slightly above the real axis.
\par
\par
Counting separately for real and imaginary parts we have five variables and three equations, which implies that there is two-fold freedom left over from eqs. \eqref{eq:unitot}. The general solution for \eqref{eq:unitot} should possess this freedom.
\par
We start by solving the first equation of eq. \eqref{eq:unitot}. It will be particularly useful to us here to exploit a well-known trick used for solving elastic unitarity in $d > 2$ (see e.g. \cite{Correia:2020xtr}). We write $S(s)$ in terms of the amplitude $\mathcal{T}(s)$,
\be
S(s) = 1 + {\mathcal{T}(s) \over 2 \sqrt{s (4m^2 - s)}}. 
\ee
Then, elastic unitarity $|S(s)|^2 = 1$ reads
\be
\Im \mathcal{T}(s) = {|\mathcal{T}(s)|^2 \over 4 \sqrt{s (s - 4m^2)} }.
\ee
Noting that $\Im \left[1 / \mathcal{T}(s) \right] = - \Im \mathcal{T}(s) \,/\, |\mathcal{T}(s)|^2$ we can write the above as
\be
\Im \left({1 \over \mathcal{T}(s)} \right) = - {1 \over 4 \sqrt{s (s - 4m^2)} }
\ee
The general solution to the above is then
\be
{1 \over \mathcal{T}(s)} =  - {1 \over 4 \sqrt{s (4m^2 - s)} } + {A(s) \over 4}
\ee
where $A(s)$ is some crossing symmetric real function on the real line.
\par
We then have
\be
\label{eq:TS}
\mathcal{T}(s) = {4 \over A(s) - 1/ \sqrt{(4m^2 - s) s}}, \qquad \text{ or } \qquad  S(s) = {A(s) \sqrt{(4m^2 - s) s} + 1 \over A(s) \sqrt{(4m^2 - s) s} - 1}.
\ee
Now, since $\mathcal{T}(s)$ is analytic, $A(s)$ must  be meromorphic in the physical sheet.

Poles of $A(s)$ translate into zeros of $\mathcal{T}(s)$. 
Note that if $A(s) = \text{const}$ the solution for $S(s)$ is precisely the so-called Castillejo-Dalitz-Dyson (CDD) factor. In fact, any solution to elastic unitarity, crossing and analyticity can be written as a product of CDD factors. It is straightforward to show that this amounts to \eqref{eq:TS} with rational $A(s)$.

The function $A(s)$ encondes the large freedom that the first equation of \eqref{eq:unitot} allows for. We may further fix $A(s)$ at certain values if further constraints are involved, e.g. bound-state poles or zeros. 
\par
With solution \eqref{eq:TS} in hand we can now try to solve for the form factor using the second equation in \eqref{eq:unitot}. If we write
\be
\label{eq:Fan}
\mathcal{F}(s) = -2m^2 B(s) \,e^{\alpha(s) - \alpha(0)},
\ee
where $B(s)$ is real, Watson's equation will only fix $\alpha(s)$,
\be
{\mathcal{F}(s) \over \mathcal{F}^*(s)} = S(s) \implies \Im \alpha(s) = - {i \over 2} \log S(s + i \epsilon), \qquad \text{ for } s \geq 4m^2
\ee
Making use of the solution for the S-matrix \eqref{eq:TS} we have
\be
\label{eq:Imalpha}
\Im \alpha(s) = \mathrm{arccot} \left(A(s) \sqrt{(s-4m^2) s} \right) \Theta(s - 4m^2)
\ee
It is difficult to  explicitly find a function which gives such imaginary part for a generic $A(s)$. We may still find numerically $\alpha(s)$ through a dispersion relation,
\be
\label{eq:alphadisp}
\alpha(s) - \alpha(0) =  {s \over \pi} \int_{4m^2}^\infty {\mathrm{arccot} \left(A(s') \sqrt{(s'-4m^2) s'} \right) \over s'(s' - s) } \,ds'.
\ee
Note that any analytic function could be added to the right hand side of the above. Such analytic function would drop out upon taking the imaginary part meaning that eq. \eqref{eq:Imalpha} would still be satisfied. The presence of a non-trivial analytic function,\footnote{By trivial we mean a constant piece which can be simply absorbed into $B(s)$.} i.e. a polynomial, will lead to an essential singularity for $\mathcal{F}(s)$ so we discard this possibility.
\par

\par
The real function $B(s)$ is not fixed by Watson's equation. It plays a similar role to $A(s)$ in eq. \eqref{eq:ST}, i.e. $B(s)$ is analytic in a neighbourhood of the physical region. These properties are of course fixed by whatever analyticity assumption we may have on $\mathcal{F}(s)$. Causality requirements indicate that $B(s)$ should at most be meromorphic, with poles signalling potential bound-states.
\par
We now impose the normalization 
\be
\mathcal{F}(0) = - 2m^2 \implies B(0) = 1.
\ee
With $\mathcal{F}(s)$ in hand, via eqs. \eqref{eq:Fan} and \eqref{eq:alphadisp}, we can now simply solve for the spectral density from eq. \eqref{eq:unitot}. The two-fold freedom left over from eq. \eqref{eq:unitot} is encoded into the meromorphic functions $A(s)$ and $B(s)$. $A(s)$ can be determined if some coupling is being extremized over in the S-matrix space, which appears to occur when $c_{UV}$ is minimized. $B(s)$ is determined by minimizing over $c_{UV}$ explicitly,
\be
\label{eq:cUVleaf}
c_{UV} = 12 \pi \int_{4m^2}^\infty {\rho(s) \over s^2} ds = 12 m^4  \int_{4m^2}^\infty { B^2(s) \, e^{2 \bar{\alpha}(s) } \over s^2 \sqrt{s(s-4m^2)} } \,ds,
\ee 
where we made use of the relation between $\rho(s)$ and $\mathcal{F}(s)$, eq. \eqref{eq:unitot}, and eq. \eqref{eq:Fan} for $\mathcal{F}(s)$, and where $\bar{\alpha}(s)$ is the real part of $\alpha(s) - \alpha(0)$, i.e. the principal value of \eqref{eq:alphadisp},
\be
\label{eq:baralpha}
\bar{\alpha}(s) = \Re[ \alpha(s) - \alpha(0)] =  {s \over \pi} \pint_{4m^2}^\infty {\mathrm{arccot} \left(A(s') \sqrt{(s'-4m^2) s'} \right) \over s'(s' - s) } \,ds'.
\ee

\par
As we will see in more detail, the polynomial degree of $B(s)$ must be bounded from above so that the integral \eqref{eq:cUVleaf} for $c_{UV}$ is convergent. In fact, we will see that for every case considered here $B(s)$ is at most linear in $s$ (apart from a very special case described in \ref{sec:ancuvg} where it can be quadratic). The constant coefficient in $B(s)$ is fixed by the normalization  $\mathcal{F}(0) = - 2m^2$, so we only need to minimize $c_{UV}$ over the free linear coefficient, which can be done analytically.

\subsection{$\mathbb{Z}_2$ symmetric theories}

\subsubsection{Finding the leaf}
\label{sec:leafcarve}

We are interested in finding the allowed space of the following parameters
\be
\label{eq:Lambda12}
    \Lambda \equiv -\mathcal{T}(2), \qquad \Lambda^{(2)} \equiv \lim_{s\to 2} \frac{\partial^2 }{\partial s^2} \mathcal{T}(s).
\ee
where $S$ and $\mathcal{T}$ are related by eq. \eqref{eq:ST},
\be
\label{eq:ST2}
S(s) = 1 + {\mathcal{T}(s) \over 2 \sqrt{s(4m^2-s)}}.
\ee

Elastic unitarity for the S-matrix reads
\be
|S(s)|^2 = 1, \;\; \text{ for } \; \; s \geq 4m^2 \vee s \leq 0.
\ee
Crossing symmetry further implies 
\be
S(s) = S(4m^2 - s).
\ee

\par
Now, we start by assuming that the S-matrix that extremizes the bounds is elastic and has the minimal number of CDD zeros. We have two parameters that we need to fix, $\Lambda$ and $\Lambda^{(2)}$, so we need at least two CDD factors,
\be
\label{eq:aaS}
S = S_{a_-} S_{a_+},
\ee
with
\be
S_a = {a - \sqrt{s(4m^2-s)} \over a + \sqrt{s(4m^2-s)}}.
\ee
Let us now set $m = 1$ as choice of units. From eqs. \eqref{eq:ST2} and \eqref{eq:aaS} we then solve for the amplitude \eqref{eq:TS} to find
\be
\label{eq:TAx}
A(s) = -{1 + { a_+ a_- \over s(4-s)}\,  \over a_ + + a_-},
\ee
We now fix $a_-$ and $a_+$ in terms of \eqref{eq:Lambda12} (or \eqref{eq:LL2} in the main text):
\be
\label{eq:apm}
a_\pm = - \frac{2 \Lambda ^2 \pm 8 \sqrt{\Lambda ^3-32 \Lambda \Lambda^{(2)} +64 (\Lambda^{(2)})^2}}{(\Lambda -16) \Lambda +32 \Lambda^{(2)}},
\ee
so that
\be
\label{eq:A}
A(s) = \left({1 \over 4} - {4 \over \Lambda} + {8 \left(\Lambda^{(2)} \right) \over \Lambda^2}\right) \;+ \; {1 - 32 \Lambda^{(2)} / \Lambda^2 \over s (4 - s)} .
\ee

\begin{figure}[h]
    \centering
    \includegraphics[scale=0.7]{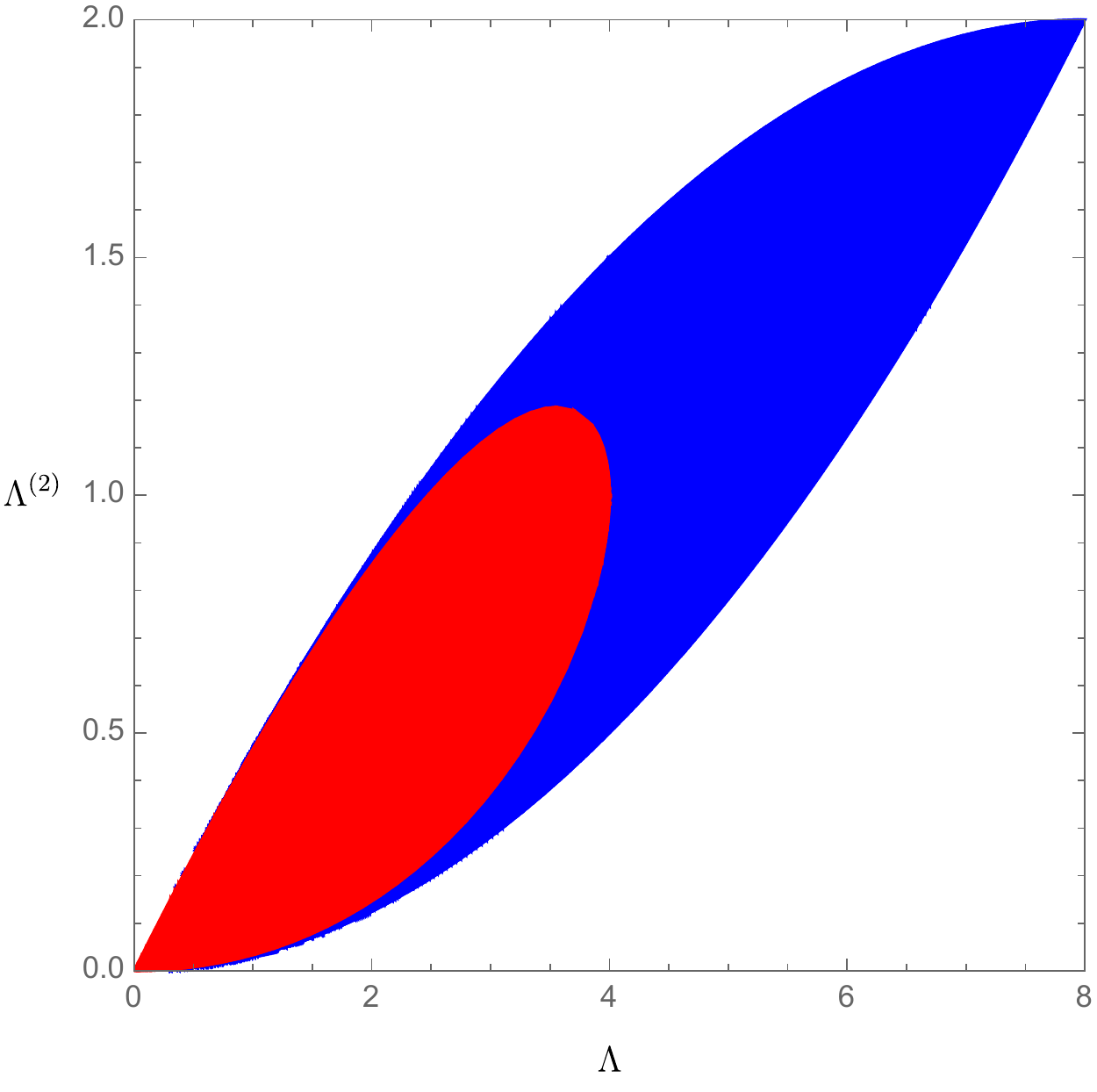}
    \caption{Allowed space for $(\Lambda, \Lambda^{(2)})$ such that the amplitude \eqref{eq:TAx} has no poles on the physical sheet. The blue region corresponds to the case where both CDD zeros are real, and is given by $\left(0<\Lambda <4\land \left(\frac{\Lambda ^2}{32}<\Lambda^{(2)}\leq \frac{\Lambda }{4}-\frac{1}{8} \sqrt{4 \Lambda ^2-\Lambda ^3}\lor \frac{1}{8} \sqrt{4 \Lambda ^2-\Lambda ^3}+\frac{\Lambda }{4}\leq \Lambda^{(2)}<\frac{1}{32} \left(16 \Lambda -\Lambda ^2\right)\right)\right)\lor \left(4\leq \Lambda <8\land \frac{\Lambda ^2}{32}<\Lambda^{(2)}<\frac{1}{32} \left(16 \Lambda -\Lambda ^2\right)\right)$. The red region corresponds to the case where the CDD zeros are complex, and is given by $0<\Lambda <4\land \frac{\Lambda }{4}-\frac{1}{8} \sqrt{4 \Lambda ^2-\Lambda ^3}<\Lambda^{(2)}<\frac{1}{8} \sqrt{4 \Lambda ^2-\Lambda ^3}+\frac{\Lambda }{4}$. Their union is given by eq. \eqref{eq:leaf_an}.}
    \label{leaf_app}
\end{figure}

\par
What remains to impose is the analyticity constraint. Concretely, for what values of $\Lambda$ and $\Lambda^{(2)}$ does $\mathcal{T}(s)$ or $S(s)$ not develop any other singularities on the physical sheet besides the physical cuts for $s \geq 4m^2$ and $s \leq 0$? We want $a_-$ and $a_+$ to be such that the denominator in eq. \eqref{eq:ST} is never zero on the physical sheet, i.e. that no pole develops and that the CDD factors remain CDD zeros.
\par
Note that the square root function only covers half of the complex plane. On the principal sheet it covers the right half plane, i.e. the real part of the square root function is always positive. This means that if 
\be
\label{eq:inel}
\mathrm{Re} \, a_- > 0 \qquad \text{and} \qquad  \mathrm{Re} \, a_+ > 0
\ee
we are sure that the denominator of \eqref{eq:TS} is never zero. Therefore, imposing \eqref{eq:inel} on \eqref{eq:apm} we find the allowed range for $\Lambda$ and $\Lambda^{(2)}$.
\par
So here we can split the problem in two. If the discriminant of \eqref{eq:apm} is positive then $a_\pm$ is real, meaning that if $a_\pm > 0$ the S-matrix will have two real zeros on the physical sheet. This region is represented in blue in fig. \ref{leaf_app}. 
\par
If the discriminant of \eqref{eq:apm} is negative then $a_\pm$ will be complex. However, if $\Re a_\pm > 0$, i.e. $(\Lambda - 16) \Lambda + 32 \Lambda^{(2)} < 0$, then the denominator of \eqref{eq:TS} will never be zero on the physical sheet. Instead, the numerator will have (complex) zeros. This region is represented in red in fig. \ref{leaf_app}.

\par
The union of these regions is the leaf in the main text, fig. \ref{dual_leaf}. Explicitly, in agreement with eqs. (A.8) and (A.10) of \cite{Chen:2022nym},
\be
\label{eq:leaf_an}
0<\Lambda <8 \;\; \land \; \; \frac{\Lambda ^2}{32}<\Lambda^{(2)}<  {16 \Lambda -\Lambda ^2 \over 32}
\ee

The optimal amplitude on the lower bound, i.e. when $\Lambda^{(2)} = {\Lambda^2 \over 32}$, and on the upper bound, i.e. when $\Lambda^{(2)} = {16 \Lambda -\Lambda ^2 \over 32}$ is given by eq. \eqref{eq:TAx} with, respectively,
\be
\label{eq:Aedge}
\text{Lower bound: } A(s) = {1 \over 2}\left(1 - {8 \over \Lambda}\right), \qquad \text{Upper bound: } A(s) = {2 \over s(4-s)} \left(1 - {8 \over \Lambda} \right)
\ee
It is straightforward to see that multiplying the 2-CDD ansatz \eqref{eq:aaS} by any further CDD zero will lead to values of $\Lambda$ and $\Lambda^{(2)}$ inside the leaf, and the same holds true if inelasticity is considered. So we conclude that the bounds are optimal.

\par
\subsubsection{$c_{UV}$ at the edges of the leaf}
\par
We can now find the two-particle stress tensor form factor $\mathcal{F}(s)$ on the edges of the leaf. We plug eqs. \eqref{eq:Aedge} for $A(s)$ at the edges of the leaf into eqs. \eqref{eq:Fan} and \eqref{eq:alphadisp}.
\par

We want minimize $c_{UV}$ over $B(s)$. First note that $B(s)$ must not have any poles outside the real cut, as it would be inconsistent with the analyticity assumptions for the form factor $\mathcal{F}(s)$. In principle it may have poles, e.g. at $s \to 4m^2$ or somewhere along the real axis, as long as as the integral \eqref{eq:cUVleaf} converges, which can only be the case if $e^{2 \bar{\alpha}(s)}$ cancels such poles. While we could not find an explicit form for $\bar{\alpha}(s)$ for either of the lower or uppers bounds \eqref{eq:Aedge} we do observe numerically that $e^{2 \bar{\alpha}(s)}$ is never zero, indicating that $B(s)$ cannot have poles and therefore be holomorphic, i.e. a polynomial.\footnote{ The only way we could make $e^{2 \bar{\alpha}(s)}$ have a zero was to have $A(s)$ be zero at some point, making the $\arccot$ function jump by $\pi$. However, $A(s)$ is sign definite at the edges of the leaf \eqref{eq:Aedge}. } Finally, we can constrain the polynomial degree of $B(s)$ according to the convergence of the integral \eqref{eq:cUVleaf}. 

\par
\textbf{Lower edge.} On the lower edge, we find that, according to eq. \eqref{eq:Imalpha}, $\Im \alpha(\infty) = \mathrm{arccot} \left(- \infty \right) = 0$, meaning that $\alpha(s \to \infty) \to \text{const}$ and $e^{2 \bar{\alpha}(s)} \to 1$. We conclude that the integrand of \eqref{eq:cUVleaf} will go as $\sim B^2(s) s^{-3}$ and, therefore, $B(s)$ must be a constant. Otherwise, the integral will not converge. Since $B(0) = 1$ we are led to the following unique choice on the lower edge,
\be
\text{Lower edge:} \qquad B(s) = 1.
\ee
So the form factor is entirely specified by $A(s)$,
\be
\mathcal{F}(s) = - 2m e^{\alpha(s) - \alpha(0)}.
\label{eq:F_loweredge}
\ee
As a simple check we may explicitly compute $\mathcal{F}(s)$ and $c_{UV}$ at the tips of the leaf, $\Lambda = 0$ and $\Lambda = 8$. In the former case we have $A(s) = {1 \over 2}\left(1 - {8 \over \Lambda}\right) \to - \infty$, for which $\mathrm{arccot}(-\infty) = 0$ and $\Im \alpha(s) = 0$. In this case eq. \eqref{eq:alphadisp} gives $\alpha(s) - \alpha(0) = 0$ and we have  $\mathcal{F}(s) = - 2m^2$ and $c_{UV} = 1$.
\par
For $\Lambda = 8$ instead we have $A(s) = 0^-$, meaning $\mathrm{arccot}(0^-) = - {\pi \over 2}$. From eq. \eqref{eq:alphadisp} it follows that 
\be
\alpha(s) - \alpha(0) = {1 \over 2} \log\left(1 - {s \over 4m^2} \right)
\ee
and 
\be
\label{eq:freefermionF}
\mathcal{F}(s) = - m \sqrt{4m^2 - s}, \qquad c_{UV} = 12m^4 \int_{4m^2}^\infty  \sqrt{s - 4m^2 \over s} {ds \over s^2} =  {1 \over 2}.
\ee
For $0 < \Lambda < 8$ we perform both integrals \eqref{eq:baralpha} and \eqref{eq:alphadisp} numerically. In fig. \ref{leaf_down} we show the optimal $S(s)$ and $\mathcal{F}(s)$ at $\Lambda = 2.4$.
\par

\textbf{Upper edge.} Let us now consider the upper edge of the leaf. In this case we have $A(s) = {2 \over s(4-s)} \left(1 - {8 \over \Lambda} \right) > 0$ across the integration region $s > 4m^2$ in \eqref{eq:alphadisp}. Now we find that $e^{2 \bar{\alpha}(s)} \sim s^{-1}$ when $s \to \infty$ so that the integrand of \eqref{eq:alphadisp} goes as $\sim B^2(s) s^{-4}$. We see that $B(s)$ can at most be linear in $s$ to ensure convergence of the integral. Therefore,
\be
\text{Upper edge:} \qquad B(s) = 1 - b \, s, 
\ee
where $b$ is a free real parameter.
\par
Plugging into \eqref{eq:cUVleaf} we have
\be
\label{eq:In}
c_{UV}(b) = I_0 - 2 b \,I_1 + b^2 \,I_2, \qquad \text{ with } \qquad I_n \equiv 12 m^4  \int_{4m^2}^\infty  {s^n \, e^{2 \bar{\alpha}(s)} \over s^2 \sqrt{s(s-4m^2)} } \,ds,
\ee
We now minimize the central charge with respect to $b$,
\be
\label{eq:bmin}
{d c_{UV} \over d b} = 0 \qquad \implies \qquad b_{min} = {I_1 \over I_2}, \qquad c_{UV}(b_{min}) = I_0 - {I_1^2 \over I_2}.
\ee
We may again test these expressions for the trivial cases $\Lambda = 0, 8$. For the former we have $A(s) \to \infty$, for which again $\Im \alpha(s) = 0$ and $\alpha(s) - \alpha(0) = 0$. In this case $I_2 \to \infty$ and we'll trivially get $b_{min} \to 0$ and $c_{UV}(b_{min}) = I_0 = 1$. For $\Lambda \to 8$ we have $A(s) \to 0^+$, for which $\mathrm{arccot}(0^+) = + {\pi \over 2}$. From eq. \eqref{eq:alphadisp} it follows that 
\be
\alpha(s) - \alpha(0) = -{1 \over 2} \log\left(1 - {s \over 4m^2} \right)
\ee
and
\be
\label{eq:leaf_upperedge_an}
e^{2 \bar{\alpha}(s)} = \left(1 - {s \over 4m^2} \right)^{-1}, \qquad
\mathcal{F}(s) = - 4m^2 {1 - b s \over \sqrt{4m^2 - s}}
\ee
Plugging into \eqref{eq:bmin} we find\footnote{Technically, in this limiting case, $I_n$ does not converge due to the end-point singularity at $s \to 4m^2$. However, upon regulating, the ratio $I_1 / I_2$ is finite and is given by $1 / 4m^2$.}
\be
\label{eq:bmin_upperedge}
b_{min} = {1 \over 4m^2},
\ee
i.e. that the polynomial $B(s)$ softens the singularity of integrand at $s \to 4m^2$,
and we recover eqs. \eqref{eq:freefermionF}.
\par
For $0 < \Lambda < 8$ we find $\bar{\alpha}(s)$ numerically eq. \eqref{eq:baralpha}. Then, we plug the result into eq. \eqref{eq:bmin} and compute the $I_n$'s numerically. In fig. \ref{leaf_up} we show the optimal $S(s)$ and $\mathcal{F}(s)$ for $\Lambda = 2.4$.

\par
\subsubsection{$c_{UV}$ in the interior of the leaf}
\par
To find $c_{UV}$ in the interior of the leaf we have to first find the optimal S-matrix. At the edges of the leaf the S-matrix is uniquely fixed because $\Lambda^{(2)}$ is either maximal (upper edge) or minimal (lower edge). In the interior of the leaf we consider extremizing over $\Lambda^{(4)}$, with
\be
\label{eq:Lambda124}
    \Lambda \equiv -\mathcal{T}(2), \qquad \Lambda^{(2)} \equiv \lim_{s\to 2} \frac{\partial^2 }{\partial s^2} \mathcal{T}(s), \qquad \Lambda^{(4)} \equiv \lim_{s\to 2} \frac{\partial^4 }{\partial s^4} \mathcal{T}(s),
\ee
under the hypothesis that $c_{UV}$ minimization for fixed $(\Lambda, \Lambda^{(2)})$ extremizes $\Lambda^{(4)}$.
\par
We can proceed as in section \ref{sec:leafcarve} and consider three CDD zeros which can be fixed in terms of the three parameters $(\Lambda, \Lambda^{(2)}, \Lambda^{(4)})$. Then explore the parameter-space for which these zeros remain zeros and do not turn into poles. Avoiding further technical details we find that
\be
\label{eq:Lambda4}
\frac{3 \left(-\Lambda ^3+24 \Lambda ^2-128 \Lambda  \Lambda^{(2)}+256 (\Lambda^{(2)})^2\right)}{128 (8-\Lambda )}\leq \Lambda^{(4)}\leq\frac{3 \left(-\Lambda ^3+128 \Lambda  \Lambda^{(2)}- 256 (\Lambda^{(2)})^2\right)}{128 \Lambda },
\ee
in agreement with eq. (A.12) of \cite{Chen:2022nym} which determined these bounds using the Schwarz-Pick theorem.
\par
We observe that the optimal amplitude coming from $c_{UV}$ minimization is the one for which $\Lambda^{(4)}$ is minimal, where $\Lambda^{(4)}$ assumes the lower bound of \eqref{eq:Lambda4}. The optimal amplitude at the lower bound is specified by
\be
\label{eq:A4}
A(s) = -\frac{4 (8-\Lambda)^2}{64 \Lambda +\Lambda ^2 \left(s^2-4 s-4\right)-32 \Lambda^{(2)} (s-2)^2},
\ee
for which the S-matrix $S(s)$ follows via \eqref{eq:TS}.  For $A(s)$ given by eq. \eqref{eq:A4} we have the same asymptotics of the upper edge of the leaf, $e^{2 \bar{\alpha}(s)} \sim s^{-1}$, so $B(s)$ is at most linear: $B(s) = 1 - b s$. As in the previous section, the form factor $\mathcal{F}(s)$ and the minimal $c_{UV}$ follow from eqs. \eqref{eq:Fan}, \eqref{eq:alphadisp}, \eqref{eq:In} and \eqref{eq:bmin}. In fig. \ref{leaf_int} we compare the analytical result with the output of dual optimization for $\Lambda = 4.56$ and $\Lambda^{(2)}=0.85$.

\subsection{Ising Field Theory}

Here the analyticity assumptions consist of a zero for the S-matrix $S(s)$ at $s = m^2(1-x)$, where $x$ is related to the magnetic field, and a pole at $s = m^2$, and their crossing-symmetric counterparts. Similarly, the form factor $\mathcal{F}(s)$, has a pole at $s= m^2$.
\par

\subsubsection{Minimization of the central charge $c_{UV}$}
The first problem consists of minimizing the central charge $c_{UV}$,
\be
\label{eq:cUV_IFT}
  c_{UV} = 12\pi \left( m^{-4} |\mathcal{F}_1^\Theta|^2 + \int_{4m^2}^\infty ds \frac{\rho(s)}{s^2} \right) 
\ee
where $\mathcal{F}_1^\Theta$ is the 1-particle form-factor, which is a constant. The residue of the 2-particle form factor $F(s)$ at $s \to m^2$ is given by $g \mathcal{F}_1^\Theta$, where $g$ is the cubic coupling, i.e. $- g^2$ is the residue of the S-matrix pole at $s \to m^2$.

The minimal choice for an S-matrix with the above properties, and which is also purely elastic, is the product of one CDD pole at $s = m^2$ and one CDD zero at $s = m^2(1-x)$,
\be
\label{eq:maximalS}
S_\pm(s) = \pm {m^2 \sqrt{(1-x)(3+x)} - \sqrt{s(4m^2-s)} \over m^2 \sqrt{(1-x)(3+x)} + \sqrt{s(4m^2-s)} } \cdot { \sqrt{3}  m^2 + \sqrt{s(4m^2-s)}  \over \sqrt{3} m^2  - \sqrt{s(4m^2-s)} }.
\ee
Now, near the pole we must have $S(s) \sim - {g^2 \over s - m^2}$, i.e. $S(s)$ must have a negative residue. Only $S_-(s)$ satisfies this requirement. We again observe that $S_-(s)$ matches with the optimal S-matrix in fig. \ref{S0v2} coming from minimizing the central charge. 
\par
By writing $S_-(s)$ in the form \eqref{eq:TS} we find (in units where $m=1$)
\be{}
\label{eq:Agx}
A(s) = - \frac{\sqrt{3}- \sqrt{(1-x) (x+3)}}{(s-4) s+ \sqrt{3(1-x) (x+3)}},
\ee
From eq. \eqref{eq:TS} we see that, near the pole, we have
\be
T(s \to m^2) \sim - {g^2 \over s - m^2},
\ee
with
\be
\label{eq:g2}
g^2  = \frac{4\sqrt{3} \left(3 - \sqrt{3(1-x) (x+3)}\right)^2}{x (x+2)} > 0
\ee
Now, given $A(s)$, the form factor $F(s)$ follows from eqs. \eqref{eq:Imalpha} and \eqref{eq:Fan}. The function $B(s)$ must contain the simple pole of $F(s)$ when $s \to m^2 = 1$ but no other singularity. We must therefore have
\be
B(s) = {P(s) \over  1 - s},
\ee
where $P(s)$ is a polynomial, for which $B(0) = 1$ implies $P(0) = 1$. If, furthermore,
\be
F(s \to 1) \sim -{g_F \over s - 1},
\ee
where $g_F \equiv g \mathcal{F}_1^\Theta$, we must have, from \eqref{eq:Fan},
\be
\label{eq:P1}
P(1) = -  { g_F \, e^{\alpha(0) - \alpha(1)} \over 2}.
\ee
To constrain further $P(s)$ we consider the central charge,
\be
\label{eq:cUVIFT}
c_{UV} = 12\pi {g_F^2 \over g^2} + 12 \int_{4}^\infty \frac{P^2(s)\,e^{2 \bar{\alpha}(s)}}{(1-s)^2\,s^2 \sqrt{s(s-4)}} \, ds.
\ee
Similarly to the previous section, we see that convergence of the above integral bounds the degree of $P(s)$. According to eq. \eqref{eq:Imalpha}, we see that $\Im \alpha(s \to \infty) = \arccot(0^-) = - {\pi \over 2}$ meaning that $e^{2 \bar{\alpha}(s)} \sim s$, when $s \to \infty$. The integrand in \eqref{eq:cUVIFT} will then go like $\sim P^2(s) s^{-4}$. In order for the integral to converge, we see that $P(s)$ must be at most linear,
\be
\label{eq:Plin}
P(s) = 1 - b s
\ee
where $b$ is some real constant parameter, which can be fixed in terms of $g_F$ via equation \eqref{eq:P1},
\be
\label{eq:b}
b = 1 +  { g_F \, e^{\alpha(0) - \alpha(1)} \over 2},
\ee
for which the form factor reads,
\be
\label{eq:Fan2}
\mathcal{F}(s) = - 2  e^{\alpha(s) - \alpha(0)} + { g_F s \over 1 - s} e^{\alpha(s) - \alpha(1)}
\ee
and the central charge \eqref{eq:cUVIFT} can be expressed as
\be
\label{eq:cUVgF}
c_{UV} = I_0 + g_F \left(e^{- \bar{\alpha}(1)} I_1 \right) + g_F^2 \left( {12 \pi \over g^2} + { e^{- 2 \bar{\alpha}(1)} \over 4} I_2  \right),
\ee
with
\be
I_n \equiv 12 \int_4^\infty {e^{2 \bar{\alpha}(s)} \over (s-1)^n s^{2 - n} \sqrt{s(s-4)}} ds.
\ee
We now minimize the central charge $c_{UV}$ with respect to $g_F$,
\be
\label{eq:cUVmin_analytic}
{d c_{UV}\over d g_F} = 0 \quad \implies \quad g_{F,min} = - {2 I_1 \over I_2 \,e^{- \bar{\alpha}(1)} + 48 \pi \, g^{-2}\, e^{\bar{\alpha}(1)}}, \qquad c_{UV, min} = I_0 - { I_1^2 \over I_2  + 48 \pi \, g^{-2}\, e^{2\bar{\alpha}(1)}},
\ee
with $g^2$ given by eq. \eqref{eq:g2}, and $\bar{\alpha}(s)$ given by eqs. \eqref{eq:baralpha} and \eqref{eq:Agx}.

\subsubsection{Bounds on $g_F$ for fixed central charge $c_{UV} = 1/2$ and maximal $g$}
\label{sec:ancuvg}

For a fixed central charge $c_{UV} = 1/2$ we find two possible roots for $g_F$ from eq. \eqref{eq:cUVgF}, 
\be
\label{eq:gFroots}
g_F^{(\pm)} = {- e^{- \bar{\alpha}(1)} I_1 \pm \sqrt{e^{- 2\bar{\alpha}(1)} I_1^2 - (48 \pi  g^{-2} +  e^{- 2 \bar{\alpha}(1)} I_2 )(I_0 - {1 \over 2}) }  \over 2 I_0 - 1},
\ee
where $g^2$ is maximal and given by \eqref{eq:g2} in terms of the zero $x$. For $x = 1/10$ we find
\be
\label{eq:gFroots01}
g_F^{(-)} = -0.011028..., \qquad g_F^{(+)} = 0.004658... .
\ee
These correspond to the leftmost and rightmost corners in fig. \ref{region_IFT}.

Note that fig. \ref{region_IFT} seems to indicate that there exists a line of solutions with the maximal coupling connecting the two extrema \eqref{eq:gFroots}, whereas our previous analysis only provides two isolated points for an S-matrix with maximal coupling. The solution to this apparent paradox lies in the possibility of having an S-matrix with cubic coupling arbitrarily close to the maximal one, but with different asymptotics, such that $P(s)$ in \eqref{eq:cUVIFT} can now be quadratic with a free parameter, instead of just linear as in \eqref{eq:Plin}.
\par
Take for example 
\be
S(s) =  S_-(s)\, {a_1 + a_2 \,s(4-s) - \sqrt{s(4-s)} \over a_1 + a_2 \, s(4-s) + \sqrt{s(4-s)} },
\ee
where $S_-(s)$ is the maximal S-matrix given by eq. \eqref{eq:maximalS}. Now, for $a_1 \to \infty$ we have $S(s) \to S_-(s)$. On the other hand, it is not hard to see that solving for $A(s)$ using eq. \eqref{eq:TS} leads to the following asymptotics
\be
A(s \to \infty) \to -\frac{\sqrt{3} -3 \sqrt{31}/10 -1/a_2}{ s^2}+O\left(\frac{1}{s^3}\right)
\ee
which only depends on $a_2$. In particular, if we take $a_2$ large enough we get back the previous case, for which $A(s \to \infty) \to 0^-$.  Now, however, $a_2$ can be tuned such that $A(s \to \infty) \to 0^+$ while also keeping $S(s)$ free of further poles, i.e. for $0 < a_2 < (\sqrt{3} - 3 \sqrt{31}/10)^{-1} \approx 16.2$. In this case we have  $\Im \alpha(s \to \infty) = \arccot(0^+) = + {\pi \over 2}$ so that $e^{2 \bar{\alpha}(s)} \sim s^{-1}$. The integrand in \eqref{eq:cUVIFT} will then go like $\sim P^2(s) s^{-6}$, which allows for a quadratic degree polynomial $P(s)$. Two of the coefficients of $P(s)$ are fixed by the normalization $P(0) = 1$ and by the residue of the form factor $g_F$, the remaining coefficient is free, i.e. it allows for a line of solutions with near-maximal cubic coupling (by letting $a_1$ be arbitrarily large) as observed in fig. \ref{region_IFT}.

\subsubsection{Small magnetic field limit $h \to 0$: Bounds on $g_F$ and extra zeros}

We first find how the analytical $g_F$ scales in the small magnetic field limit,
\be
g_F(x \to 0) = x \,\tilde{g}_F.
\ee
We assume this behavior for now, but will then confirm it by explicit computation. We start by noting that in the limit $x \to 0$, the cubic coupling $g^2$ \eqref{eq:g2} goes as 
\be
g^2(x \to 0) = 2 \sqrt{3} x
\ee
Then, the central charge \eqref{eq:cUVIFT} reads
\be
\label{eq:cUVIFT2}
c_{UV} \;= \; {6\pi \over \sqrt{3}} \tilde{g}_F^2 \, x \;+ \; 12 \int_{4}^\infty \frac{P^2(s)\,e^{2 \bar{\alpha}(s)}}{(1-s)^2\,s^2 \sqrt{s(s-4)}} \, ds \; + \; O(x^2).
\ee
We still have to determine the $x \to 0$ limit of  $P(s)$ and $\bar{\alpha}(s)$. Let us first consider $\bar{\alpha}(s)$. From \eqref{eq:Agx} we have
\be
A(s) = - {x / \sqrt{3} \over 3 + s(s-4)} + O(x^2)
\ee
Plugging into \eqref{eq:Imalpha} and expanding at small $x$
\be
\label{eq:Imalpha2}
\Im \alpha(s) = \left[-{\pi \over 2} \; + \;x \, {\sqrt{s(s-4)/3} \over 3 + s(s-4)} \;+ \;O(x^2) \right] \Theta(s-4)
\ee
We find
\be
\label{eq:alphas}
\alpha(s) - \alpha(0) \;=\; {1 \over 2} \log \left(1 - {s \over 4}\right) \;+\; x \, \big[\alpha_1(s) - \alpha_1(0)\big] \; + \; O(x^2)
\ee
with
\be
\alpha_1(s) = {\beta(s) \over (s - 3)(s-1)} - {\beta(3) \over 2 (s - 3)} + {\beta(1) \over 2 (s - 1)}, \qquad \beta(s) = -{2 \over \pi} \sqrt{s(4-s) \over 3} \, \arctan\sqrt{s \over 4 - s}.
\ee
Let us now take the limit $x \to 0$ of the polynomial $P(s) = 1 - b s$ in \eqref{eq:cUVIFT2}, where $b$ is given by \eqref{eq:b}. We have
\be
b =  1 \; +  \; x \, { \tilde{g}_F \, e^{\alpha(0) - \alpha(1)} \over 2}  \; + \; O(x^2) \;=\; 1\; + \;x \, {\tilde{g}_F \over \sqrt{3} } \; + \; O(x^2) 
\ee
where we made use of \eqref{eq:alphas}.
\par
We then have
\be
{P^2(s)\,e^{2 \bar{\alpha}(s)} \over (1 - s)^2} = {s - 4 \over 4} \; + \; x \, (s - 4) \left[{ \sqrt{3} \tilde{g}_F s  \over 6(s-1)} \; + \; {\mathrm{Re}[\alpha_1(s) - \alpha_1(0)]\over 2} \right] \; + \; O(x^2).
\ee
Now, replacing the above formulas into \eqref{eq:cUVIFT2} and performing the simple integrals, we find
\be
\label{eq:cUVIFT3}
c_{UV} - {1 \over 2} \;= \; x\left[ \, 2 \sqrt{3} \pi \, \tilde{g}_F^2 \: + \; (4\sqrt{3} - 2\pi) \, \tilde{g}_F \; - \; \frac{126 - 23 \sqrt{3} \pi}{81} - {2  \over \sqrt{3} \pi} I_t  \right] \; + \; O(x^2).
\ee
where
\be
I_t \; = \; \int_4^\infty \frac{s-4 }{s^2 \left(s^2-4 s+3\right)} \log\left({\sqrt{s-4} + \sqrt{s} \over 2}\right) ds \; = \; {23\over 648}  \pi^2  - {1 \over 3}.
\ee
\par
Notice that setting $x = 0$ in \eqref{eq:cUVIFT3} leads necessarily to $c_{UV} = 1/2$, the central charge of the free fermion. Since we are studying IFT we set $c_{UV} = 1/2$ regardless, which leads to a quadratic equation for the rate $\tilde{g}_F$ with the following roots
\be
\tilde{g}_F^{(\pm)} \; = \; \frac{1}{2 \sqrt{3}} -\frac{1}{\pi } \pm {1 \over \sqrt{3} \pi} \mp {1 \over 9}\;= \;( - \, 0.1023000, + \,0.0430304 \,)
\ee
To express this in terms of the magnetic field $h$ we make use of the relation $x = 36 \sqrt{3} (\mathcal{F}_1^\sigma)^2 \, h^2$, where $\mathcal{F}_1^\sigma = 1.3578$, to find the rates,
\be
g_F^{(\pm)} = x \, \tilde{g}_F^{(\pm)} \;= \;( - \, 11.7601, + \,4.9466 \,) \, h^2.
\ee

\par
\noindent
\textbf{Extra zeros.} Let us now see what we can say about potential extra zeros of the IFT S-matrix in the small magnetic field limit $h \to 0$. 
\par
For the case of IFT at lowest order in $h^2$ we have \cite{Zamolodchikov_2011} (see also eqs. \eqref{perturbation} and \eqref{Aw})
\be
\label{eq:SIFT}
S_\text{IFT}(s) = -1-i h^2 \sqrt{(s-4) s} \left(\frac{72 (\mathcal{F}^\sigma_1)^2}{s^2-4 s+3}+ \int_4^\infty  \frac{(s'-2) \, \sigma_{2\to 3}(s') \, ds'}{\pi  \sqrt{(s'-4) s'} (s'-s) (s+s'-4)}\right) + O(h^4)
\ee
with $\sigma_{2\to 3}(s)$ given by eq. \eqref{oskour}.
\par
Let us now match eq. \eqref{eq:SIFT} with the general nonperturbative solution to unitarity and crossing symmetry. In $d=2$ unitarity reads
\be
|S(s)|^2 = 1 - \sigma_{inel}(s) 
\ee
where $\sigma_{inel}(s) > 0$ is the particle production cross-section. The general solution to this equation and crossing symmetry $S(s) = S(4m^2 - s)$ reads \cite{Tourkine:2021fqh,PhysRevD.6.2763,Paulos_2017_I}
\be
\label{eq:Sgen}
S(s) = S_{CDD}(s) \, \exp\left[\int_{4m^2}^\infty {ds' \over 2\pi i} \log(1 - \sigma_{inel}(s')) \sqrt{s (s - 4m^2)\over s' (s' - 4m^2)} \left( {1\over s' - s} + {1 \over s' - (4m^2 - s)}  \right)\right]
\ee
where $|S_{CDD}(s)|^2 = 1$. Now, $S_{CDD}(s)$ can be any product of CDD factors. In particular for IFT, it should include the CDD pole at $s = m^2 = 1$ and the CDD zero \eqref{eq:zero}. So we take
\be
S_{CDD}(s) = S_{-}(s) \, {\sqrt{a (4 - a)} - \sqrt{s (4 - s)} \over \sqrt{a (4 - a)} + \sqrt{s (4 - s)} } \, {\sqrt{b (4 - b)} - \sqrt{s (4 - s)} \over \sqrt{b (4 - b)} + \sqrt{s (4 - s)} } \, \cdots
\ee
with $S_{-}(s)$ given by eq. \eqref{eq:maximalS} and the remaining zeros $a, b, \dots$ we wish to constrain as $h \to 0$. 
\par
Letting $\sigma_{inel} = h^2 \sigma_{2 \to 3} + O(h^4)$ and expanding \eqref{eq:Sgen} at small $h$, assuming $a = a_0 h^{-2}  + O(h^{-4})$, and likewise for $b$ and etc, we find
\be
S(s) - S_\text{IFT}(s) = 2 h^2 \sqrt{s(s-4)} \,\big( |a_0| + |b_0| + \dots \big) + O(h^4)
\ee
Since $S(s)$ and $S_\text{IFT}(s)$ must match at order $h^2$ we see that $a_0 = b_0 = \dots = 0$. 
\par
In case the zeros are complex we must have $a = b^*$ so that $S_{CDD}$ is real. Letting $a_0 = \alpha + i \beta$, and $b_0 = \alpha - i \beta$ we have
\be
|a_0| + |b_0| = \sqrt{a_0^2} + \sqrt{b_0^2} = \sqrt{\alpha^2 - \beta^2 + 2 i \alpha \beta} + \sqrt{\alpha^2 - \beta^2 - 2 i \alpha \beta} = 2 |\alpha|
\ee
Meaning that $\alpha = 0$, but the imaginary part $\beta$ of the zeros is unconstrained.
\par
Therefore, the (real part of the) extra zeros must go to infinity at least as $\sim h^{-4}$ when $h \to 0$.

\section{Dual Bootstrap of the Sine-Gordon model}
\label{sec:SG}
\begin{figure}[h]
    \centering
    \includegraphics[scale=1]{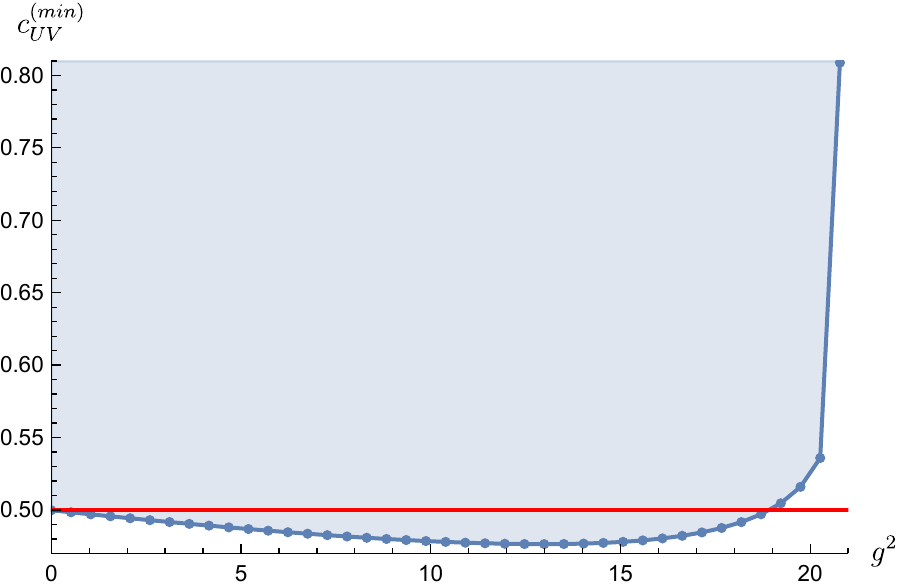}
    \caption{Lower bound on the central charge of the UV CFT for different values of the cubic coupling $g$ between two particles of mass $m_1=1$ and one particle of mass $m_2 = \sqrt{3}$. We used $N=50$ and the bound extends until $g^2=12 \sqrt{3}$ after which the problem is unfeasible. The red line at $c_{UV}=1/2$ is added for convenience.}
    \label{cUVSG}
\end{figure}

In this appendix we reproduce one of the results of \cite{Karateev_2020}, namely the one presented on their Figure 3, but with the dual setup. The assumed spectrum consists of 2 particles of mass $m_1=1$ and $m_2 = \sqrt{3}$ to target the Sine-Gordon theory. Only one pole in the scattering amplitude is assumed, the one corresponding to an exchange of the particle $m_2$. This residue $g$ is fixed and the central charge is minimized, which leads to Figure \ref{cUVSG}.

The maximal value of $g^2$ for which the problem is feasible and the corresponding minimal central charge are
\begin{equation}
\label{g2cUVSG}
    g^2 = 12 \sqrt{3}, \qquad c_{UV} = 0.808823...
\end{equation}
and they correspond to the cubic coupling and the contribution to the central charge of the Sine-Gordon model from  the particle $m_2$ and states with two particles $m_1$.

The advantage of our dual setup is that the bound on $c_{UV}$ is more accurate than what was presented in \cite{Karateev_2020}. Indeed at $g=0$ we have for sure the free Majorana fermion that is allowed and the corresponding central charge of $1/2$ should not be excluded. This is in agreement with our bound.

\begin{figure}[h]
    \centering
    \includegraphics[scale=1.5]{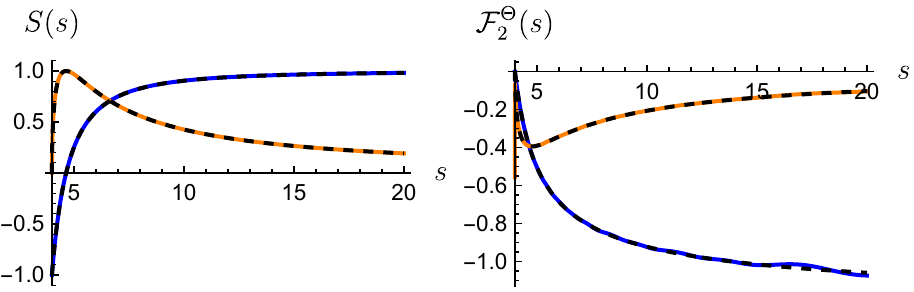}
    \caption{Scattering amplitude (on the left) and form factor (on the right) for the optimal $c_{UV}$ when $g^2 = 12\sqrt{3}$. The blue lines are the real parts and the orange lines are the imaginary parts. They correspond to the analytical solutions of the Sine-Gordon model with $m_1=1$ and $m_2 = \sqrt{3}$, that are plotted in dashed lines and given by eq. (4.18) and (4.19) in \cite{Karateev_2020}. We used $N=50$.}
    \label{S_F_SG}
\end{figure}

The optimal scattering amplitude and form factor are presented on Figure \ref{S_F_SG}, and correspond also to the analytical solutions in the Sine-Gordon model. 

\section{Dual optimization problems}
\label{sec:dual_problems}
In this appendix we derive various generalizations of the example in section \ref{sec:dual} and present the dual optimization problems.

\subsection{$\mathbb{Z}_2$ symmetric theories}
Here the goal is to minimize the central charge on top of the leaf parametrized by \eqref{analytic_leaf}. We do not assume any pole in the S-matrix and form factor. We have 
\begin{equation}
    \begin{split}
        \mathcal{L} =  12\pi \int_4^\infty ds \frac{\rho(s)}{s^2} &+ \int_4^\infty ds w_{\mathcal{T}}(s)\left[ \mathcal{T}(s)+\Lambda - \int_4^\infty \frac{dz}{\pi} \text{Im} \mathcal{T}(z)\left( \frac{1}{z-s} + \frac{1}{z+s-4}- \frac{2}{z-2} \right) \right] \\
        &+ \int_4^\infty ds w_{\mathcal{F}}(s) \left[\mathcal{F}(s)-\mathcal{F}(0) - \int_4^\infty \frac{dz}{\pi} \text{Im} \mathcal{F}\left( \frac{1}{z-s} - \frac{1}{z} \right) \right]\\
        &+ \int_4^\infty ds \operatorname{Tr}\mathbb{\Lambda}(s) \cdot B(s)  + b\left(\int_4^\infty \frac{dz}{\pi} \operatorname{Im}\mathcal{T}(z) \frac{4}{(z-2)^3}-\Lambda^{(2)}\right)
    \end{split}
\end{equation}
The dual scattering function is defined
\begin{equation}
    \begin{split}
        &W_{\mathcal{T}}(s) \equiv \int_4^\infty \frac{dz}{\pi} w_{\mathcal{T}}(z) \left( \frac{1}{z-s} - \frac{1}{z+s-4} +\frac{2}{s-2}\right), \\
        & \operatorname{Im}W_\mathcal{T}(s) = w_\mathcal{T}(s), \qquad \operatorname{Re}W_\mathcal{T}(s) = -P\int_4^\infty \frac{dz}{\pi} w_{\mathcal{T}}(z) \left( \frac{1}{s-z} + \frac{1}{s-(4-z)} -\frac{2}{s-2} \right),
    \end{split}
\end{equation}
where the last term is not eliminated as in the example in section \ref{sec:dual} since we do not extremize over the primal variable $\mathcal{T}(2)$ which is fixed to be $-\Lambda$.

The elimination of the primal variable $\mathcal{T}$ constrains the dual variables to obey
\begin{equation}
    \lambda_4 = \frac{\mathcal{N}_2}{2}\left( W_\mathcal{T} + \frac{4b}{\pi(s-2)^3} \right).
\end{equation}
We therefore get

\begin{mdframed}
\underline{Dual Problem (Minimization of the central charge on top of the leaf)}
\begin{equation}
\underset{\{\lambda_1, W_\mathcal{T}, W_\mathcal{F},b\}}{\text{Maximize}}\left[ \int_4^\infty ds \left( 2\lambda_1 + 2 \operatorname{Im}W_\mathcal{F} + \mathcal{N}_2 \operatorname{Re}W_\mathcal{T} +\Lambda \operatorname{Im}W_\mathcal{T} \right)
       +(1- \Lambda^{(2)}) b \right]
\end{equation}
Constrained by  
\begin{equation}
    \begin{pmatrix}
    \lambda_1& \frac{\mathcal{N}_2}{2} W_\mathcal{T} + \frac{2 \mathcal{N}_2b}{\pi(s-2)^3} & i \frac{\sqrt{\mathcal{N}_2}}{4} W_\mathcal{F} \\ \frac{\mathcal{N}_2}{2} W_\mathcal{T}^*+\frac{2 \mathcal{N}_2b}{\pi(s-2)^3}  & \lambda_1 & -i \frac{\sqrt{\mathcal{N}_2}}{4} W_\mathcal{F}^* \\ -i \frac{\sqrt{\mathcal{N}_2}}{4} W_\mathcal{F}^* &  i \frac{\sqrt{\mathcal{N}_2}}{4} W_\mathcal{F} & -\frac{6}{s^2}
    \end{pmatrix} \preccurlyeq 0, \quad \forall s \in [4,\infty),
\end{equation}
\end{mdframed}
where $\mathbb{\Lambda} \preccurlyeq 0$ means $-\mathbb{\Lambda} \succcurlyeq 0$.

\subsubsection*{Non-Linear dual for $\mathbb{Z}_2$ symmetric theories}
In the pure S-Matrix context the dual problem is simpler and in particular we can get the leaf by computing the extreme values for $\Lambda^{(2)} \equiv \mathcal{T}''(2)$ and then the extreme values for $\Lambda \equiv \mathcal{T}(2)$ with fixed $\Lambda^{(2)}$ within the allowed range.

For the first problem the lagrangian is 
\begin{equation}
    \mathcal{L} = \int_4^\infty \frac{dz}{\pi} \operatorname{Im}\mathcal{T}(z) \frac{4}{(z-2)^3} + \int_4^\infty ds \left( \lambda(s) U(s) + w(s) \mathcal{A}(s) \right),
\end{equation}
where $w$ and $\lambda \geq 0$ are Lagrange multipliers and  $\mathcal{A}$ is the usual analyticity and crossing constraint for the scattering amplitude and 
\begin{equation}
    U(s) \equiv 2 \operatorname{Im}\mathcal{T}(s) - \frac{|\mathcal{T}(s)|^2}{\mathcal{N}_2(s)}
\end{equation}
is the unitarity constraint in the pure S-Matrix setup. Using the equations of motion for $\mathcal{T}$ to eliminate it, we get the dual problem 
\begin{equation}
    \mathcal{L} = \int_4^\infty ds \mathcal{N}_2(s) \left[\sqrt{ (\operatorname{Im}W(s) )^2 + \left(\operatorname{Re}W(s)+\frac{4}{\pi (s-2)^3} \right)^2} + \operatorname{Re}W(s)+\frac{4}{\pi (s-2)^3} \right].
\end{equation}

For the second problem we start with the lagrangian 
\begin{equation}
    \mathcal{L} = \mathcal{T}(2) + b\left(\Lambda^{(2)} - \int_4^\infty \frac{dz}{\pi} \operatorname{Im}\mathcal{T}(z) \frac{4}{(z-2)^3} \right) + \int_4^\infty ds \left( \lambda(s) U(s) + w(s) \mathcal{A}(s) \right).
\end{equation}
Eliminating the primal variable $\mathcal{T}$ we get 
\begin{equation}
    \begin{split}
    \mathcal{L} &= \mathcal{T}(2) \left(1- \int_4^\infty ds \operatorname{Im}W(s) \right) + b\Lambda^{(2)}\\
    & + \int_4^\infty ds \mathcal{N}_2(s) \left[\sqrt{ (\operatorname{Im}W(s) )^2 + \left(\operatorname{Re}W(s)-\frac{4b}{\pi (s-2)^3} \right)^2} + \operatorname{Re}W(s)-\frac{4b}{\pi (s-2)^3} \right]
    \end{split}
\end{equation}

To summarize we have 

\begin{mdframed}
\underline{Dual Problem ($\Lambda^{(2)}$ maximization)}
\begin{equation}
\label{nonlindual1}
\begin{split}
\underset{\{W(s)\}}{\text{Minimize}} & \left(  \int_4^\infty ds \mathcal{N}_2(s) \left[\sqrt{ (\operatorname{Im}W(s) )^2 + \left(\operatorname{Re}W(s)-\frac{4}{\pi (s-2)^3} \right)^2} \right. \right. \\  & \left. \left. + \operatorname{Re}W(s)-\frac{4}{\pi (s-2)^3} \right]  \right)
\end{split}
\end{equation}
\end{mdframed}

\begin{mdframed}
\underline{Dual Problem ($\Lambda^{(2)}$ minimization)}
\begin{equation}
\begin{split}
\underset{\{W(s)\}}{\text{Minimize}} & \left(  \int_4^\infty ds \mathcal{N}_2(s) \left[-\sqrt{ (\operatorname{Im}W(s) )^2 + \left(\operatorname{Re}W(s)-\frac{4}{\pi (s-2)^3} \right)^2} \right. \right. \\  & \left. \left. + \operatorname{Re}W(s)-\frac{4}{\pi (s-2)^3} \right]  \right)
\end{split}
\end{equation}
\end{mdframed}

\begin{mdframed}
\underline{Dual Problem ($\Lambda$ maximization with fixed $\Lambda^{(2)}$)}
\begin{equation}
\begin{split}
\underset{\{b,W(s)\}}{\text{Minimize}} & \left( b\Lambda^{(2)} +  \int_4^\infty ds \mathcal{N}_2(s) \left[\sqrt{ (\operatorname{Im}W(s) )^2 + \left(\operatorname{Re}W(s)-\frac{4b}{\pi (s-2)^3} \right)^2} \right. \right. \\  & \left. \left. + \operatorname{Re}W(s)-\frac{4b}{\pi (s-2)^3} \right]  \right)
\end{split}
\end{equation}
Constrained by 
\begin{equation}
   \int_4^\infty ds \operatorname{Im}W(s) = 1
\end{equation}
\end{mdframed}

\begin{mdframed}
\underline{Dual Problem ($\Lambda$ minimization with fixed $\Lambda^{(2)}$)}
\begin{equation}
\begin{split}
\underset{\{b,W(s)\}}{\text{Minimize}} & \left( b\Lambda^{(2)} +  \int_4^\infty ds \mathcal{N}_2(s) \left[-\sqrt{ (\operatorname{Im}W(s) )^2 + \left(\operatorname{Re}W(s)-\frac{4b}{\pi (s-2)^3} \right)^2} \right. \right. \\  & \left. \left. + \operatorname{Re}W(s)-\frac{4b}{\pi (s-2)^3} \right]  \right)
\label{nonlindual2}
\end{split}
\end{equation}
Constrained by 
\begin{equation}
   \int_4^\infty ds \operatorname{Im}W(s) = 1
\end{equation}
\end{mdframed}

Following \cite{Guerrieri_2020}, for the $\Lambda^{(2)}$ maximization we use the ansatz 
\begin{equation}
    W(s) = \frac{1}{s(4-s)} \sum_{n=0}^N a_n \left(\rho^n(s,2)-\rho^n(4-s,2) \right),
\end{equation}
and for the $\Lambda$ maximization we use 
\begin{equation}
    W(s) = \frac{1}{s(4-s)} \left( \frac{8}{\pi (s-2)} +\sum_{n=0}^N a_n \left(\rho^n(s,2)-\rho^n(4-s,2) \right) \right),
\end{equation}
with $\rho(s,s_0)$ defined by \eqref{rhomapeq}.

\subsection{Ising Field Theory }

\subsubsection*{Zero in the S-Matrix at $s=1-x$ }
\label{sec:zero_dual}
To implement the zero in the S-matrix we can use the subtraction
\begin{equation}
\begin{split}
    \mathcal{T}(s)-\mathcal{T}(1-x)& = -g^2 \left( \frac{1}{s-1}+ \frac{1}{3-s}  -\frac{1}{1-x-1} -\frac{1}{3-(1-x)} \right) \\
    &- \int_4^\infty \frac{dz}{\pi} \operatorname{Im}T(z) \left( \frac{1}{z-s} + \frac{1}{z+s-4} -\frac{1}{z-1+x} -\frac{1}{z+1-x-4} \right),
\end{split}
\end{equation}
and we will fix $\mathcal{T}(1-x) = i \mathcal{N}_2(1-x)$ so that $S(1-x)=0$.
The dual scattering function is now defined as
\begin{equation}
    W_\mathcal{T}(s) = \int_4^\infty \frac{dz}{\pi} w_\mathcal{T}(z) \left( \frac{1}{z-s} -\frac{1}{z+s-4} + \frac{1}{s+x-1}+ \frac{1}{s-x-3} \right).
\end{equation}
Its ansatz therefore needs to have poles at $s=1-x$ and $s=3+x$ with residues $\int \operatorname{Im}W_\mathcal{T} / \pi$ and to be odd under crossing.
To moreover satisfy the fact that the objective must be integrable we can make the ansatz\footnote{In principle the coefficient $a$ should be fixed by the constraint on the residues. A direct computation gives
\begin{equation}
     \frac{1}{\pi} \int_4^\infty \operatorname{Im}W_\mathcal{T} = a  \frac{1}{4}\left( \frac{1}{3+x}+ \frac{1}{1-x} \right) = \frac{a }{(3+x)(1-x)}.
\end{equation}
On the other hand the residue is given by
\begin{equation}
    \underset{s= 1-x}{\operatorname{Res} W_\mathcal{T}(s)} = \frac{a}{(1-x)(4-1+x)},
\end{equation}
and the constraint is therefore satisfied for all $a$ that can be a free parameter.}
\begin{equation}
\label{ansatzzero}
     W_\mathcal{T}(s)= \frac{a}{s(4-s)}\left( \frac{1}{s+x-1} + \frac{1}{s-x-3} \right)+ \frac{1}{s(4-s)} \sum_{n=1}^{N} f_n ( \rho(s,2)^n- \rho(4-s,2)^n),
\end{equation}
where the second term is the ansatz that we make when we don't impose the zero, defined by \eqref{WT_nopole}.

The only modification to the lagrangian induced from the addition of the zero comes from the subtraction of $\mathcal{T}(1-x)$ which gives the new term
\begin{equation}
    \mathcal{L} \supset \int_4^\infty ds w_{\mathcal{T}}(s) \left(-\mathcal{T}(1-x) \right) = -i \mathcal{N}_2(1-x) \int_4^\infty \operatorname{Im} W_\mathcal{T}.
\end{equation}
In summary to implement the zero in the $S$-Matrix, it suffices to derive the problem without the zero and at the end we modify the lagrangian as
\begin{equation}
    \mathcal{L} \to   \mathcal{L} -i \mathcal{N}_2(1-x) \int_4^\infty \operatorname{Im} W_\mathcal{T},
\end{equation}
and use \eqref{ansatzzero} for $W_\mathcal{T}$.

\subsubsection*{Minimization of the central charge}

An interesting problem is the minimization of the UV central charge when there is the zero in the S-Matrix. We start with the lagrangian
\begin{equation}
    \mathcal{L} = 12\pi\left( \frac{g_F^2}{g^2} + \int_4^\infty \frac{\rho}{s^2} \right) + \int_4^\infty ( \mathcal{A}_\mathcal{T}w_\mathcal{T} + \mathcal{A}_\mathcal{F}w_\mathcal{F}  + \operatorname{Tr}\mathbb{\Lambda} B),
\end{equation}
and primal variables $\mathcal{T}$, $\mathcal{F}$ and $\rho$ are eliminated as before. It remains to eliminate $g_F$ and $g^2$. For the former, a functional variation with respect to $g_F$ yields
\begin{equation}
    g_F = - \frac{W_\mathcal{F}(1)}{24}g^2.
\end{equation}
Then, variation of $g^2$ yields the constraint
\begin{equation}
    48 W_\mathcal{T}(1) = W_\mathcal{F}(1)^2.
\end{equation}
It was appearing in the lagrangian as $g^2(W_\mathcal{T}(1) - W_\mathcal{F}(1)^2/48)$, with positive $g^2$, and we are minimizing over the primal variables $g^2$, so it can be implemented linearly with the positive semidefinite constraint
\begin{equation}
    \begin{pmatrix}
    W_\mathcal{T}(1) & W_\mathcal{F}(1) \\ W_\mathcal{F}(1) & 48
    \end{pmatrix} \succcurlyeq 0.
\end{equation}
Therefore 

\begin{mdframed}
\underline{Dual Problem (Minimization of the central charge $c_{UV}$ with a zero $S(1-x)=0)$}
\begin{equation}
\underset{\{\lambda_1, W_\mathcal{T}, W_\mathcal{F}\}}{\text{Maximize}}\left[ \int_4^\infty ds \left( 2\lambda_1 + 2 \operatorname{Im}W_\mathcal{F} + \mathcal{N}_2 \operatorname{Re}W_\mathcal{T} -i\mathcal{N}_2(1-x) \operatorname{Im}W_\mathcal{T} \right)
        \right]
\end{equation}
Constrained by 
\begin{equation}
    \begin{pmatrix}
    \lambda_1& \frac{\mathcal{N}_2}{2} W_\mathcal{T} & i \frac{\sqrt{\mathcal{N}_2}}{4} W_\mathcal{F} \\ \frac{\mathcal{N}_2}{2} W_\mathcal{T}^* & \lambda_1 & -i \frac{\sqrt{\mathcal{N}_2}}{4} W_\mathcal{F}^* \\ -i \frac{\sqrt{\mathcal{N}_2}}{4} W_\mathcal{F}^* &  i \frac{\sqrt{\mathcal{N}_2}}{4} W_\mathcal{F} & -\frac{6}{s^2}
    \end{pmatrix} \preccurlyeq 0, \quad \forall s \in [4,\infty), \qquad \begin{pmatrix}
    W_\mathcal{T}(1) & W_\mathcal{F}(1) \\ W_\mathcal{F}(1) & 48
    \end{pmatrix} \succcurlyeq 0.
\end{equation}
\end{mdframed}

\subsubsection*{Bounds in the $(g^2,g \mathcal{F}_1^\Theta)$ plane}

We will know derive the problem to get bounds on $g_F \equiv g \mathcal{F}_1^\Theta$ and $g^2$. We start by getting absolute bounds on $g_F$ and then we optimize $g^2$ for fixed values of $g_F$. We will impose the zero in the S-Matrix at the end.

The lagrangian can be written
\begin{equation}
    \begin{split}
        \mathcal{L} = g_F &+ \int_4^\infty ds w_{\mathcal{T}}(s)\left[ \mathcal{T}(s)-\mathcal{T}(2) +g^2\left( \frac{1}{s-1} + \frac{1}{3-s}-2 \right) \right.\\&  \left.- \int_4^\infty \frac{dz}{\pi} \text{Im} \mathcal{T}(z)\left( \frac{1}{z-s} + \frac{1}{z+s-4}- \frac{2}{z-2} \right) \right] \\
        &+ \int_4^\infty ds w_{\mathcal{F}}(s) \left[\mathcal{F}(s)-\mathcal{F}(0) +g_F\left( \frac{1}{s-1} + 1 \right)- \int_4^\infty \frac{dz}{\pi} \text{Im} \mathcal{F}\left( \frac{1}{z-s} - \frac{1}{z} \right) \right]\\
        &+ \int_4^\infty ds \operatorname{Tr}\mathbb{\Lambda}(s) \cdot B(s) \\
        &+ c_\rho \left[ 12\pi \left( \frac{g_F^2}{g^2} + \int_4^\infty ds \frac{\rho(s)}{s^2} \right)-c_{UV} \right],
    \end{split}
\end{equation}
where the last term constrains the UV central charge to be $c_{UV}$.
The extremization over the primal variable $\mathcal{T}(2)$ already gives 
\begin{equation}
    \int_4^\infty ds w_{\mathcal{T}}(s) =0.
\end{equation}

We can now define the dual scattering function 
\begin{equation}
        W_{\mathcal{T}}(s) \equiv \int_4^\infty \frac{dz}{\pi} w_{\mathcal{T}}(z) \left( \frac{1}{z-s} - \frac{1}{z+s-4} \right),
\end{equation}
and the dual form factor function 
\begin{equation}
        W_{\mathcal{F}}(s) \equiv \int_4^\infty \frac{dz}{\pi} w_{\mathcal{F}}(z)  \left(\frac{1}{z-s} + \frac{1}{s} \right).
\end{equation}

The analyticity constraints can then considerably be simplified. We get
\begin{equation}
    \begin{split}
        \mathcal{L} = g_F &+ \int_4^\infty ds \left( \operatorname{Im}(\mathcal{T} W_\mathcal{T}) + \operatorname{Im}(\mathcal{F} W_\mathcal{F})+ 2 \operatorname{Im}W_\mathcal{F} + \operatorname{Tr} \mathbb{\Lambda}(s)\cdot B(s)\right)\\
        &+ c_\rho \left[ 12\pi \left( \frac{g_F^2}{g^2} + \int_4^\infty ds \frac{\rho(s)}{s^2} \right)-c_{UV} \right] + \pi g^2 W_\mathcal{T}(1) +\pi g_F W_\mathcal{F}(1)
    \end{split}
\end{equation}

We are now ready to eliminate the primal variables. Varying with respect to $\mathcal{F}, \mathcal{T}$ and $\rho$ we get
\begin{equation}
    \lambda_4 = \frac{\mathcal{N}_2}{2}W_\mathcal{T}, \qquad \lambda_6 = i\frac{\sqrt{\mathcal{N}_2}}{4}W_\mathcal{F}, \qquad \lambda_3 = -\frac{6 C_\rho}{s^2}.
\end{equation}
The lagrangian is reduced to 
\begin{equation}
    \mathcal{L} = g_F + \int_4^\infty ds \left( 2\lambda_1 + 2 \operatorname{Im}W_\mathcal{F} + \mathcal{N}_2 \operatorname{Re}W_\mathcal{T} \right)
        + c_\rho \left( 12\pi \frac{g_F^2}{g^2}- c_{UV} \right) + \pi g^2 W_\mathcal{T}(1) +\pi g_F W_\mathcal{F}(1).
\end{equation}

To finalize the elimination of primal variables we need to do variations over $g_F$ and $g^2$. The former gives the equation
\begin{equation}
     g_F = -\frac{g^2}{24 \pi C_\rho}(1 + \pi W_\mathcal{F}(1) ),
\end{equation}
that yields the lagrangian
\begin{equation}
     \mathcal{L} = -\frac{g^2}{48 \pi C_\rho}(1+\pi W_\mathcal{F}(1))^2+ \int_4^\infty ds \left( 2\lambda_1 + 2 \operatorname{Im}W_\mathcal{F} + \mathcal{N}_2 \operatorname{Re}W_\mathcal{T} \right)
       - c_{UV} C_\rho + \pi g^2 W_\mathcal{T}(1).
\end{equation}
Now extremizing over $g^2$ we get the constraint
\begin{equation}
    C_\rho = \frac{ (1+\pi W_\mathcal{F}(1))^2}{48 \pi^2 W_\mathcal{T}(1)}.
\end{equation}
Therefore the lagrangian becomes
\begin{equation}
     \mathcal{L} = \int_4^\infty ds \left( 2\lambda_1 + 2 \operatorname{Im}W_\mathcal{F} + \mathcal{N}_2 \operatorname{Re}W_\mathcal{T} \right)
       - c_{UV} \frac{ (1+\pi W_\mathcal{F}(1))^2}{48 \pi^2 W_\mathcal{T}(1)}.
\end{equation}
All the primal variables are eliminated and we're ready to extremize over dual variables. The last term is not linear in those variables, and it's convenient to introduce a new variable $u$ such that 
\begin{equation}
    \begin{pmatrix} -u & 1+\pi W_\mathcal{F}(1) \\ 1+\pi W_\mathcal{F}(1) & -48\pi^2 W_\mathcal{T}(1) \end{pmatrix} \succcurlyeq 0. 
\end{equation}
We can finally add the zero in the scattering amplitude as described above and formulate the dual problem that can be implemented in SDPB : 

\begin{framed}
\underline{Dual Problem ($g_F$ maximization and $S(m^2(1-x))=0$)}
\begin{equation}
\underset{\{\lambda_1, W_\mathcal{T}, W_\mathcal{F},u\}}{\text{Minimize}}\left[ \int_4^\infty ds \left( 2\lambda_1 + 2 \operatorname{Im}W_\mathcal{F} + \mathcal{N}_2 \operatorname{Re}W_\mathcal{T} -i \mathcal{N}_2(1-x) \operatorname{Im}W_{\mathcal{T}}\right)
       - c_{UV} u \right]
\end{equation}
Constrained by 
\begin{equation}
    \begin{pmatrix}
    \lambda_1& \frac{\mathcal{N}_2}{2} W_\mathcal{T} & i \frac{\sqrt{\mathcal{N}_2}}{4} W_\mathcal{F} \\ \frac{\mathcal{N}_2}{2} W_\mathcal{T}^* & \lambda_1 & -i \frac{\sqrt{\mathcal{N}_2}}{4} W_\mathcal{F}^* \\ -i \frac{\sqrt{\mathcal{N}_2}}{4} W_\mathcal{F}^* &  i \frac{\sqrt{\mathcal{N}_2}}{4} W_\mathcal{F} & -\frac{6u}{s^2}
    \end{pmatrix} \succcurlyeq 0, \quad \forall s \in [4,\infty), \quad \begin{pmatrix} -u & 1+\pi W_\mathcal{F}(1) \\ 1+\pi W_\mathcal{F}(1) & -48\pi^2 W_\mathcal{T}(1) \end{pmatrix} \succcurlyeq 0,
\end{equation}
and \begin{equation}
    \int_4^\infty ds W_\mathcal{T}(s) =0.
    \label{imWT}
\end{equation}
\end{framed}

The minimization is similar, we just have to maximize the dual objective instead of minimizing and adjust the constraint on $u$ so that it becomes the correct form when the dual objective is maximized. The dual matrix $\mathbb{\Lambda}(s)$ needs also to be negative instead of positive, or equivalently $-\mathbb{\Lambda}$ is positive. We get

\begin{framed}
\underline{Dual Problem ($g_F$ minimization and $S(m^2(1-x))=0$)}
\begin{equation}
\underset{\{\lambda_1, W_\mathcal{T}, W_\mathcal{F},u\}}{\text{Maximize}}\left[ \int_4^\infty ds \left( 2\lambda_1 + 2 \operatorname{Im}W_\mathcal{F} + \mathcal{N}_2 \operatorname{Re}W_\mathcal{T} -i \mathcal{N}_2(1-x) \operatorname{Im}W_{\mathcal{T}}\right)
       - c_{UV} u \right]
\end{equation}
Constrained by 
\begin{equation}
    \begin{pmatrix}
    \lambda_1& \frac{\mathcal{N}_2}{2} W_\mathcal{T} & i \frac{\sqrt{\mathcal{N}_2}}{4} W_\mathcal{F} \\ \frac{\mathcal{N}_2}{2} W_\mathcal{T}^* & \lambda_1 & -i \frac{\sqrt{\mathcal{N}_2}}{4} W_\mathcal{F}^* \\ -i \frac{\sqrt{\mathcal{N}_2}}{4} W_\mathcal{F}^* &  i \frac{\sqrt{\mathcal{N}_2}}{4} W_\mathcal{F} & -\frac{6u}{s^2}
    \end{pmatrix} \preccurlyeq 0, \quad \forall s \in [4,\infty), \quad \begin{pmatrix} u & 1+\pi W_\mathcal{F}(1) \\ 1+\pi W_\mathcal{F}(1) & 48\pi^2 W_\mathcal{T}(1) \end{pmatrix} \succcurlyeq 0,
\end{equation}
and \begin{equation}
    \int_4^\infty ds W_\mathcal{T}(s) =0.
\end{equation}
\end{framed}

Now we have to maximize the cubic coupling $g^2$ with fixed $g_F$ and we start with the lagrangian
\begin{equation}
    \begin{split}
        \mathcal{L} = g^2 &+ \int_4^\infty ds w_{\mathcal{T}}(s)\left[ \mathcal{T}(s)-\mathcal{T}(2) +g^2\left( \frac{1}{s-1} + \frac{1}{3-s}-2 \right) \right. \\ & \left. - \int_4^\infty \frac{dz}{\pi} \text{Im} \mathcal{T}(z)\left( \frac{1}{z-s} + \frac{1}{z+s-4}- \frac{2}{z-2} \right) \right] \\
        &+ \int_4^\infty ds w_{\mathcal{F}}(s) \left[\mathcal{F}(s)-\mathcal{F}(0) +g_F\left( \frac{1}{s-1} + 1 \right)- \int_4^\infty \frac{dz}{\pi} \text{Im} \mathcal{F}\left( \frac{1}{z-s} - \frac{1}{z} \right) \right]\\
        &+ \int_4^\infty ds \operatorname{Tr}\mathbb{\Lambda}(s) \cdot B(s) \\
        &+ c_\rho \left[ 12\pi \left( \frac{g_F^2}{g^2} + \int_4^\infty ds \frac{\rho(s)}{s^2} \right)-c_{UV} \right] + C_g(g_F-g_F^*).
    \end{split}
\end{equation}
The primal variables are eliminated as before, and we arrive at 
\begin{equation}
\begin{split}
    \mathcal{L} &= g^2 + \int_4^\infty ds \left( 2\lambda_1 + 2 \operatorname{Im}W_\mathcal{F} + \mathcal{N}_2 \operatorname{Re}W_\mathcal{T} \right)
        + c_\rho \left( 12\pi \frac{g_F^2}{g^2}- c_{UV} \right) \\&+ \pi g^2 W_\mathcal{T}(1) +\pi g_F W_\mathcal{F}(1) + C_g(g_F-g_F^*).
\end{split}
\end{equation}
The variations with respect to $g_F$ and then $g^2$ give
\begin{equation}
    g_F = -\frac{g^2}{24 \pi C_\rho}\left( \pi W_\mathcal{F}(1) +C_g \right), \qquad C_\rho = \frac{(\pi W_\mathcal{F}(1)+C_g)^2}{48\pi +48\pi^2W_\mathcal{T}(1)}.
\end{equation}
Introducing a new variable $u$ to linearize the objective as before, we get
\begin{mdframed}
\underline{Dual Problem ($g^2$ maximization with fixed $g_F = g_F^*$ and $S(m^2(1-x))=0$)}
\begin{equation}
\underset{\{\lambda_1, W_\mathcal{T}, W_\mathcal{F}, C_g,u\}}{\text{Minimize}}\left[ \int_4^\infty ds \left( 2\lambda_1 + 2 \operatorname{Im}W_\mathcal{F} + \mathcal{N}_2 \operatorname{Re}W_\mathcal{T} -i \mathcal{N}_2(1-x) \operatorname{Im}W_{\mathcal{T}}\right)
       - c_{UV} u -C_gg_F^*\right]
\end{equation}
Constrained by 
\begin{equation}
    \begin{pmatrix}
    \lambda_1& \frac{\mathcal{N}_2}{2} W_\mathcal{T} & i \frac{\sqrt{\mathcal{N}_2}}{4} W_\mathcal{F} \\ \frac{\mathcal{N}_2}{2} W_\mathcal{T}^* & \lambda_1 & -i \frac{\sqrt{\mathcal{N}_2}}{4} W_\mathcal{F}^* \\ -i \frac{\sqrt{\mathcal{N}_2}}{4} W_\mathcal{F}^* &  i \frac{\sqrt{\mathcal{N}_2}}{4} W_\mathcal{F} & -\frac{6u}{s^2}
    \end{pmatrix} \succcurlyeq 0, \quad \forall s \in [4,\infty), 
\end{equation}
\be
\begin{pmatrix} -u & \pi W_\mathcal{F}(1) + C_g \\ \pi W_\mathcal{F}(1)+C_g & -48\pi - 48\pi^2 W_\mathcal{T}(1) \end{pmatrix} \succcurlyeq 0,
\ee 
and \begin{equation}
    \int_4^\infty ds W_\mathcal{T}(s) =0.
\end{equation}
\end{mdframed}

Again the minimization is similar and we get

\begin{mdframed}
\underline{Dual Problem ($g_T^2$ minimization with fixed $g_F = g_F^*$ and $S(m^2(1-x))=0$)}
\begin{equation}
\underset{\{\lambda_1, W_\mathcal{T}, W_\mathcal{F}, C_g,u\}}{\text{Maximize}}\left[ \int_4^\infty ds \left( 2\lambda_1 + 2 \operatorname{Im}W_\mathcal{F} + \mathcal{N}_2 \operatorname{Re}W_\mathcal{T} -i \mathcal{N}_2(1-x) \operatorname{Im}W_{\mathcal{T}}\right)
       - c_{UV} u -C_gg_F^*\right]
\end{equation}
Constrained by 
\begin{equation}
    \begin{pmatrix}
   - \lambda_1& -\frac{\mathcal{N}_2}{2} W_\mathcal{T} & -i \frac{\sqrt{\mathcal{N}_2}}{4} W_\mathcal{F} \\ -\frac{\mathcal{N}_2}{2} W_\mathcal{T}^* & -\lambda_1 & i \frac{\sqrt{\mathcal{N}_2}}{4} W_\mathcal{F}^* \\ i \frac{\sqrt{\mathcal{N}_2}}{4} W_\mathcal{F}^* &  -i \frac{\sqrt{\mathcal{N}_2}}{4} W_\mathcal{F} & \frac{6u}{s^2}
    \end{pmatrix} \succcurlyeq 0, \quad \forall s \in [4,\infty), 
\end{equation}
\begin{equation}
    \begin{pmatrix} u & \pi W_\mathcal{F}(1) + C_g \\ \pi W_\mathcal{F}(1)+C_g & 48\pi + 48\pi^2 W_\mathcal{T}(1) \end{pmatrix} \succcurlyeq 0,
\end{equation}
and \begin{equation}
    \int_4^\infty ds W_\mathcal{T}(s) =0.
\end{equation}
\end{mdframed}

\subsection{Two poles : minimization of the central charge}
We now consider the case with two particles having masses $m_1=1$ and $m_2 = \sqrt{3}$. There is only one pole in the scattering amplitude and in the form factor corresponding to an exchange of the second particle between two of the lightest particles. The pole terms in the dispersion relations are slightly modified and read 
\begin{equation}
\begin{split}
    \mathcal{T}(s)-\mathcal{T}(2) &= -g_{112}^2\left( \frac{1}{s-3} + \frac{1}{1-s}+2 \right)+ \int_4^\infty \frac{dz}{\pi} \text{Im} \mathcal{T}(z)\left( \frac{1}{z-s} + \frac{1}{z+s-4}- \frac{2}{z-2} \right), \\
\mathcal{F}(s)-\mathcal{F}(0) &= -g_F\left( \frac{1}{s-3} + \frac{1}{3} \right)+ \int_4^\infty \frac{dz}{\pi} \text{Im} \mathcal{F}\left( \frac{1}{z-s} - \frac{1}{z} \right).
\end{split}
\end{equation}

We therefore have the lagrangian
\begin{equation}
    \begin{split}
        \mathcal{L} &=  12\pi \left( \frac{g_F^2}{9 g_{112}^2} + \int_4^\infty ds \frac{\rho(s)}{s^2} \right) \\ &+ \int_4^\infty ds w_{\mathcal{T}}(s)\left[ \mathcal{T}(s)-\mathcal{T}(2) +g_{112}^2\left( \frac{1}{s-1} + \frac{1}{3-s}-2 \right) \right. \\ & \left. - \int_4^\infty \frac{dz}{\pi} \text{Im} \mathcal{T}(z)\left( \frac{1}{z-s} + \frac{1}{z+s-4}- \frac{2}{z-2} \right) \right] \\
        &+ \int_4^\infty ds w_{\mathcal{F}}(s) \left[\mathcal{F}(s)-\mathcal{F}(0) +g_F\left( \frac{1}{s-1} + 1 \right)- \int_4^\infty \frac{dz}{\pi} \text{Im} \mathcal{F}\left( \frac{1}{z-s} - \frac{1}{z} \right) \right]\\
        &+ \int_4^\infty ds \operatorname{Tr}\mathbb{\Lambda}(s) \cdot B(s).
    \end{split}
\end{equation}

Again the steps are similar and we obtain the dual problem

\begin{mdframed}
\underline{Dual Problem ($c_{UV}$ minimization with $m_1=1, m_2 = \sqrt{3}$ and $ g_{112}$ fixed)}
\begin{equation}
\underset{\{\lambda_1, W_\mathcal{T}, W_\mathcal{F},u\}}{\text{Maximize}}\left[ \int_4^\infty ds \left( 2\lambda_1 + 2 \operatorname{Im}W_\mathcal{F} + \mathcal{N}_2 \operatorname{Re}W_\mathcal{T} \right)
       - 3\frac{g_{112}^2}{16\pi} u + \pi g_{T}^2W_\mathcal{T}(3)\right]
\end{equation}
Constrained by 
\begin{equation}
    \begin{pmatrix}
    \lambda_1& \frac{\mathcal{N}_2}{2} W_\mathcal{T} & i \frac{\sqrt{\mathcal{N}_2}}{4} W_\mathcal{F} \\ \frac{\mathcal{N}_2}{2} W_\mathcal{T}^* & \lambda_1 & -i \frac{\sqrt{\mathcal{N}_2}}{4} W_\mathcal{F}^* \\ -i \frac{\sqrt{\mathcal{N}_2}}{4} W_\mathcal{F}^* &  i \frac{\sqrt{\mathcal{N}_2}}{4} W_\mathcal{F} & -\frac{6u}{s^2}
    \end{pmatrix} \preccurlyeq 0, \quad \forall s \in [4,\infty), 
\end{equation}
\begin{equation}
    \begin{pmatrix} 1 & \pi W_\mathcal{F}(3) \\ \pi W_\mathcal{F}(3) & u \end{pmatrix} \succcurlyeq 0,
\end{equation}
and \begin{equation}
    \int_4^\infty ds W_\mathcal{T}(s) =0.
\end{equation}
\end{mdframed}

\section{Numerical implementation}
\label{sec:numerics}
\begin{figure}[h]
\centering
\includegraphics[scale=0.5]{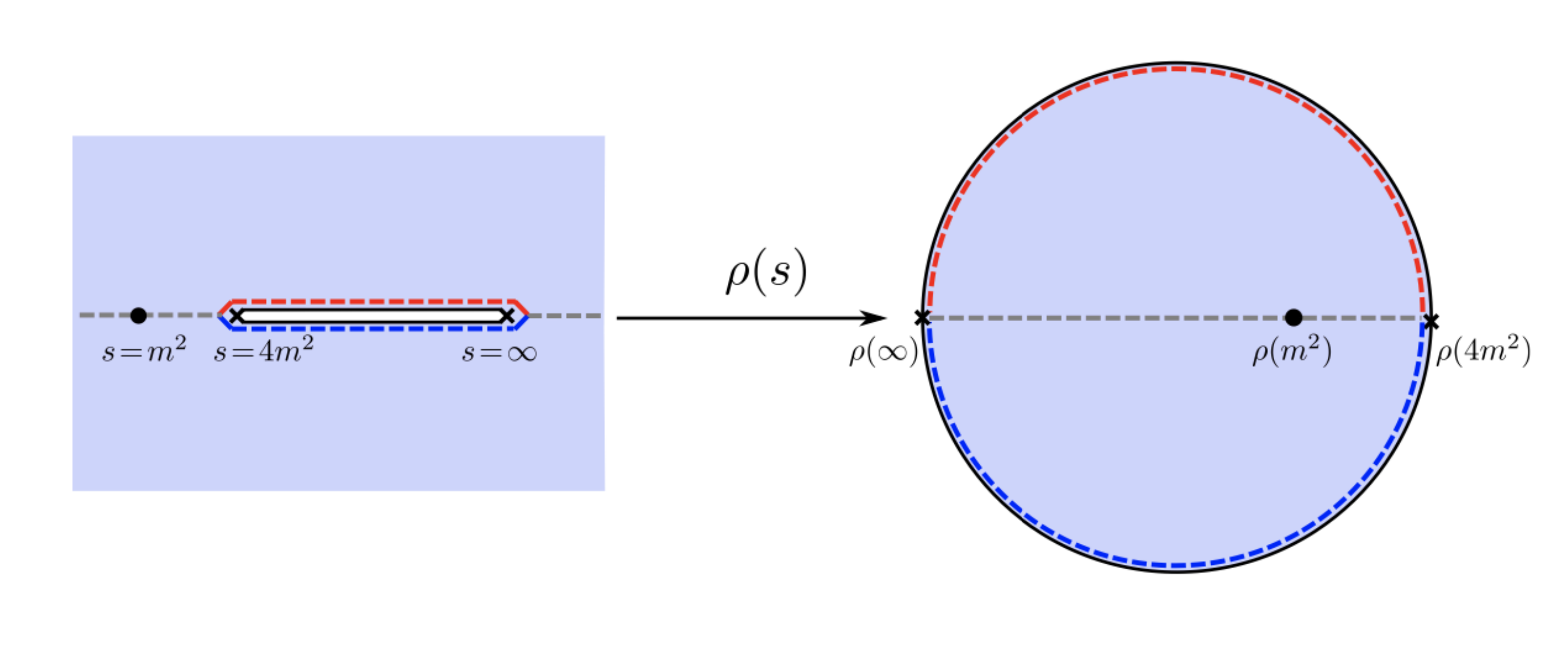}
\caption{Map $\rho(s)$ from the $s$ complex plane to the unit disk given in eq.\eqref{rhomapeq}. Figure  taken from \cite{Paulos:2017fhb}.}
\label{rhomap}
\end{figure}

We are now almost ready to use our linear dual formulation to get non perturbative bounds on different quantities. The last step is to implement the dual problem numerically. The dual functions $W_\mathcal{T}$ and $W_\mathcal{F}$ have branch cuts on the segment $[4, \infty)$ and it is therefore useful to use the new variable

\begin{equation}
\label{rhomapeq}
    \rho(s,s_0) \equiv \frac{ \sqrt{4-s_0} + \sqrt{4-s}}{ \sqrt{4-s_0} - \sqrt{4-s}}
\end{equation}
that maps the $s$ complex plane in the unit disk with the cut mapped on the boundary and that we depict on Figure \ref{rhomap} for $s_0=2$. Then
any function that is analytic apart for some poles and cuts starting at $s=4$ can be expressed as the sum of the pole terms
to which we add a Taylor series in the variable $\rho$.

Accounting for the fact that $W_\mathcal{T}$ is odd under crossing and $\mathcal{N}_2 \operatorname{Re}W_\mathcal{T}$ must be integrable, we make the ansatz
\begin{equation}
    W_\mathcal{T}(s) = \frac{1}{s(4-s)} \sum_{n=1}^{N} f_n ( \rho(s,2)^n- \rho(4-s,2)^n).
    \label{WT_nopole}
\end{equation}
In the case of $\mathbb{Z}_2$ odd theories, $W_\mathcal{T}$ also has a pole as can be seen from eq.\eqref{WTZ2}. To account for this singularity and match the residue, the ansatz is modified as
\begin{equation}
    W_\mathcal{T}(s) = -\frac{8}{\pi s(s-2)(s-4)}+\frac{1}{s(4-s)} \sum_{n=1}^{N} f_n ( \rho(s,2)^n- \rho(4-s,2)^n).
\end{equation}
As derived in \eqref{sec:zero_dual} when there is a zero in the scattering amplitude at $s=1-x$ the ansatz becomes 
\begin{equation}
     W_\mathcal{T}(s)= \frac{f_0}{s(4-s)}\left( \frac{1}{s+x-1} + \frac{1}{s-x-3} \right)+ \frac{1}{s(4-s)} \sum_{n=1}^{N} f_n ( \rho(s,2)^n- \rho(4-s,2)^n).
      \label{WT_pole}
\end{equation}
The second dual function $W_\mathcal{F}$ must have an integrable imaginary part, and the definition (\eqref{dualWF} implies the existence of a pole at $s=0$ with residue $\int \operatorname{Im}W_\mathcal{F} / \pi$. We therefore propose the ansatz 
\begin{equation}
\begin{split}
    W_\mathcal{F}(s) &= \frac{1}{s-4} \sum_{n=0}^{N} g_n \rho(s,0)^n + h_0 + \frac{1}{\pi s}\int_4^\infty ds \operatorname{Im} W_\mathcal{F}(s) \\
    &= \sum_{n=0}^{N} g_n \left( \frac{ \rho(s,0)^n}{s-4} - \frac{(-1)^n}{s} \right) + h_0,
\end{split}
\end{equation}
where we were careful to include the contribution of the pole term in the imaginary part  
\begin{equation}
    \frac{1}{4-s} = \frac{1}{4-\bar{s}-i\epsilon} = {\rm PV}\frac{1}{4-\bar{s}} + i \pi \delta(\bar{s}-4),
\end{equation}
where we wrote explicitely $s =\bar{s} + i\epsilon, \bar{s} \in \mathbb{R}$.
Finally the remaining dual variable $\lambda_1$ is a real function that needs to be integrable. We use the ansatz 
\begin{equation}
    \lambda_1(s) = \frac{1}{s \sqrt{s-4}} \left( \sum_{n=0}^{1}a_n \frac{ \rho(s,0)^n+ \rho(s,0)^{n,*}}{2} + \sum_{n=1}^{N} b_n \frac{ \rho(s,0)^n- \rho(s,0)^{n,*}}{2 i} \right).
    \label{lambda1}
\end{equation}
In some cases we observed the numerical convergence is improved if the ansatz for $\lambda_1$ is modified as 
\begin{equation}
    \Tilde{\lambda}_1(s) = \lambda_1(s) + \frac{1}{s^2} \sum_{n=1}^{10} c_n \frac{ \rho(s,0)^n- \rho(s,0)^{n,*}}{2i}.
    \label{lambda1improved}
\end{equation}
In this work we use \eqref{WT_pole} and \eqref{lambda1improved} in section \ref{sec:leaf},  and \eqref{WT_nopole} and \eqref{lambda1} in section \ref{sec:IFT} .

To implement the semipositive definite constraint on the interval $[4m^2,\infty)$, we discretize it on a Chebyshev grid with 200 points. We also checked, without rigorous analysis, that increasing this number does not change the results.  Then we optimize numerically over the free parameters $a_n,b_n,f_n,g_n ,h_0$ and eventually $c_n$ using SDPB to solve the dual optimization problem.

\section{Form factor perturbation theory }

\label{sec:perturbation_theory}

In this appendix we present our perturbative calculation for the one particle form factor $\mathcal{F}_1^\Theta$ and then we review Zamolodchikov's result for the perturbative scattering amplitude \cite{Zamolodchikov_2011}.\footnote{Form factor perturbation theory was first proposed in \cite{Delfino:1996xp} and was applied in different contexts (see e.g. \cite{Delfino:1995zk,Delfino:2005bh}).}

The Ising Field Theory has been extensively studied e.g. in \cite{Zamo_2013, Fonseca:2006au, Zamolodchikov_2011} for non integrable directions. The starting point for perturbation theory will be the action 
\begin{equation}
    S_{IFT} = S_{CFT}^{(c=1/2)} + \frac{m}{2\pi} \int d^2x \epsilon(x) +h \int d^2 x \sigma(x),
    \label{IFTaction}
\end{equation}
where the scaling dimensions of the operators $\epsilon$ and $\sigma$ are respectively 
\begin{equation}
 \label{scalingdimension}
    \Delta_{\epsilon} = 1, \qquad \Delta_\sigma = \frac{1}{8}.
\end{equation}

\subsection{Perturbation theory for $\mathcal{F}_1^\Theta$}
\label{sec:perturb_thy}

In the  interacting theory the trace of the stress energy tensor can be deduced from \eqref{IFTaction} and \eqref{scalingdimension} and takes the form\footnote{
 When $h=0$, one can check that the normalization for $\Theta$ is compatible with $\mathcal{F}_2^\Theta(s=0) = -2m^2$ and the CFT normalization of the operator $\epsilon$. On the one hand, we have \begin{equation}
    \langle \Theta(x) \Theta(0) \rangle_0 = \left( \frac{m}{2\pi} \right)^2 \langle \epsilon(x) \epsilon(0) \rangle \xrightarrow[x \to 0]{} \left( \frac{m}{2\pi} \right)^2 \frac{1}{|x|^2}\,.
\end{equation}
On the other hand, using that only the $n=2$ form factor $\mathcal{F}_2^\Theta = i  m\sqrt{s-4m^2}$  is non vanishing when $h=0$ we also have
\begin{equation}
    \langle \Theta(x) \Theta(0) \rangle_0 = \int_{4m^2}^\infty ds \frac{|\mathcal{F}_2^\Theta(s)|^2}{4\pi \sqrt{s(s-4m^2)}}\Delta_E(x,s) \xrightarrow[x \to 0]{} \left( \frac{m}{2\pi} \right)^2 \frac{1}{|x|^2},
\end{equation}
showing that the normalizations are compatible.} 
\begin{equation}
    \Theta(x) = \frac{m}{2\pi} \epsilon(x) + \frac{15h}{8}  \sigma(x).
\end{equation}

We can split the IFT action in the thermal deformation part $S_0$ and treat the magnetic deformation as a perturbation, which reads
\begin{equation}
    S = S_0 + h \int d^2x \sigma(x).
\end{equation}
\subsubsection*{First approach}
Our goal is now to use the two point function of the trace of the stress energy tensor to extract the one particle form factor $\mathcal{F}_1^\Theta$. To this end we can use the Euclidean spectral decomposition 
\begin{equation}
\label{2ptfctTheta}
\begin{split}
    \langle \Theta(x) \Theta(0) \rangle_c &= \int_0^\infty ds \rho_\Theta(s) \Delta_E(x,s) \\
    &= |\mathcal{F}_1^\Theta|^2 \Delta_E(x,m^2) + \mathcal{O}(e^{-2mx}),
\end{split}
\end{equation}
where we used to explicit form of the Euclidean propagator and its asymptotic form
\begin{equation}
   \label{euclidean_spectral}
    \Delta_E(x,m^2) = \frac{1}{2\pi} K_0(mx) \underset{x \to \infty}{\sim} \frac{1}{2\pi} \sqrt{\frac{\pi}{2mx}} e^{-mx}.
\end{equation}

On the other hand the two point function \eqref{2ptfctTheta} can be written as a path integral
\begin{equation}
    \langle \Theta(x) \Theta(0) \rangle = \frac{1}{Z} \int [\mathcal{D} \phi 
    ] e^{-S_0}\left(1-h \int d^2y \sigma(y) +\frac{h^2}{2} \int d^2 y d^2z \sigma(y) \sigma(z) + \mathcal{O}(h^3) \right) \Theta(x) \Theta(0).
\end{equation}
We can therefore get a perturbative expression for this two point function in terms of correlation functions in the theory with $h=0$, which we denote as $\langle ... \rangle_0$. We get
\begin{equation}
\label{twoptfctTheta2}
\begin{split}
     \langle \Theta(x) \Theta(0) \rangle_c &= \left( \frac{m}{2\pi} \right)^2 \langle \epsilon(x) \epsilon(0) \rangle_0 + \left( \frac{15 h}{8} \right)^2 \langle \sigma(x) \sigma(0) \rangle_0\\
     &-  \frac{15mh^2}{8\pi} \int d^2y \langle \epsilon(x) \sigma(0) \sigma(y) \rangle_0 \\
     &+ \left( \frac{m^2 h^2}{8\pi^2} \right) \int d^2 y d^2 z  \langle \epsilon(x) \epsilon(0) \sigma(y) \sigma(z) \rangle_0+ \mathcal{O}(h^3).
\end{split}
\end{equation}
Comparing \eqref{2ptfctTheta} and \eqref{twoptfctTheta2}, the one particle form factor can be computed as 
\begin{equation}
\label{f1theta}
    |\mathcal{F}_1^\Theta|^2 = h^2\left[ \left( \frac{15}{8} \right)^2 G^{(2)} -  \frac{15m}{8 \pi} G^{(3)} +  \frac{m^2}{8 \pi^2} G^{(4)} \right]+ \mathcal{O}(h^3),
\end{equation}
where the coefficients $G^{(i)}$ are defined as 
\begin{equation}
\label{largexlimit}
    \begin{split}
        & G^{(2)} \equiv \lim_{|x|\to \infty} \frac{1}{\Delta_E(x,m^2)}\langle \sigma(x) \sigma(0) \rangle_0, \\
        & G^{(3)} \equiv \lim_{x\to \infty} \frac{1}{\Delta_E(x,m^2)}\int d^2y \langle \epsilon(x) \sigma(0) \sigma(y) \rangle_0, \\
        & G^{(3)} \equiv \lim_{x\to \infty} \frac{1}{\Delta_E(x,m^2)}\int d^2y d^2z \langle \epsilon(x) \epsilon(0) \sigma(y) \sigma(z) \rangle_0 .
    \end{split}
\end{equation}

The two point function contribution is immediately obtained by using again the spectral representation
\begin{equation}
    G^{(2)} =\lim_{x\to \infty} \frac{1}{\Delta_E(x,m^2)} \int_0^\infty \rho_\sigma(s) \Delta_E(x,s) =  |\mathcal{F}_1^\sigma|^2.
\end{equation}
To evaluate the 3 and 4 point functions in the thermal deformation we can use the known form factors \cite{Mussardo:1281256, PhysRevD.19.2477}
\begin{equation}
\begin{split}
    & \mathcal{F}_2^\epsilon(p_1,p_2)  = -2\pi  \sqrt{4m^2- s_{12}},\\
    & \mathcal{F}_3^\sigma(p_1,p_2,p_3)  =  2\mathcal{F}_1^\sigma \sqrt{ \frac{4m^2 -s_{12}}{s_{12}}}\sqrt{ \frac{4m^2-s_{13}}{s_{13}}}\sqrt{ \frac{4m^2-s_{23}}{s_{23}}},\\
    & \mathcal{F}_1^\sigma =2^{1/12}e^{-1/8}m^{1/8}A_G^{3/2}  \approx 1.3578...,
\end{split}
\end{equation}
where $A_G$ is the Glaisher's constant and we use the $s$ Mandelstam variable defined as
\begin{equation}
    s_{12} \equiv -(p_1+p_2)^2= 2m^2 +2(E_{\bm{p}_1}E_{\bm{p}_2} - \bm{p}_1\bm{p}_2).
\end{equation}
We will also need the parity reversed 2-momentum $\Tilde{p}$ and the rest frame 2-momentum $\bar{k}$ defined as 
\begin{equation}
    p \equiv \begin{pmatrix} E_{\bm{p}} \\ \bm{p} \end{pmatrix}, \qquad \Tilde{p} \equiv \begin{pmatrix} E_{\bm{p}} \\ -\bm{p} \end{pmatrix}, \qquad \bar{k} \equiv \begin{pmatrix} m \\ 0 \end{pmatrix}.
\end{equation}

We can evaluate the correlation functions written as a time ordered vacuum   correlator. Without loss of generality, we choose $x$ to be pointing in the Euclidean time direction and we parametrize $y$ as 
\begin{equation}
    x = \begin{pmatrix} |x| \\ 0 \end{pmatrix}, \qquad y = \begin{pmatrix} y^1 \\ y^2 \end{pmatrix}.
\end{equation}
Then, we can write
\begin{equation}
\label{3ptfct}
\begin{split}
    \int d^2y \langle \epsilon(x) \sigma(0) \sigma(y) \rangle_0 &= \int dy^2 \int_{-\infty}^0 dy^1 \bra{0} \epsilon(x) \sigma(0) \sigma(y) \ket{0} \\& +\int dy^2 \int_{0}^{|x|} dy^1 \bra{0} \epsilon(x) \sigma(y) \sigma(0)  \ket{0}\\ &+\int dy^2 \int_{|x|}^\infty dy^1 \bra{0} \sigma(y) \epsilon(x) \sigma(0)  \ket{0}.
    \end{split}
\end{equation}
Using that the operators can be translated as \footnote{With these conventions,  $P_1$ is the  hamiltonian and $P_2$ is the spatial momentum and both are hermitian operators on the physical Hilbert space.}
\begin{equation}
    \mathcal{O}(y) = e^{P_1 y^1-iP_2y^2}\mathcal{O}(0) e^{-P_1 y^1 + iP_2y^2},
\end{equation}
the first term in \eqref{3ptfct} reads
\begin{equation}
\begin{split}
    G^{(3)}_{1} &\equiv \int dy^2 \int_{-\infty}^0 d y^1 \bra{0} \epsilon(x) \sigma(0) \sigma(y)  \ket{0} \\&= \int dy^2 \int_{-\infty}^0 dy^1 \int d \Phi_{12} d\Phi_3 \bra{0} \epsilon(0) \ket{p_1,p_2}_{in} {}_{in}\bra{p_1,p_2} \sigma(0) \ket{p_3}_{out}  {}_{out}\bra{p_3} \sigma(0) \ket{0}  \\ & \quad \times \quad  e^{-|x|(E_{\bm{p}_1} + E_{\bm{p}_2})} e^{y^1 E_{\bm{p}_3}} e^{-iy^2 \bm{p}_3 } + \mathcal{O}(e^{-2m|x|})\\
    &= \int d\Phi_{12} \frac{1}{4m^2}  e^{-|x| (E_{\bm{p}_1} + E_{\bm{p}_2})} \mathcal{F}_2^{\epsilon}(p_1,p_2) \mathcal{F}_3^{\sigma,*}(\bar{k},-p_1,-p_2) \mathcal{F}_1^\sigma + \mathcal{O}(e^{-2m|x|})= \mathcal{O}(e^{-2m|x|}).
\end{split}
\end{equation}
In the first line we inserted a complete set of states between each operator, using \eqref{completenessH} and the notation
\begin{align}
    d\Phi_{12\dots n} =\frac{1}{n!} \frac{d{\bm p}_1}{(2\pi) 2 E_{\bm{p}_1}}\dots
    \frac{d{\bm p}_n}{(2\pi) 2 E_{\bm{p}_n}}\,.
\end{align}
Between $\epsilon(x)$ and $\sigma(0)$ only two particle states can contribute because $\epsilon$ can only create two particles from the vacuum.
Between $\sigma(0)$ and $\sigma(y)$, there are many states that contribute to the correlator but only the single particle state can hope to survive the limit \eqref{largexlimit}.
In the second line we integrated over $y^2$ to get a spatial delta function that we used to integrate over $\bm{p}_3$, and we also integrated over $y^1$. 
We also used  \cite{Karateev_2020}
\begin{align}
   {}_{out}\bra{p_1,\dots, p_n} \mathcal{O}(0) \ket{k_1,\dots , k_m}_{in}  =\mathcal{F}_{m+n}^{\mathcal{O}}(p_1,\dots, p_n,-k_1, \dots, -k_m)\,,
\end{align}

for hermitian operators $\mathcal{O}$.
This contribution does not survive the limit \eqref{largexlimit}.

We now treat the second term in \eqref{3ptfct} that reads
\begin{equation}
\begin{split}
    G^{(3)}_{2} &\equiv \int dy^2 \int_{0}^{|x|} d y^1 \bra{0} \epsilon(x) \sigma(y) \sigma(0)  \ket{0} \\&= \int dy^2 \int_{0}^{|x|} dy^1\int d \Phi_{12} d\Phi_3 \bra{0} \epsilon(0) \ket{p_1,p_2}_{in} {}_{in}\bra{p_1,p_2} \sigma(0) \ket{p_3}_{out}  {}_{out}\bra{p_3} \sigma(0) \ket{0}  \\ & \quad \times \quad  e^{-|x|(E_{\bm{p}_1} + E_{\bm{p}_2})} e^{y^1(E_{\bm{p}_1}+E_{\bm{p}_2}-E_{\bm{p}_3})} e^{-iy^2(\bm{p}_1 + \bm{p}_2 - \bm{p}_3)} + \mathcal{O}(e^{-2m|x|})\\
    &= \int d\Phi_{12} \frac{1}{2E_{\bm{p}_1 + \bm{p}_2}} \frac{1}{E_{\bm{p}_1}+E_{\bm{p}_2}-E_{\bm{p}_1+\bm{p}_2}} 
    e^{-|x| E_{\bm{p}_1 + \bm{p}_2}} \mathcal{F}_2^\epsilon(p_1,p_2) \mathcal{F}_3^{\sigma,*}(p_3,-p_1,-p_2) \mathcal{F}_1^\sigma \\ & + \mathcal{O}(e^{-2m|x|}).
\end{split}
\end{equation}
where $p_3=(E_{\bm{p}_1 + \bm{p}_2}, \bm{p}_1 + \bm{p}_2)$.
The steps are identical as above and we now get a non-zero contribution to the limit \eqref{largexlimit}. A similar calculation shows that the third term \eqref{3ptfct} is
\begin{equation}
\begin{split}
    G^{(3)}_{3} &\equiv \int dy^2 \int_{|x|}^\infty dy^1 \bra{0} \sigma(y) \epsilon(x)  \sigma(0)  \ket{0} = \int d\Phi_1 \frac{1}{2m^2} e^{-E_{\bm{p}_1}|x|} (\mathcal{F}_1^\sigma)^2 \mathcal{F}_2^{\epsilon,*}(p_1,-\bar{k})
\end{split}
\end{equation}

To evaluate the $x\to \infty$ limit we can use that the $x$ dependence in the integrand only comes from the exponential. If the integral is only over $\bm{p}$ we have
\begin{align}
  &  \lim_{x\to \infty} \frac{1}{\Delta_E(x,m^2)}\int d\bm{p} f(\bm{p}) e^{-|x| E_{\bm{p}}} =
  \nonumber\\
  =& \lim_{x\to \infty} \frac{1}{\Delta_E(x,m^2)}\int \frac{d\bm{p}}{\sqrt{|x|}} f\left(\frac{\bm{p}}{\sqrt{|x|}}\right) e^{-|x|(m + \bm{p}^2/(2m|x|)) + \mathcal{O}(\bm{p}^3/\sqrt{|x|})} = 4\pi f(0),
\end{align}
where we rescaled the integration variable by $1/\sqrt{|x|}$ and expanded the energy with its Taylor series.
When we have an integral over $\bm{p}_1$ and $\bm{p}_2$ and the exponential damping contains $E_{\bm{p}_1 + \bm{p}_2}$, we get
\begin{equation}
     \lim_{x\to \infty} \frac{1}{\Delta_E(x,m^2)}\int d\bm{p}_1d\bm{p}_2 f(\bm{p}_1,\bm{p}_2) e^{-|x| E_{\bm{p}_1 + \bm{p}_2}} = 4\pi \int d\bm{p} f(\bm{p}, -\bm{p}).
\end{equation}
This result is obtained by first changing variables to $\bm{p}_1 +\bm{p}_2$ and $\bm{p}_1-\bm{p}_2$ and then rescaling as above.

Using these results we can take the limit \eqref{largexlimit} and get the contribution of the 3 point function that reads
\begin{equation}
\begin{split}
\label{3ptresult}
    G^{(3)} &= 4\pi   \frac{1}{8\pi m^3} \mathcal{F}_2^{\epsilon,*}(\bar{k},-\bar{k}) (\mathcal{F}_1^\sigma)^2  \\ &- 4\pi i \int d\bm{p} \frac{1}{2(2\pi 2E_{\bm{p}})^2}\frac{1}{2m} \frac{1}{2E_{\bm{p}}-m} \mathcal{F}_2^\epsilon( p,\Tilde{p}) \mathcal{F}_1^\sigma \mathcal{F}_3^{\sigma,*}(\bar{k},-p,-\Tilde{p})
    \end{split}
\end{equation}

The contributions from the 4 point function are similar. We have 
\begin{equation}
    \int d^2 y d^2 z \langle \epsilon(x) \epsilon(0) \sigma(y) \sigma(z) \rangle_0 = 2 \int dy^2 dz^2 \int_{-\infty}^\infty dy^1 \int_{-\infty}^{y^1} dz^1 \langle \epsilon(x) \epsilon(0) \sigma(y) \sigma(z) \rangle_0.
\end{equation}
Splitting up the integrals in time-ordered correlators, we get four contributing terms
\begin{equation}
\begin{split}
    G^{(4)}_1 &\equiv \int dy^2 dz^2\int_{0}^{|x|} dy^1 \int_{-\infty}^{0} dz^1 \bra{0} \epsilon(x) \sigma(y) \epsilon(0) \sigma(z) \ket{0}\\
    &= \int d\Phi_2 \frac{1}{2m^2} \frac{1}{E_{\bm{p}_1}+E_{\bm{p}_2}-E_{\bm{p}_1+\bm{p}_2}}  e^{-|x| (E_{\bm{p}_1 + \bm{p}_2})} \mathcal{F}_2^\epsilon(p_1,p_2) \mathcal{F}^{\sigma,*}_3(p_1,p_2,-p_3) \mathcal{F}_1^\sigma \mathcal{F}_2^\epsilon( -p_3,\bar{k}),
\end{split}
\end{equation}
where $p_3=(E_{\bm{p}_1 + \bm{p}_2}, \bm{p}_1 + \bm{p}_2)$,
\begin{equation}
\begin{split}
    G^{(4)}_2 &\equiv \int dy^2 dz^2\int_{0}^{|x|} dy^1 \int_{0}^{y^1} dz^1 \bra{0} \epsilon(x) \sigma(y) \sigma(z) \epsilon(0)  \ket{0}\\
    &= \int d\Phi_2 d\Phi_2' \frac{1}{2E_{\bm{p}_1+\bm{p}_2}} \frac{1}{E_{\bm{p}_1}+E_{\bm{p}_2}-E_{\bm{p}_1+\bm{p}_2}} \frac{1}{E_{\bm{p}_1}+E_{\bm{p}_2}-E_{\bm{p}'_1+\bm{p}'_2}} e^{-|x| (E_{\bm{p}_1 + \bm{p}_2})} \\& (2\pi) \delta(\bm{p}_1 + \bm{p}_2 -\bm{p}'_1 -\bm{p}'_2)  \quad \times \quad  \mathcal{F}_2^\epsilon(p_1,p_2)\mathcal{F}_2^{\epsilon,*}(p_1',p_2')\mathcal{F}^{\sigma}_3(p_1',p_2',-p_3')  \mathcal{F}^{\sigma,*}_3(p_1,p_2,-p_3),
\end{split}
\end{equation}
where $p_3=(E_{\bm{p}_1 + \bm{p}_2}, \bm{p}_1 + \bm{p}_2)$ and 
$p_3'=(E_{\bm{p}_1' + \bm{p}_2'}, \bm{p}_1' + \bm{p}_2')$,
\begin{equation}
\begin{split}
    G^{(4)}_3 &\equiv \int dy^2 dz^2 \int_{|x|}^\infty dy^1 \int_{-\infty}^{0} dz^1 \bra{0} \sigma(y) \epsilon(x)  \epsilon(0) \sigma(z) \ket{0}\\& = \int d\Phi_1 \frac{1}{4m^2} e^{-|x|E_{\bm{p}_1}} (\mathcal{F}_1^\sigma)^2 |\mathcal{F}_2^\epsilon(-p_1,\bar{k})|^2,
    \end{split}
\end{equation}
and finally
\begin{equation}
\begin{split}
    G^{(4)}_4 &\equiv \int dy^2 dz^2\int_{|x|}^{\infty} dy^1 \int_{0}^{|x|} dz^1 \bra{0} \sigma(y)\epsilon(x) \sigma(z) \epsilon(0)  \ket{0}\\
    &= \int d\Phi_2 \frac{1}{2m^2} \frac{1}{E_{\bm{p}_1}+E_{\bm{p}_2}-E_{\bm{p}_1+\bm{p}_2}}  e^{-|x| (E_{\bm{p}_1 + \bm{p}_2})} \mathcal{F}_2^\epsilon(p_1,p_2) \mathcal{F}^{\sigma,*}_3(p_1,p_2,-p_3) \mathcal{F}_1^\sigma \mathcal{F}_2^\epsilon( -p_3,\bar{k}) \\&= G^{(4)}_1,
\end{split}
\end{equation}
where $p_3=(E_{\bm{p}_1 + \bm{p}_2}, \bm{p}_1 + \bm{p}_2)$.
We can take the $|x|\to \infty$ limit and we get for the total 4 point function contribution 
\begin{equation}
\label{4ptresult}
\begin{split}
    G^{(4)} &= 32\pi \int d \bm{p}\frac{1}{2(2\pi 2E_{\bm{p}})^2} \frac{1}{4m^3} \frac{1}{2E_{\bm{p}}-2m} \mathcal{F}_2^\epsilon(p, \Tilde{p}) \mathcal{F}_3^{\sigma,*}(p,\Tilde{p},-\bar{k}) \mathcal{F}_2^\epsilon(-\bar{k},\bar{k}) \mathcal{F}_1^\sigma \\
    &+ 16\pi \int d\bm{p}_1 d\bm{p}_2 \frac{1}{2(2\pi 2E_{\bm{p}_1})^2}\frac{1}{2(2\pi 2E_{\bm{p}_2})^2} \frac{1}{2m}\frac{1}{2E_{\bm{p}_1}-m}\frac{1}{2E_{\bm{p}_2}-m} (2\pi) \\ & \quad \times \quad \mathcal{F}_2^\epsilon(p_1,\Tilde{p}_1) \mathcal{F}_2^{\epsilon,*}(p_2, \Tilde{p}_2)\mathcal{F}_3^{\sigma}(p_2,\Tilde{p}_2, -\bar{k})\mathcal{F}_3^{\sigma,*}(p_1, \Tilde{p}_1,-\bar{k})\\
    &+ 16\pi \frac{1}{2\pi 2m} \frac{(\mathcal{F}_1^\sigma)^2}{4m^4} |\mathcal{F}_2^\epsilon(-\bar{k},\bar{k})|^2.
\end{split}
\end{equation}

The integrals in \eqref{3ptresult}) and \eqref{4ptresult} can be evaluated and we find 
\begin{equation}
    \begin{split}
      G^{(3)} &= |\mathcal{F}_1^\sigma |^2\left(2\pi + 2(\sqrt{3}-1)\pi -1 \right), \\
      G^{(4)} &=  2|\mathcal{F}_1^\sigma |^2\left[\left(2(\sqrt{3}-1)\pi -1\right)^2 + 8\pi \left((\sqrt{3}-1)\pi -\frac{1}{2}\right) + 4\pi^2\right].
    \end{split}
\end{equation}
Plugging those results in \eqref{f1theta} we obtain 
\begin{equation}
\begin{split}
\label{perturbativenumber}
     |\mathcal{F}_1^\Theta| &=  h  |\mathcal{F}_1^\sigma| \sqrt{ \left( \frac{15}{8} -\sqrt{3}+\frac{1}{2\pi} \right)^2 } + \mathcal{O}(h^2)\\ &= h  |\mathcal{F}_1^\sigma| \times 0.302104... + \mathcal{O}(h^2)
    \end{split}
\end{equation}

\subsubsection*{Second approach}
This computation can be verified by doing an independent derivation. Our strategy is to follow the lines of perturbative quantum mechanics and generalize them to quantum field theory. We split the total hamiltonian as
\begin{equation}
    H = H_0 + hV, \qquad V = \int d\bm{x} \sigma(x),
\end{equation}
where $H_0$ is the hamiltonian of the free Majorana fermion described by the theory at $h=0$. We denote by $\ket{\psi}$ the eigenstates of the full hamiltonian $H$ with eigenvalues $E$ that can be expanded as
\begin{equation}
    \ket{\psi}= \ket{\psi}^{(0)} + h \ket{\psi}^{(1)} + \mathcal{O}(h^2), \qquad E= E^{(0)} + h E^{(1)}+\mathcal{O}(h^2),
\end{equation}
where $\ket{\psi}^{(0)}$ is an eigenstate of $H_0$ with eigenvalue $E^{(0)}$. In QFT a convenient basis for those states are the multi-particle states $\ket{p_1...p_n}$.  The form factor $\mathcal{F}_1^\Theta$ reads
\begin{equation}
\label{perturbativeQMF1theta}
\begin{split}
   \mathcal{F}_1^\Theta &= _{out}\bra{p} \Theta(0) \ket{0}_{in} = \left( _{out}^{(0)} \bra{p}  +  _{out}^{(1)}\bra{p} h \right) \left( \frac{m}{2 \pi} \epsilon(0) + \frac{15h}{8} \sigma(0) \right) \left( \ket{0}^{(0)}_{in}  + h \ket{0}^{(1)}_{in} \right) + \mathcal{O}(h^2) \\
   &= h \left[ \frac{m}{2\pi}\left( _{out}^{(0)}\bra{p} \epsilon(0) \ket{0}_{in}^{(1)} + _{out}^{(1)}\bra{p} \epsilon(0) \ket{0}_{in}^{(0)}\right) + \frac{15}{8} \mathcal{F}_1^\sigma \right] + \mathcal{O}(h^2).
\end{split}
\end{equation}
The problem we need to solve is now to compute the first correction to the energy eigenstates $\ket{p}^{(1)}$ and $\ket{0}^{(1)}$. For this we expand the Schrödinger equation 
\begin{equation}
    H \ket{\psi} = E \ket{\psi} \implies (H_0-E^{(0)}) \ket{\psi}^{(1)} = (E^{(1)} - V) \ket{\psi}^{(0)},
\end{equation}
where the implication follows by comparing terms in first order in $h$. Multiplying the equation by $^{(0)}\bra{\psi}$ and using that the form factors of $\sigma$ are zero for an even number of particles we get $E^{(1)}=0$. Inserting the identity \eqref{completenessH} we get
\begin{equation}
\label{eq40}
 \sumint \ket{p_1...p_n} (E_n^{(0)}-E^{(0)}) \braket{p_1...p_n}{\psi}^{(1)} = - \sumint \ket{p_1...p_n} \bra{n} V \ket{\psi}^{(0)},
 \end{equation}
 where $E_n \equiv E_{\bm{p}_1}+...+E_{\bm{p}_n}$. Here we are already at order $h$ so we can consider the states $\ket{p_1...p_n}$ as eigenstates of $H_0$. We need to be careful with degeneracies. For our problem $\ket{\psi}^{(0)}$ will either be a one particle state or the vacuum. Due to Lorentz invariance we can set $p=\bar{k}=(m,\bm{0})$ so that $E_{\bm{p}}^{(0)}=m$ will never be degenerate. The vacuum is assumed to have zero energy and is also never degenerate. If $\ket{p_1...p_n} \neq \ket{\psi}^{(0)}$, we can then invert \eqref{eq40} to get 
 \begin{equation}
     \braket{p_1...p_n}{\psi}^{(1)} = - \frac{\bra{p_1...p_n}V \ket{\psi}^{(0)}}{E_n^{(0)}-E^{(0)}}.
 \end{equation}
 This gives all the components of $\ket{\psi}^{(1)}$ except along $\ket{\psi}^{(0)}$. Assuming that the norm is conserved, ie $\braket{\psi}{\psi} = ^{(0)}\braket{\psi}{\psi}^{(0)}$ we get $^{(0)}\braket{\psi}{\psi}^{(1)}=0$. Therefore we found 
 \begin{equation}
 \label{statecorrection}
     \ket{\psi}^{(1)} = \sumint_{ n \neq \psi} \ket{p_1...p_n} \frac{ \bra{p_1...p_n}V \ket{\psi}^{(0)}}{E^{(0)}-E_n^{(0)}},
 \end{equation}
where we defined 
\begin{equation}
\sumint_{ n \neq \psi} \ket{p_1...p_n} \equiv \sumint \ket{p_1...p_n} - \frac{ \ket{\psi}^{(0)}}{{}^{(0)}\braket{\psi}{\psi}^{(0)}} \qquad \text{such that} \qquad  ^{(0)}\bra{\psi}\sumint_{ n \neq \psi} \ket{p_1...p_n} = 0.
\end{equation}
The result \eqref{statecorrection} is the direct generalization of the familiar quantum mechanical result. We can now proceed to the evaluation of the matrix elements in \eqref{perturbativeQMF1theta}. First we have 
\begin{equation}
    ^{(0)}_{out} \bra{ \bar{k}} \epsilon(0) \ket{0}^{(1)}_{in} = -\int \frac{d\bm{p}_1}{(2\pi) 2E_{\bm{p}_1}} {}_{out}\bra{\bar{k}} \epsilon(0) \ket{p_1}_{int} \frac{_{in} \bra{p_1}V \ket{0}_{out}}{E_{\bm{p}_1}},
\end{equation}
where we dropped the 0 superscripts in the RHS because everything is in the free theory sector and we used that $\epsilon$ only interpolates between free particles and $\ket{0}_{in} = \ket{0}_{out}$ in free theory. To continue we compute
\begin{equation}
    \bra{p_1}V \ket{0}_{out} = \int d\bm{x} e^{-i \bm{x} \bm{p}_1} \mathcal{F}_1^\sigma = (2\pi) \delta(\bm{p}_1) \mathcal{F}_1^\sigma,
\end{equation}
where we used the time independence of the Hamiltonian to set $t=0$ in $\sigma(t,\bm{x})$ and $\mathcal{F}_1^{\sigma,*} = \mathcal{F}_1^\sigma$. Using this spatial delta function to perform the integral we get 
\begin{equation}
    ^{(0)}_{out} \bra{ \bar{k}} \epsilon(0) \ket{0}^{(1)}_{in} = -\frac{1}{2m^2} \mathcal{F}_2^\epsilon(\bar{k},-\bar{k}) \mathcal{F}_1^\sigma = \frac{2\pi}{m} \mathcal{F}_1^\sigma.
\end{equation}
The second term in \eqref{perturbativeQMF1theta} is computed by similar methods and we get 
\begin{equation}
\begin{split}
     ^{(1)}_{out} \bra{ \bar{k}} \epsilon(0) \ket{0}^{(0)}_{in} &= -\frac{1}{2} \int \frac{d\bm{p}_1}{(2\pi) 2E_{\bm{p}_1}} \frac{1}{2E_{\bm{p}_1}} \frac{1}{2E_{\bm{p}_1}-m} \mathcal{F}_3^{\sigma,*}(p_1, \Tilde{p}_1, - \bar{k}) \mathcal{F}_2^\epsilon(p_1, \Tilde{p}_1) \\&= \left(1 -2\pi(\sqrt{3}-1)\right) \mathcal{F}_1^\sigma.
     \end{split}
\end{equation}
Plugging those results back in \eqref{perturbativeQMF1theta} we get 
\begin{equation}
    \mathcal{F}_1^\Theta = h \mathcal{F}_1^\sigma \left( \frac{15}{8} -\sqrt{3}+\frac{1}{2\pi} \right),
\end{equation}
which is identical to our previous result \eqref{perturbativenumber}.

\subsection{Perturbation theory for $S(s)$}
 The first order correction to the scattering amplitude in presence of a small magnetic field was computed in \cite{Zamolodchikov_2011}. More precisely they derived the term $S^{(1)}(s)$ in 
\begin{equation}
    S(s) = -1 + h^2S^{(1)}(s) + \mathcal{O}(h^4).
    \label{316}
\end{equation}
We introduce the change of variables 
\begin{equation}
    s \mapsto w(s) \equiv \frac{s(s-4m^2)}{4m^4}, \qquad w \mapsto s(w) \equiv 2m^2(1+\sqrt{1+w}).
    \label{variableperturb}
\end{equation}
The correction $S^{(1)}$ can then be written 
\begin{equation}
    S^{(1)}(s) = -\frac{iA(w(s))}{\sqrt{w(s)}},
    \label{perturbation}
\end{equation}
where  
\begin{equation}
    A(w) = \frac{rw}{w+ \frac{3}{4}} + w\int_{45/4}^{\infty}\frac{dv}{2\pi} \frac{\sigma_{2\to 3}(s(v))}{(v-w)\sqrt{v}},
    \label{Aw}
\end{equation}
where $\sigma_{2\to 3}$ is the part of the inelastic cross section giving the total probability to scatter 2 particles and end up with 3 particles, and $s(v)$ is the function defined in \eqref{variableperturb}.
The numerator of the pole term is given by
\begin{equation}
    r = 36 |\mathcal{F}_1^\sigma|^2,
\end{equation}
and the inelastic scattering cross section can be written as
\begin{equation}
\begin{split}
    \sigma_{2\to 3}(s) &= B(s) I(s) \\
    B(s) &= \frac{4 |\mathcal{F}_1^\sigma|^2}{\pi} \frac{ (\sqrt{s} + 2)^{\frac{5}{2}}(2\sqrt{s}-1)^4(\sqrt{s}-3)^3}{(\sqrt{s}-2)^{\frac{3}{2}}(\sqrt{s}+1)(\sqrt{s}-1)^{\frac{5}{2}}(\sqrt{s}+3)^{\frac{3}{2}}s^{\frac{3}{2}}} \\
    I(s) &= \int_{-1}^{1}dt \left(\frac{1-\mu t^2}{1-\nu t^2} \right)^2 \frac{ \sqrt{1-t^2}}{(1-\lambda t^2)^{\frac{5}{2}}} \\
    \lambda &= \frac{(\sqrt{s}+1)(\sqrt{s}-3)^3}{(\sqrt{s}-1)(\sqrt{s}+3)^3}, \qquad \mu = \frac{(\sqrt{s}-2)(2\sqrt{s}+1)^2}{(\sqrt{s}+2)(2\sqrt{s}-1)^2}\lambda, \qquad \nu = \frac{\sqrt{s}+2}{\sqrt{s}-2}\lambda.
\end{split}
\label{oskour}
\end{equation}

From this result we can find out a perturbative expression for the residue $g^2$ of the scattering amplitude 
\begin{equation}
    \mathcal{T}(s) = -\frac{g^2}{s-m^2} + ... \implies S(s) = -\frac{i}{\mathcal{N}_2}\frac{g^2}{s-m^2} + ...
\end{equation}
where the dots denote all terms that are not the pole term.
Using \eqref{variableperturb} we have $\mathcal{N}_2 = 4\sqrt{w} $. Therefore we get
\begin{equation}
\begin{split}
    S(s) &= -\frac{ih^2 A(w(s))}{\sqrt{w(s)}} +... = -ih^2r \frac{4}{\mathcal{N}_2}\frac{s(s-4m^2)}{s(s-4m^2)+3m^4}+ ... \\
    \implies \frac{g^2}{s-m^2}+... &= 4h^2r\left(1-\frac{3}{s(s-4m^2)+3m^4} \right)+ ... = 12h^2r \frac{1}{(s-m^2)(s-3m^2)}+... \\& = \frac{6h^2r}{s-m^2}+...
\end{split}
\end{equation}
Therefore we have 
\begin{equation}
    g^2 = 6rh^2 + \mathcal{O}(h^4).
    \label{gsq_perturbation}
\end{equation}
We can also get a relation between $h^2$ and the position of the zero in the S-matrix parametrized by $x$. Indeed setting $S(m^2(1-x)) = 0$ we get 
\begin{equation}
    -1 - 4m^4ih^2\frac{A(s=m^2(1-x))}{\sqrt{m^2(1-x)}{\sqrt{m^2(1-x)-4m^2}}} = 0,
\end{equation}
where we abused the notation by setting $A(s) \equiv A(w(s))$. Solving for $h^2$ we have 
\begin{equation}
    h^2 = -\frac{1}{2}\sqrt{1-x}\sqrt{3+x} \frac{1}{A(s= m^2(1-x))}.
    \label{hfield}
\end{equation}
If $x$ is sufficiently small we can expand around $x=0$ to get
\begin{equation}
    h \approx \frac{3^{-1/4}}{ 6 \mathcal{F}_1^\sigma} \sqrt{x}.
\end{equation}

\section{Integral representation for $c_{UV}$: c-sum rule}
\label{sec:csum}

Here we review the argument from \cite{PhysRevLett.60.2709} to derive a sum rule relating the two point function of the trace of the stress energy tensor to the central charge of the UV CFT.

In Euclidean space and complex coordinates the conservation of the stress energy tensor becomes
\begin{equation}
    \bar{\partial} T + \frac{\pi}{2} \partial \Theta = 0,
\end{equation}
where we defined $T(z, \bar{z}) \equiv 2\pi T_{zz}(z, \bar{z})$ and $\bar{T}(z, \bar{z}) \equiv 2\pi T_{\bar{z}\bar{z}}(z, \bar{z})$. This gives the following relations between two point functions
\begin{equation}
     \langle \bar{\partial} T(z, \bar{z}) T(0,0) \rangle = -\frac{\pi}{2}  \langle \partial \Theta(z, \bar{z}) T(0,0) \rangle, \qquad 
     \langle \bar{\partial} T(z, \bar{z}) \Theta(0,0) \rangle = -\frac{\pi}{2}  \langle \partial \Theta(z, \bar{z}) \Theta(0,0) \rangle.
     \label{relT}
\end{equation}

On the other hand we know how $T$ and $\bar{T}$ transform under rotations, and it constrains the two point functions to take the form
\begin{equation}
        \langle T(z, \bar{z}) T(0,0) \rangle = \frac{F(z \bar{z})}{z^4}, \qquad
         \langle T(z, \bar{z}) \Theta(0,0) \rangle = \frac{G(z \bar{z})}{z^3 \bar{z}}, \qquad
          \langle \Theta(z, \bar{z}) \Theta(0,0) \rangle = \frac{H(z \bar{z})}{z^2 \bar{z}^2},
          \label{invrot}
\end{equation}
where $F$, $G$ and $H$ are unknown functions that do not transform under rotations. Furthermore invariance under translations also gives
\begin{equation}
     \langle T(z, \bar{z}) \Theta(0,0) \rangle =  \langle \Theta(z, \bar{z}) T(0,0) \rangle.
\end{equation}

Comparing \eqref{relT} and \eqref{invrot} we get that the functions $F,G$ and $H$ must obey
\begin{equation}
    z \bar{z} F' + \frac{\pi}{2}( z \bar{z} G' - 3G) = 0, \qquad z \bar{z} G' - G + \frac{\pi}{2}( z \bar{z} H' - 2H) = 0.
    \label{FGHrel}
\end{equation}
We now define the C function by
\begin{equation}
    C \equiv 2F - 2\pi G - \frac{3\pi^2}{2}H.
    \label{defC}
\end{equation}
Taking a derivative and multiplying by $z\bar{z}$, a direct comparison with \eqref{FGHrel} gives
\begin{equation}
    z \bar{z} C'(z \bar{z}) = -3 \pi^2 H.
    \label{c_eq}
\end{equation}

In the UV CFT the OPE of the stress tensor takes the form
\begin{equation}
    \label{dOz}
    T(z)T(w) = \frac{c/2}{(z-w)^4} + 2 \frac{T(w)}{(z-w)^2} + \frac{ \partial T(w)}{z-w} + ... \quad,
\end{equation}
where the coefficient $c$ of the most singular term is called the central charge of the $2D$ CFT.

With this OPE in hands it is a simple task to compute the two point functions in \eqref{invrot} in the UV CFT, or equivalently in the short distance asymptotic regime $z \bar{z} \to 0$. We get
\begin{equation}
    F_{UV} = \frac{c}{2}, \qquad G_{UV}=H_{UV}=0,
    \label{F_UV}
\end{equation}
where we used $\Theta=0$ in the CFT, as follows from invariance under global scale transformations. Going back to Euclidean cartesian coordinates we therefore have 
 \begin{equation}
     \int_0^\infty dr^2 C'(r^2) = C(\infty)-C(0) = C_{IR}-C_{UV} = c_{IR} - c_{UV},
 \end{equation}
 where in the last equality we used the definition of $C$ \eqref{defC} and the expressions of $F$,$G$ and $H$ in the CFT \eqref{F_UV}. If we assume a massive theory the IR CFT is trivial and we have $c_{IR}=0$. We can therefore use \eqref{c_eq} and the last expression to get 
\begin{equation}
    c_{UV} = - \int_0^\infty dr^2 C'(r^2) = \int_0^\infty 2r dr 3 \pi^2 \frac{H(r^2)}{r^2} = 3\pi \int d^2x x^2 \langle \Theta(x) \Theta(0) \rangle.
    \label{256}
\end{equation}
Then using the Euclidean spectral representation we get
\begin{equation}
\begin{split}
    \int d^2x x^2 \langle \Theta(x) \Theta(0) \rangle &= \int_0^\infty ds \int d^2x x^2 \rho(s) \int \frac{d^2p}{(2\pi)^2} e^{i p \cdot x} \frac{1}{p^2+s}.
\end{split}
\end{equation}
The $x$ integral can be easily evaluated by using
\begin{equation}
    \int d^2x x^2 e^{ipx} = -(2\pi)^2 \nabla_p^2 \delta^{(2)}(p),
\end{equation}
and then we integrate by parts two times to perform the $p$ integral. We  get
\begin{equation}
    \int d^2p \frac{1}{p^2+s} \nabla_p^2 \delta^{(2)}(p) =  \nabla_p^2 \frac{1}{p^2+s} \bigg|_{p=0} = -\frac{4}{s^2}.
\end{equation}

This finally yields the sum-rule for the central charge of the UV CFT in terms of the spectral density of the trace of the stress energy tensor 
\begin{equation}
    c_{UV} = 12\pi \left( m^{-4} |\mathcal{F}_1^\Theta|^2 + \int_{4m^2}^\infty ds \frac{\rho(s)}{s^2} \right),
    \label{c_sum_rule}
\end{equation}
where we used \eqref{rho_parts} to integrate the delta function contribution to the spectral density.

\section{Normalization of the stress energy tensor form factors}
\label{sec:normF}

Here we derive a normalization condition that needs to be satisfied by $\mathcal{F}_2^\Theta(s)$.
We follow closely \cite{Karateev_2020}. Because of Lorentz invariance the most general expression for the 2 particle form factor of the full stress tensor is 
\begin{equation}
    \mathcal{F}^{T^{\mu \nu}}_2(p_1,p_2) = a_1 p^\mu p^\nu + a_2 q^\mu q^\nu + a_3 p^\mu q^\nu + a_4 p^\nu q^\mu + a_5 \eta^{\mu \nu},
\end{equation}
where $p \equiv p_1 + p_2$ and $q \equiv p_1-p_2$.
The symmetry condition $T^{\mu \nu} = T^{\nu \mu} $ gives $a_3 = a_4$ and  conservation gives
\begin{equation}
    \partial_\mu T^{\mu \nu} = 0 \implies (p_1+p_2)_\mu \mathcal{F}^{T^{\mu \nu}}_2(p_1,p_2) = 0,
\end{equation}
which yields $a_1 p^2 = -a_5$ and $a_3=0$. Therefore the most general form for the form factor of the stress tensor is 
\begin{equation}
    \label{generalFmunu}
    \mathcal{F}^{T^{\mu \nu}}_2(p_1,p_2) = A(s) ( p^\mu p^\nu - p^2 \eta^{\mu \nu}) + B(s) q^\mu q^\nu,
\end{equation}
where $A$ and $B$ are Lorentz invariant. However the two terms in \eqref{generalFmunu} are linearly dependent as it can be seen for example by going to the center of mass frame and writing the two terms explicitly. Therefore there is no loss of generality in setting $A(s)=0$, which leads to
\begin{equation}
 \mathcal{F}^{T^{\mu \nu}}_2(s) = B(s) q^\mu q^\nu, \qquad \mathcal{F}^{\Theta}_2(s) = (s-4m^2)B(s),
  \label{fnorm1}
\end{equation}
where we  used $q^2=s-4m^2$.

On the other hand the stress energy tensor is normalized when acting on one particle states such that 
\begin{equation}
    P^\mu \ket{ p} = \int dx T^{0\mu}(x) \ket{p} = p^\mu \ket{p}.
\end{equation}
Hence we get
\begin{equation}
\begin{split}
    \bra{p_1} P^\mu \ket{p_2} &= p_2^\mu (2\pi) 2E_{\bm{p}_2} \delta(\bm{p}_2-\bm{p}_1) \\
    &= \mathcal{F}^{T^{0\mu}}_2(p_1,-p_2) \int dx e^{ix\cdot(p_1-p_2)} =  \mathcal{F}^{T^{0\mu}}_2(p_1,-p_2) (2\pi)  \delta(\bm{p}_1-\bm{p}_2),
    \end{split}
    \label{expressions}
\end{equation}
where in the last equality the exponential with time components simply gives 1 because the particles have the same mass.

Analytic continuing \eqref{fnorm1} we have
\begin{equation}
\label{fTB}
    \mathcal{F}^{T^{0\mu}}_2(p_1,-p_2) = B(s-4m^2)(E_{\bm{p}_1}+E_{\bm{p}_2}) (p_1^\mu+p_2^\mu).
\end{equation}
Subtracting the expressions multiplying the spatial delta function in the first and second line of \eqref{expressions} we get
\begin{equation}
    \left( B(s-4m^2) (p_1^\mu +p_2^\mu) - p_2^\mu  \right) \delta(\bm{p}_1-\bm{p}_2) = 0,
\end{equation}
where we used \eqref{fTB}. Evaluating this for $\mu=0$ at $s=4m^2$ yields 
\begin{equation}
    B(s=0) = \frac{1}{2}.
    \label{fnorm2}
\end{equation}
Combining \eqref{fnorm1} and \eqref{fnorm2} we therefore derived the implication of the normalization of the stress tensor on the 2 particle form factor of the trace, which reads
\begin{equation}
    \mathcal{F}_2^\Theta(s=0) = -2m^2.
    \label{F2norm}
\end{equation}

\end{appendices}

\bibliographystyle{JHEP}
\bibliography{refs} 


\end{document}